\documentclass{aa}
\pdfoutput=1
\usepackage{times}
\usepackage{amsmath}
\usepackage{amssymb}
\usepackage{graphicx}
\usepackage[normal]{caption}
\usepackage[labelfont=bf]{caption}
\usepackage{subcaption}
\usepackage{tablefootnote}
\usepackage{natbib}
\usepackage{colortbl}
\usepackage[usenames,dvipsnames]{xcolor}
\usepackage{enumitem} 
\usepackage{makecell}
\usepackage{tabularx}
\bibpunct{(}{)}{;}{a}{}{,} 

\begin{document}

\title{\emph{HERschel}\thanks{{\it Herschel} is an ESA space observatory with science instruments provided by European-led Principal Investigator consortia and with important participation from NASA.} Observations of Edge-on Spirals (HEROES)}
\subtitle{II. Tilted-ring modelling of the atomic gas disks}
\author{F.~Allaert\inst{1}
\and G.~Gentile\inst{1,2}
\and M.~Baes\inst{1}
\and G.~De Geyter\inst{1}
\and T.M.~Hughes\inst{1,3}
\and F.~Lewis\inst{4,5}
\and S.~Bianchi\inst{6}
\and I.~De Looze\inst{1,7}
\and J.~Fritz\inst{8}
\and B. W.~Holwerda\inst{9}
\and J.~Verstappen\inst{10}
\and S.~Viaene\inst{1}}
\institute{Sterrenkundig Observatorium, Universiteit Gent, Krijgslaan 281, B-9000 Gent, Belgium\\ \email{Flor.Allaert@UGent.be}
\and Department of Physics and Astrophysics, Vrije Universiteit Brussel, Pleinlaan 2, 1050 Brussels, Belgium
\and Instituto de F\'{i}sica y Astronom\'{i}a, Universidad de Valpara\'{i}so, Avda. Gran Breta\~{n}a 1111, Valpara\'{i}so, Chile
\and Faulkes Telescope Project, Cardiff University, The Parade, Cardiff CF24 3AA, Cardiff, Wales
\and Astrophysics Research Institute, Liverpool John Moores University, IC2, Liverpool Science Park, 146 Brownlow Hill, Liverpool L3 5RF, UK
\and Osservatorio Astrofisico di Arcetri – INAF, Largo E. Fermi 5, 50125 Firenze, Italy
\and Institute of Astronomy, University of Cambridge, Madingley Road, Cambridge, CB3 0HA, UK
\and Centro de Radioastronom\'{i}a y Astrof\'{i}sica, CRyA, UNAM, Campus Morelia, A.P. 3-72, C.P. 58089, Michoac\'{a}n, Mexico 
\and University of Leiden, Sterrenwacht Leiden, Niels Bohrweg 2, NL-2333 CA Leiden, The Netherlands
\and Kapteyn Astronomical Institute, University of Groningen, Postbus 800, NL-9700 AV Groningen, the Netherlands
}
\date{}
\abstract{Edge-on galaxies can offer important insights in galaxy evolution as they are the only systems where the distribution of the different components can be studied both radially and vertically. The HEROES project was designed to investigate the interplay between the gas, dust, stars and dark matter (DM) in a sample of 7 massive edge-on spiral galaxies.}
{In this second HEROES paper we present an analysis of the atomic gas content of 6 out of 7 galaxies in our sample. The remaining galaxy was recently analysed according to the same strategy. The primary aim of this work is to constrain the surface density distribution, the rotation curve and the geometry of the gas disks in a homogeneous way. In addition we identify peculiar features and signs of recent interactions.}
{We construct detailed tilted-ring models of the atomic gas disks based on new GMRT 21-cm observations of NGC\,973 and UGC\,4277 and re-reduced archival H{\sc{i}} data of NGC\,5907, NGC\,5529, IC\,2531 and NGC\,4217. Potential degeneracies between different models are resolved by requiring a good agreement with the data in various representations of the data cubes.}
{From our modelling we find that all but one galaxy are warped along the major axis. In addition, we identify warps along the line of sight in three galaxies. A flaring gas layer is required to reproduce the data only for one galaxy, but (moderate) flares cannot be ruled for the other galaxies either. A coplanar ring-like structure is detected outside the main disk of NGC\,4217, which we suggest could be the remnant of a recent minor merger event. We also find evidence for a radial inflow of 15 $\pm$ 5 km s$^{-1}$ in the disk of NGC\,5529, which might be related to the ongoing interaction with two nearby companions. For NGC\,5907, the extended, asymmetric and strongly warped outer regions of the H{\sc{i}} disk also suggest a recent interaction. In contrast, the inner disks of these three galaxies (NGC\,4217, NGC\,5529 and NGC\,5907) show regular behaviour and seem largely unaffected by the interactions. Our models further support earlier claims of prominent spiral arms in the disks of IC\,2531 and NGC\,5529. Finally, we detect a dwarf companion galaxy at a projected distance of 36 kpc from the center of NGC\,973.}
{}
\maketitle

\section{Introduction}

Galaxies are dynamic, continually evolving systems in which the various components never cease to interact and influence each other's evolution. Primordial atomic gas condenses into clouds of molecular gas, which then cool and collapse to form stars. At very low metallicities, stars can even form directly from the atomic gas \citep[e.g.][]{krumholz12}. Metals are produced by nucleosynthesis in stellar cores and are injected into the ISM through stellar winds and supernova explosions. These metals then either mix with the interstellar gas or condense to form small dust grains through various processes \citep[see e.g.][]{dwekscalo80,kozasa91,dwek98,zhukovska13,zhukovska14}. On the other hand, dust grains are also continuously destroyed in the ISM, for example by supernova shock waves or processing by UV or X-ray photons, hot electrons or cosmic rays \citep[e.g.][]{jones04,micelotti10,jones11,bocchio12}. In their turn, dust grains stimulate star formation by catalysing the formation of H$_{2}$ molecules on their surface and shielding the molecular gas from the dissociating stellar UV radiation \citep{gould63,hollenbach71,cazaux02}. Dust grains are also efficient coolants of hot gas through inelastic collisions with thermal particles \citep[e.g.][]{dalgarno72,ostriker73}. In addition to this `closed-box' evolution, galaxies also accrete external gas through merger events or gas infall from the intergalactic medium \citep[e.g.][]{sancisi08,ocvirk08,keres09a,diteodoro14}, while supernovae and active galactic nuclei (AGN) can expel material from a galaxy \citep[e.g.][]{keres09b,alatalo11,lagos13,sharma13}. On top of this, according to the $\Lambda$CDM model, all these processes take place within a halo of dark matter \citep[see e.g.][]{zwicky37,trimble87,sahni05}, of which the nature and distribution, and the very existence, are still a matter of debate \citep[e.g.][]{milgrom83,deblok10,feng10,famaey12}. In order to uncover how galaxies form and evolve, it is therefore crucial to study all the components of a galaxy and the links between them, both locally and on global scales.
\par
Over the years many authors have investigated the connections between the different galaxy components. These studies, however, mostly targeted low to moderately inclined galaxies, where the radial distribution of the components can be derived easily and where structural features such as spiral arms or bars are clearly visible. Edge-on galaxies can offer an important complementary contribution to our understanding of galaxy evolution. Indeed, this is the only orientation where the vertical distribution of the different components can be studied. Moreover, due to line of sight (LOS) projection effects, the stars, gas and dust can be detected out to large galactocentric radii and low column densities. Another advantage of edge-on galaxies is that they allow the study of dust both in extinction and in emission. Indeed, in edge-on galaxies the dust often shows up as dark lanes in optical images. This opens up the possibility to also constrain the dust content by fitting detailed radiative transfer models to optical and NIR images \citep{xilouris97,xilouris99,bianchi07,baes11,delooze12,degeyter13,schechtman13,degeyter14}.
\par
Taking advantage of the unprecedented capabilities of the {\it Herschel} Space Observatory \citep{pilbratt10}, the {\it HERschel} Observations of Edge-on Spirals (HEROES) project was designed to map the distribution of the dust, gas, stars and dark matter in a sample of seven nearby, massive edge-on spiral galaxies and study the links between these components. This will extend the recent study of the ISM of NGC\,891 by \cite{hughes14} to a larger sample of galaxies. New {\it Herschel} observations of the HEROES galaxies and a qualitative and quantitative analysis of them are presented in Paper I of the project \citep{verstappen13}. 
\par
In this second HEROES paper, we present a detailed analysis of the atomic gas content of the galaxies in our sample, based on a combination of new and archival interferometric H{\sc{i}} 21-cm observations. This analysis is crucial to the HEROES project for different reasons. Indeed, apart from the amount and distribution of the atomic gas in the disk of a galaxy, interferometric H{\sc{i}} observations yield a great deal of additional information. As the H{\sc{i}} gas is generally the most extended (visible) component, past or ongoing galaxy interactions are often most clearly or even solely visible in the atomic gas, for example through large-scale warps, a disrupted H{\sc{i}} morphology, tidal structures or the presence of companion galaxies in the same velocity range \citep[e.g.][]{kamphuis92,yun94,sancisi99,holwerda11}. In the current work, NGC\,5529 is a nice example of this. In addition to this morphological information, interferometric H{\sc{i}} observations also allow to constrain the kinematics of a galaxy out to large radii, which in turn is necessary to derive the distribution of the dark matter. In order to unambiguously extract all the information that is available in the observations, we perform a detailed tilted-ring modelling of the full data cube of each galaxy in our sample. The gas disk of each galaxy is represented as a set of concentric rings characterised by a number of geometric and kinematic parameters, which are optimised to give the best possible agreement with the data cube. While time-consuming, this is the best strategy to resolve the degeneracies between different models of the atomic gas disk, which are especially present for an edge-on geometry.
\par
The H{\sc{i}} content in some of our target galaxies has already been analysed in varying degrees of detail in the past (see Section \ref{sec:results}). However, these studies often overlooked important information in the data cube or did not adapt their methods for an edge-on geometry. In order to obtain homogeneous and accurate results for our entire sample, we re-analyse these galaxies along with the others.
\par
This work is the second step in our effort to obtain a detailed overview of the distribution and the properties of the different constituents of the galaxies in the HEROES sample. As a following step, we are gathering CO line observations to also constrain the molecular gas component. Our team is currently also fitting radiative transfer models to a set of optical and near-infrared (NIR) images of each galaxy to constrain the distribution of the dust and the stars (Mosenkov et al., in prep). Taking into account the distributions of the gas and the stars, the rotation curves obtained in this work will then be modelled to derive the distribution of the dark matter (Allaert et al., in prep). Finally, bringing together our analyses of the various components, we will thoroughly investigate the connections between them, both radially and vertically.
\par
This paper is outlined as follows: in Section \ref{sec:obs} we describe the sample selection criteria, the observations and the data reduction. In Section \ref{sec:strategy} we explain the general modelling strategy and our method to estimate the uncertainties. In Section \ref{sec:results} we present the results of our modelling and a short discussion of the main features for each galaxy. In Section \ref{sec:flare} a constant scale height is weighed against a flared gas layer in the light of theoretical expectations. The effect on the gas disks of interactions with satellite galaxies is briefly discussed in Section \ref{sec:discussion} and, finally, our conclusions are presented in Section \ref{sec:conclusions}.

\section{Observations and data reduction}
\label{sec:obs}

\begin{table*}[t]
\caption{H{\sc{i}} observations}
\label{table:observations}
\centering
\begin{tabular}{l c c c c c c c}
\hline \hline\noalign{\smallskip}
Galaxy & Telescope & Array & Date & Bandwidth & Number of & Time on & Reference \\
 & & & & (MHz) & channels & source (h) & \\
\hline\noalign{\smallskip}
NGC\,973 & GMRT & & Aug 2011 & 4.2 & 256 & 5.7 & 1 \\
UGC\,4277 & GMRT & & Jul 2012 & 16.7 & 512 & 6.2 & 1 \\
IC\,2531 & ATCA & 6.0A & Apr 1996 & 8.0 & 512 & 9.9 & 2 \\
IC\,2531 & ATCA & 6.0B & Sep 1996 & 8.0 & 512 & 10.0 & 3 \\
IC\,2531 & ATCA & 1.5D & Mar 1997 & 8.0 & 512 & 9.1 & 3 \\
IC\,2531 & ATCA & 750A & Jan 2002 & 8.0 & 512 & 9.6 & 4 \\
NGC\,4217 & WSRT & Not mentioned & Aug 1990 & 5.0 & 63 & 12.0 & 5 \\
NGC\,5529 & WSRT & Maxi-Short & May 2001 & 10.0 & 128 & 11.9 & 3 \\
NGC\,5907 & VLA & Modified C & Aug 1997 & 2 $\times$ 3.1 & 2 $\times$ 32 & 4.7 & 6 \\
\hline
\end{tabular}
\tablebib{
(1) This work; (2) \cite{bureau97}; (3) \cite{kregel04}; (4) \cite{obrien10a}; (5) \cite{verheijen01}; (6) \cite{shang98}.
}
\end{table*}

\subsection{The sample}

The HEROES project targets a sample of seven massive, edge-on spiral galaxies. A detailed description of the sample selection is given in Paper I, so we restrict ourselves to a short summary here. One of the goals of the HEROES project is to map the dust emission in the FIR/submm with the \textit{Herschel} telescope and compare this to the predictions from detailed radiative transfer models fitted to optical/NIR images. To ensure sufficient spatial resolution in all the \textit{Herschel} bands, the galaxies were selected to have an optical diameter of at least 4$\arcmin$. They also need to have a well-defined and regular dust-lane to allow radiative transfer modelling. The latter criterion automatically rules out all galaxies with an inclination that deviates more than a few degrees from edge-on. Fitting radiative transfer models to optical/NIR images of galaxies was already done in the past by \cite{xilouris97, xilouris99} and \cite{bianchi07} with their respective codes. Their combined samples therefore formed the starting point for the HEROES sample. After applying our selection criteria, eight galaxies remained, of which NGC\,891 was omitted due to overlap with another \textit{Herschel} program. An overview of the general properties of the HEROES galaxies can be found in Table 1 of Paper I.
\par
Although part of the HEROES sample, NGC\,4013 will not be treated in this paper. The H{\sc{i}} content of this galaxy was recently extensively modelled by \cite{zschaechner15}, according to the same strategy as followed in this work. In summary, they find that the gas disk of NGC\,4013 has both a substantial edge-on warp and a LOS warp. In addition, the gas disk is flared, with the scale height increasing from 280 pc in the center to about 1 kpc at radii larger than 7 kpc. Finally they report a lag (decrease of rotational velocity with increasing scale height) that becomes more shallow with radius. No extraplanar H{\sc{i}} gas is found, despite the presence of extraplanar dust and diffuse ionised gas.

\subsection{Observations}

NGC\,973 and UGC\,4277 were observed with the Giant Metrewave Radio Telescope (GMRT) in August 2011 and July 2012, respectively. The GMRT consists of 30 45-m antennas located at a fixed position, yielding only one possible antenna configuration. A central area of approximately one square kilometer contains 14 antennas, with the shortest baselines measuring about 100 m. The other antennas are placed in a roughly Y-shaped configuration with three arms of about 14 km long, giving a maximum baseline of 25 km. In L-band (1000-1500 MHz), each GMRT antenna provides two orthogonal, linearly polarised signals. Both galaxies were observed for about 8 hours, including the observation of standard flux and phase calibrator sources. NGC\,973 was observed with a bandwidth of 4.2 MHz, divided into 256 channels. For UGC\,4277, a bandwidth of 16.7 MHz divided into 512 channels was used.
\par
For the remaining galaxies discussed in this paper, we used literature data from the Very Large Array (VLA), Westerbork Synthesis Radio Telescope (WSRT) and Australia Telescope Compact Array (ATCA). A brief overview of these observations is given in Table~\ref{table:observations}. We refer the reader to the original papers for further details of the data.
\par
The data for NGC\,5907 and NGC\,5529 were obtained from the VLA and WSRT archive respectively. For IC\,2531, the flagged and calibrated H{\sc{i}} data used in the series of papers by \cite{obrien10a,obrien10b,obrien10c,obrien10d} and in his own work \citep{peters13} were kindly made available by S. Peters. These consist of a combination of data sets obtained between 1992 and 2002 with different configurations of the ATCA. Upon inspection of the individual data sets, however, it became evident that two of them were severely corrupted, so we did not include them in our final data cube. Part of these data were also used by \cite{kregel04,kregel04b}. Finally, M. Verheijen was kind to provide a fully reduced and deconvolved data cube of NGC\,4217 from his WSRT observations \citep{verheijen01}.

\subsection{Data reduction}
\label{sec:datareduction}

\begin{table*}[t]
\caption{Parameters of the final data cubes used in the modelling}
\label{table:cubes}
\centering
\begin{tabular}{l c c c c c}
\hline \hline\noalign{\smallskip}
Galaxy & \multicolumn{2}{c}{Synthesised beam FWHM} & Velocity resolution & RMS noise & mJy/beam to Kelvin \\
 & (arcsec) & (kpc) & (km s$^{-1}$) & (mJy beam$^{-1}$) & (K) \\
\hline\noalign{\smallskip}
NGC\,973 & 24.5 $\times$ 15.3 & 7.5 $\times$ 4.7 & 6.9 & 1.6 & 1.62\\ 
UGC\,4277 & 15.6 $\times$ 9.8 & 5.8 $\times$ 3.6 & 6.9 & 1.0 & 3.96\\ 
IC\,2531 & 16.0 $\times$ 9.2 & 2.8 $\times$ 1.6 & 6.6 & 1.1 & 4.12\\ 
NGC\,4217 & 18.0 $\times$ 13.0 & 1.7 $\times$ 1.2 & 33.0 & 1.0 & 2.59\\ 
NGC\,5529 & 23.6 $\times$ 14.0 & 5.7 $\times$ 3.4 & 16.5 & 0.6 & 1.83\\ 
NGC\,5907 & 14.8 $\times$ 13.3 & 1.2 $\times$ 1.1 & 20.6 & 0.3 & 3.08\\ 
\hline
\end{tabular}
\end{table*}

The reduction of the GMRT and VLA data (NGC\,973, UGC\,4277 and NGC\,5907) was performed in the AIPS software package \citep{greisen03} following the standard procedure. To save time in handling the large data volumes at hand, the initial part of the reduction was performed on the `channel-0' data set (i.e. the average of the central 75$\%$ of the frequency channels). This data set was visually inspected and data corrupted by radio frequency interference, malfunctioning hardware, etc. were flagged. Next, the flux and phase calibration was performed for the channel-0 data. Calibration of VLA or GMRT data in AIPS is based on the observation of a flux calibrator at the start and the end of the observing run and periodic observations of a phase calibrator during the run. First, the observed flux calibrator was used to set the absolute flux scale. Next, the antenna gain amplitude and phase solutions were determined for both calibrator sources and the absolute flux information from the primary calibrator was transferred to the secondary calibrator. Finally, the antenna gain solutions from the calibrators were interpolated to obtain the antenna gains for the target source. After the initial reduction of the channel-0 data, the antenna gains and bad data flags were copied to the spectral line data, which were finally bandpass calibrated. Typically several iterations of flagging and calibrating were needed to achieve a satisfactory result.
\par
The WSRT data of NGC\,5529 were reduced with the Miriad software package \citep{sault95}. Although the global strategy of the data reduction is the same as in AIPS (flagging, calibrating and imaging the data), the exact calibration procedure is somewhat different. The observations of NGC\,5529 did not include any  periodic observations of a phase calibrator. Only a flux calibrator source was observed at the start and the end of the observing run. This was used to set the absolute flux scale and obtain initial antenna gain solutions, as well as to determine the bandpass corrections to be applied. To take into account the strong variations of the antenna gain phases during the observing run, the quality of the initial calibration was improved through several iterations of self-calibration. This was performed on the continuum emission to increase the signal-to-noise ratio and the derived antenna gains were subsequently copied to the spectral line data.
\par
In order to suppress potential Gibbs ringing in the spectra, a data set is usually Hanning smoothed. For NGC\,4217 this step was already performed online, i.e. during the observations. For the remaining galaxies this was not the case and the data had to be Hanning smoothed manually during the data reduction. The inevitable downside of this smoothing, however, is that adjacent channels are no longer independent. One usually accounts for this by dropping every second channel, thereby reducing the spectral resolution of the final data cube by a factor of 2. The purpose of our analysis is to fit 3D models to the H{\sc{i}} data cubes of the galaxies in our sample. This requires a good constraint on their kinematics and hence a sufficiently high velocity resolution, preferably below 20 km s$^{-1}$. For NGC\,5907 and NGC\,5529, Hanning smoothing would lead to a resolution of 41 km s$^{-1}$ and 33 km s$^{-1}$ respectively. We therefore decided to perform Hanning smoothing only for NGC\,973, UGC\,4277 and IC\,2531 and not for NGC\,5529 and NGC\,5907. For NGC\,4217 the resolution is also 33 km s$^{-1}$, but unfortunately nothing could be done about this, since Hanning smoothing was already applied during the observations.
\par
After subtracting the continuum in the {\it uv} plane, the data were imaged using a robust parameter of 0. This provides a good compromise between a high resolution and a high signal-to-noise ratio. The maximum baseline of the GMRT is approximately 25 km, which corresponds to a maximum {\it uv} distance of 119 k$\lambda$ at a wavelength of 21 cm. Using this entire {\it uv} range to image the data results in a very narrow synthesised beam and produces a data cube where the emission is effectively resolved out. To image the data of NGC\,973, the {\it uv} range was therefore limited to 40 k$\lambda$ and a Gaussian taper with a width of 10 k$\lambda$ at the 30 $\%$ level was applied. For UGC\,4277, a better result was obtained by limiting the {\it uv} range to 15 k$\lambda$ and not applying any taper.
\par
The deconvolution of the dirty beam was generally done using the CLEAN task in Miriad. This task offers a choice between the H\"ogbom \citep{hogbom74}, Clark \citep{clark80} and Steer \citep{steer84} Clean algorithms and by default determines by itself which is the most appropriate algorithm to use. After an initial Clean of the data cube down to 3 times the rms noise ($\sigma$), the data cube was smoothed to a beam size of 60$\arcsec$ and 2$\sigma$ masks were defined in the resulting cube. The original data cube was subsequently Cleaned down to 1$\sigma$ in the masked regions and the resulting CLEAN model was then used as input model for a final Cleaning of the original cube down to 3$\sigma$. Finally, the Clean components were convolved with a Gaussian beam fitted to the central lobe of the synthesised beam and restored to the residual image. A different strategy was used for the deconvolution of the cube of NGC\,5907, the closest and by far the largest (in terms of angular size) galaxy in our sample. Ordinary Clean algorithms typically represent a source by a collection of point sources, which is a poor approximation for extended sources like NGC\,5907. To overcome this problem we used the multi-scale Cleaning option in AIPS \citep{greisen09}, with an enhanced version by Bill Cotton and Fred Schwab \citep{schwab84} of the Clark Clean algorithm. Multi-scale Cleaning essentially uses a set of circular Gaussians as model components instead of point sources. In the current case three of these Gaussian components, with widths of 40$\arcsec$, 120$\arcsec$ and 360$\arcsec$, were used in combination with a point source component and each channel map was Cleaned down to 2$\sigma$.
\par
Finally, each data cube was corrected for the primary beam and transformed to barycentric radio velocities. An overview of the properties of the final data cubes is presented in Table \ref{table:cubes}.
\par
To produce total H{\sc{i}} maps of our galaxies, we first smoothed each cube to a resolution of twice the original (Clean) beam size. Masks were then generally defined in the smoothed cubes by excluding all emission that does not exceed 2$\sigma$ in three or more consecutive channels. If the velocity resolution of a data cube is low, however, three consecutive channels can already span a large velocity range. For NGC\,5529, NGC\,5907 and especially NGC\,4217, the latter criterion was therefore somewhat relaxed. After applying the masks to the original data cubes, their zeroth moment was calculated. 
\par
Following standard practice, we assume that the H{\sc{i}} 21 cm line is completely optically thin. This means that the observed 21 cm surface brightnesses can be translated directly into atomic gas column densities. In fact, it has been known for a fairly long time that this assumption does not always hold in the ISM of a galaxy. Observations of the absorption of continuum background emission by clouds of atomic hydrogen in the Milky Way have shown that H{\sc{i}} gas can indeed become self-opaque at 21 cm \citep[e.g.][]{hagen54, radhak60, gibson05}. The most detailed study of H{\sc{i}} self-opacity to date was carried out by \cite{braun09} and \cite{braun12}. They analysed high-resolution H{\sc{i}} data cubes of M31, M33 and the LMC and suggested that these galaxies contain a population of dense atomic hydrogen clouds that are self-opaque in the 21 cm line. These structures are compact, with typical sizes of the order of 100 pc. Correction factors for the total atomic gas mass of 1.30, 1.36 and 1.33 were found for M31, M33 and the LMC respectively. However, since self-opacity occurs on such small linear scales, it is very difficult to correct for this effect in a meaningful way in more distant galaxies (observed with lower physical resolution). Studies of the atomic gas content of galaxies, both for moderately inclined systems and edge-ons, therefore generally assume that the 21 cm line is optically thin.
\par
An important question to ask now is what the consequences of this assumption are in an edge-on galaxy, where each line of sight cuts through the entire plane of the disk. Fortunately, for self-absorption to be effective, H{\sc{i}} clouds must not only overlap in projected position, but also in radial velocity. Due to the rotation of a galaxy, the latter is generally not the case along a line of sight and the problem is in fact reduced to the level of the individual channel maps. Hence the effects of assuming an optically thin 21 cm line for edge-on galaxies will not be much larger than for studies of low or moderately inclined galaxies.

\section{Modelling strategy}
\label{sec:strategy}

\subsection{General strategy}

To model our H{\sc{i}} data cubes, we made use of the tilted-ring fitting software TiRiFiC \citep{jozsa07}. For each galaxy, the width of the rings making up the models was taken to be half the size of the minor beam axis, as a compromise between angular resolution and efficiency. Initial guesses for the relevant parameters were derived as follows. The inclination was taken over from the literature. The (central) position angle and systemic velocity were estimated by eye from the observed data cube. With these values as inputs, we then used the task {\it rotcur} from the GIPSY software package \citep{vanderhulst92} to perform a tilted-ring modelling of the observed velocity field. This technique is actually not well-suited for an edge-on galaxy and yielded rough initial estimates of the central postion and rotation curve. The radial surface density profile was derived with the GIPSY task {\it radial}, which essentially applies the method described by \cite{warmels88} and \cite{lucy74}. An initial value of 0.3 kpc was used for the vertical scale height and finally the velocity dispersion was estimated as the squared sum of the velocity resolution of the observations and an intrinsic velocity dispersion of 10 km s$^{-1}$ \citep{leroy08}. For NGC\,4217, the results from the analysis by \cite{verheijen01} were used as initial estimates. All models use a sech$^2$ distribution for the vertical gas distribution. This profile was originally devised for stellar disks \citep{spitzer42}, but is also often used to model gas disks \citep[see e.g.][]{moster10,gentile13}.
\par
These initial estimates were used as input to perform a first TiRiFiC fit of the data cube. A very basic geometry was imposed here, consisting of a one-component disk with a constant scale height and inclination across the entire disk. The first fit was performed with the central position, the systemic velocity, the velocity dispersion and the position angle as free global parameters, and the rotational velocity free for each ring. For warped galaxies, the position angles were then adjusted to match the warp. Subsequently, the results were used as input in a second fit, this time leaving the inclination and vertical scale height free as global parameters and fitting the surface brightness for each ring. The resulting basic model was then further refined by investigating different geometries and gradually adding extra features, each time optimising the column densities and rotational velocities (if necessary) to obtain the best possible fit to the data. The approaching and receding halves were modelled separately, with the (obvious) constraint that the central position, the systemic velocity and the velocity dispersion should be the same for both sides. In the basic geometry the inclination and scale height were also required to be the same on both sides. The detailed strategy followed for each galaxy depends on the specific features encountered and is described in the following sections.
\par
An automated $\chi^2$ minimisation is included in the TiRiFiC code, which is performed after convolving the model cube with the observational beam. Although this greatly expedites the optimisation of the parameters, its results should not be trusted blindly. With a constant noise across the maps, by construction this minimisation scheme will assign more weight to regions of higher surface brightness. However, low surface brightness features often impose the strongest constraints and are most important to discriminate between models. Also, TiRiFiC does not take into account whether a model is realistic or not. The theoretically best fitting model will often contain unphysical values and unrealistically strong radial variations of the parameters. Therefore, the geometry of the models was generally imposed by hand, and the parameter values obtained from a TiRiFiC fit were further refined by manual adjustment.
\par
The edge-on orientation of the galaxies in this work brings along a number of potential degeneracies between different models. A prime example of this is the vertical thickness of the gas disks, as seen in projection on the plane of the sky. This can be modelled by adjusting either the vertical scale height or the inclination of the disk, or a combination of both. A second important example is the interpretation of the column density peaks observed in a zeroth moment map, where each line of sight contains the sum of multiple rings at different galactocentric radii. In order to resolve these degeneracies, it is important to use all the information provided by the observations and to compare the models to the data in different projections of the data cube. Scale height and inclination effects, for example, can be disentangled by studying the shape of the emission contours in the different channel maps. Increasing the scale height simply makes the emission thicker in all the channel maps. On the other hand, decreasing the inclination moves the emission away from the major axis and gives it a V-like shape in the channel maps (see e.g. Fig. \ref{fig:N5907_models_channels} later in this work for an example of this). The strength of this effect also strongly varies within the data cube, going from hardly visible at high line of sight velocities to clearly visible close to the systemic velocity. Density peaks in a total H{\sc{i}} map can be understood by inspecting the associated major axis position-velocity diagram. The addition of the kinematical information allows to assign the correct surface density to each ring and potentially identify additional features that do not belong to the axisymmetric disk (see e.g. Figs. \ref{fig:N5529_models_mom0} and \ref{fig:N5529_pv_models} later in this work).

\subsection{Uncertainties}

Following standard practice, the modelling strategy that we use in this work is a combination of $\chi^2$ minimisation and the personal judgement of the modeller \citep[e.g.][]{zschaechner12,gentile13,kamphuis13}. As a consequence it is not possible to determine the errors on the parameters of the final models in a statistical way. We therefore estimated the uncertainty on each parameter by keeping all the other parameters fixed to their best fit values and varying the parameter in question until the model clearly deviated from the observed cube. For each parameter, there are usually certain regions of the disk and/or certain projections of the data that are especially sensitive to variations of that parameter. We focussed on these regions and projections to assess the deviation of the model from the observed cube. Errors on the central position and the position angle were determined based on the moment-0 maps. For the inclination and scale height we focussed both on the moment-0 maps and on the inner and outer channel maps, respectively. For the systemic velocity we focussed on the central part of the major axis position-velocity (XV) diagrams and the uncertainties on the velocity dispersion were determined from the spacing between the contours on the terminal edges of the XV-diagrams. Finally, errors on the rotation curve were determined as the maximum of two values: half of the difference between the approaching and the receding side, and the channel width of the data cube divided by the sine of the inclination.

\section{Modelling results}
\label{sec:results}

\begin{figure}[t]
  \resizebox{\hsize}{!}{\includegraphics{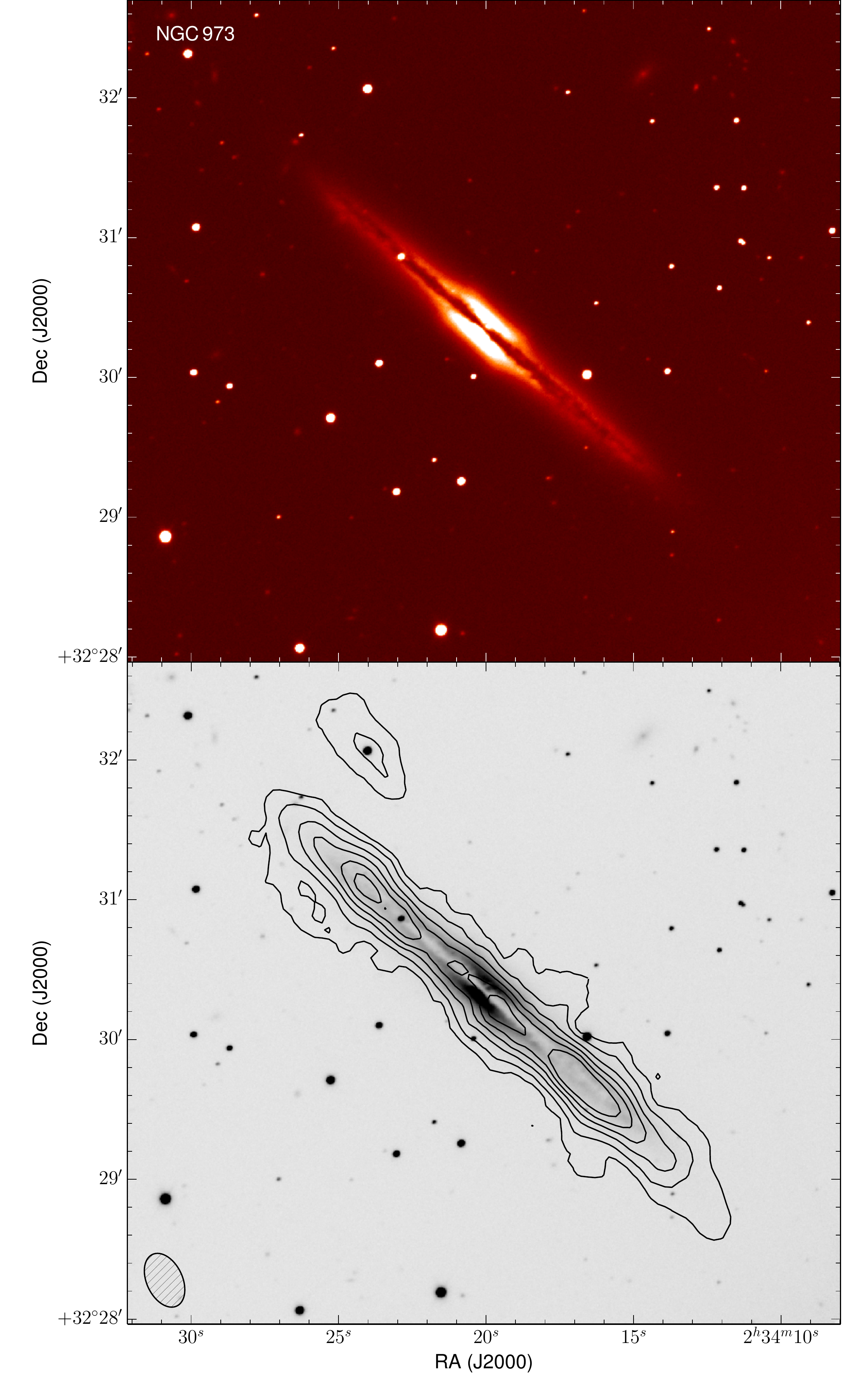}}
  \caption{Top: False colour V-band image of NGC\,973 based on observations made with the Italian Telescopio Nazionale Galileo (TNG) operated on the island of La Palma by the Fundación Galileo Galilei of the INAF (Istituto Nazionale di Astrofisica) at the Spanish Observatorio del Roque de los Muchachos of the Instituto de Astrofisica de Canarias. The data were cleaned and combined using standard data reduction techniques. Bottom: H{\sc{i}} contours overlaid on the same V-band image. The contours start at $1.76 \times 10^{20}$ atoms cm$^{-2}$ and increase as 3, 6, 9, 12 and 15 times this value. The H{\sc{i}} beam is shown in the bottom-left corner. North is up, East is to the left.}
  \label{fig:N973_V_mom0}
\end{figure}

\subsection{NGC\,973}

NGC\,973 (UGC\,2048) is located at a distance of 63.5 Mpc\footnote{The distances we use here are taken from NED, as explained in section 2.1 of Paper I.}. It is classified as an Sb or Sbc galaxy and has a boxy/peanut-shaped bulge (\citealt{lutticke00} classify it as a `bulge which is on one side boxy and on the other peanut-shaped'). After subtracting an axisymmetric model from their I-band image of NGC\,973, \cite{patsis06} find an X-shaped pattern in the central regions which they interpret as evidence for the presence of a bar. So far, the structure and kinematics of the atomic gas disk of NGC\,973 have never been studied.

\subsubsection{The data}

Our GMRT data of NGC\,973 have a total integration time of 5.7 hours, which is rather limited for a galaxy at a distance of 63.5 Mpc. Consequently the signal-to-noise ratio in the channel maps of our data cube is generally on the low side. As a result, the shape of the observed emission is rather irregular and sometimes varies significantly from one channel map to the other. Unfortunately not much can be done to improve this. We therefore do not expect our models to nicely reproduce the shape of the emission in each channel map.
\par
The zeroth moment map of the H{\sc{i}} disk of NGC\,973 is compared to a V-band image in Fig. \ref{fig:N973_V_mom0}. The NE side of the disk is the approaching half, the SW side is the receding half. Both the stellar and the H{\sc{i}} disk are remarkably flat, even in the outer regions, but some asymmetry is visible at high contour levels in the total H{\sc{i}} map. This is also visible in Fig. \ref{fig:N973_models_mom0}. Indeed, on the approaching side only one density peak is observed (at a major axis offset of +1$\arcmin$ in Fig. \ref{fig:N973_models_mom0}). This peak is mirrored on the receding side, at an offset of -1$\arcmin$, but additionally a second, more central peak is also present on this side. The maximum (integrated) surface density is also significantly higher on the receding side.
\par
We further report the detection of a small satellite galaxy slightly above the NE edge of the disk, at a distance of 36 kpc from the center of NGC\,973 (as determined from our models). We estimate its central position as RA 02\textsuperscript{h}34\textsuperscript{m}23.9\textsuperscript{s}, Dec 32$\degr$32$\arcmin$5.0$\arcsec$ and find no optical counterpart. The satellite appears in 9 consecutive (independent) channels in the data cube, between 4517 and 4572 km s$^{-1}$, although often only at the 3$\sigma$ level, and has a total H{\sc{i}} flux of 0.36 Jy km s$^{-1}$. This translates to a H{\sc{i}} mass of $3.39\times 10^{8}$ M$_{\odot}$. 
\par
For the disk of NGC\,973 we find an integrated H{\sc{i}} flux of 9.32 Jy km s$^{-1}$, corresponding to a total atomic hydrogen mass of $8.87 \times 10^{9}$ M$_{\odot}$. \cite{huchtmeier89} report a single dish flux of 8.83 Jy km s$^{-1}$ while \cite{springob05} give a corrected\footnote{Corrected for pointing offsets and source extent, not H{\sc{i}} self-absorption.} single dish flux of 13.9 Jy km s$^{-1}$. Although these flux measurements are of the same order, they show a relatively large spread, which is caused by a combination of different effects. The dominant factor, in general, is the uncertainty in the flux calibration of the different observations, which can cause differences of up to a few ten percent. Secondly, given the limited observing time, a significant reservoir of low column density gas might still be hidden in the noise in our channel maps. In addition, single dish observations often have a half power beam width (HPBW) that is smaller than the HI disk of the galaxy, causing them to miss some of the emission from the gas. To correct for this effect, \cite{springob05} assume a smooth Gaussian distribution for the HI, scaled according to the optical radius. Given the variety of morphologies shown by the atomic gas disks of galaxies, such a general correction can also be rather inaccurate for a specific galaxy.

\subsubsection{Models}
\label{sec:N973_models}

\begin{figure}[]
  \resizebox{\hsize}{!}{\includegraphics{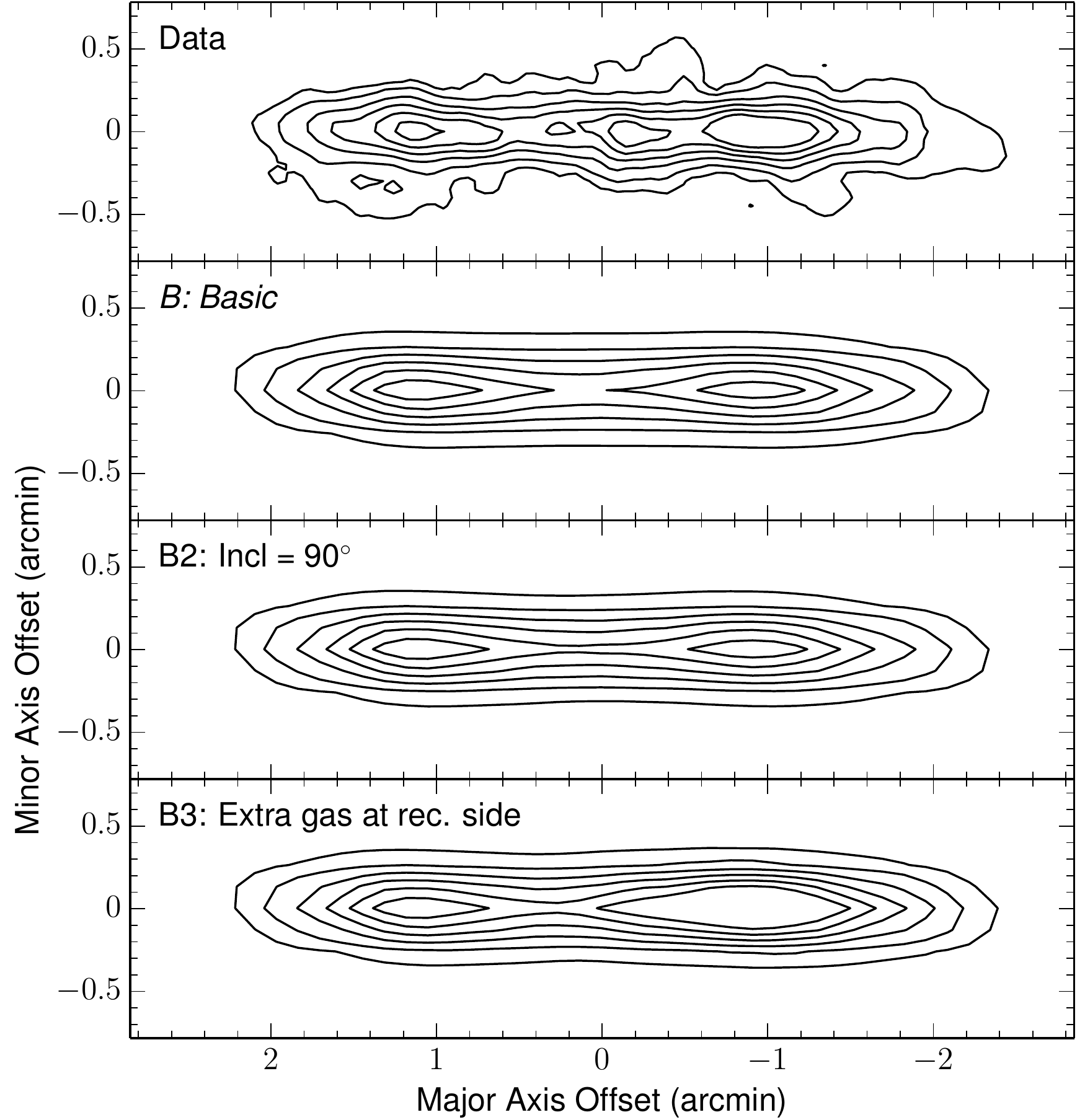}}
  \caption{Total H{\sc{i}} maps of the various models discussed here as compared to the observed total H{\sc{i}} map of NGC\,973. The \textit{B} model is our best model. Contour levels are the same is in Fig. \ref{fig:N973_V_mom0}.}
  \label{fig:N973_models_mom0}
\end{figure}

\begin{figure*}[]
\centering
   \includegraphics[width=0.825\textwidth]{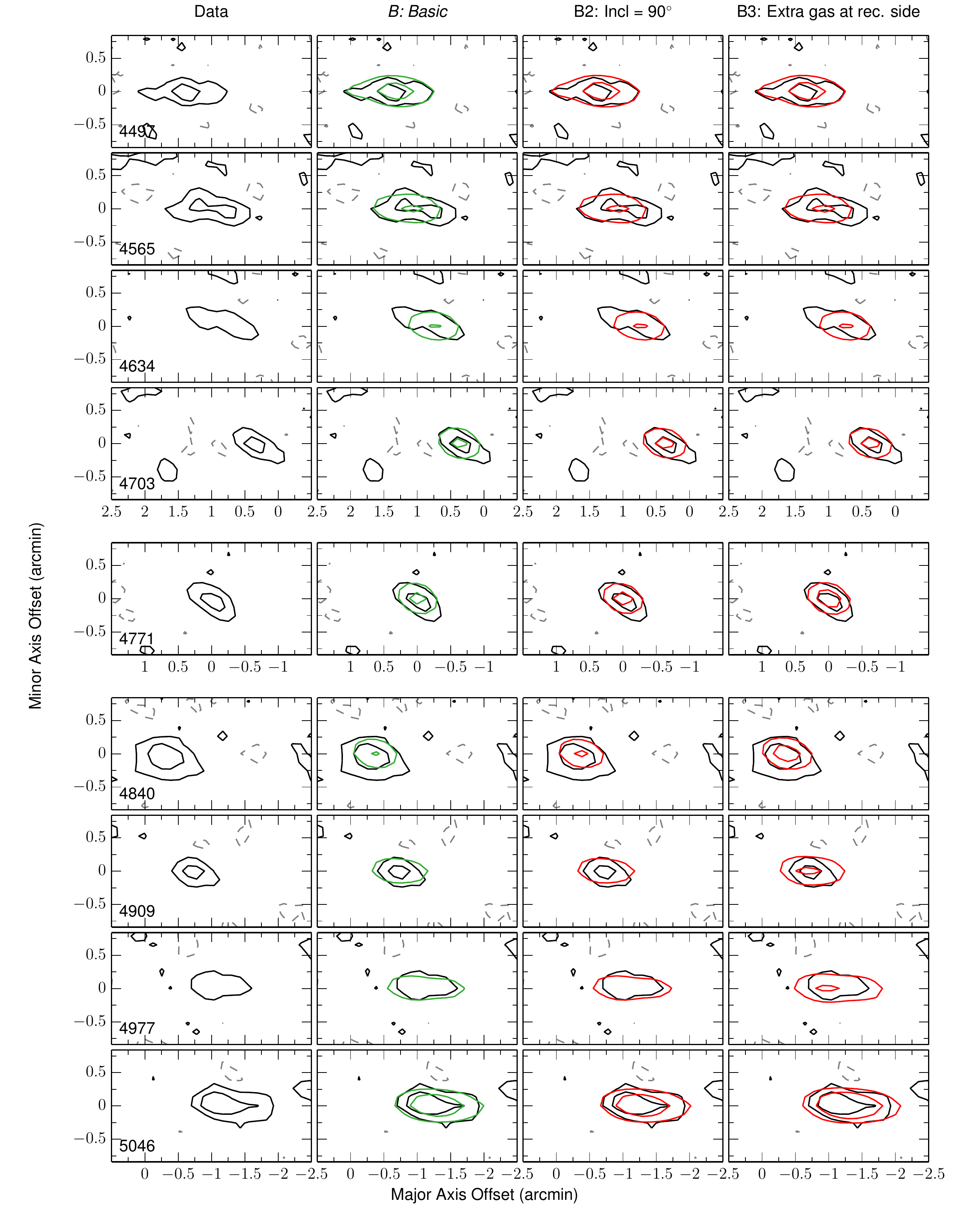}
     \caption{Representative channel maps from the observed data cube of NGC\,973 and the various models. Contour levels are -2.3, 2.3 (1.5$\sigma$) and 6.2 mJy beam$^{-1}$. The observations are shown as black contours, with the negative contours as dashed grey. The green contours show the final (\textit{B}) model. Other models are shown as red contours. The line of sight velocity is indicated in km s$^{-1}$ in the bottom left corner of the first panel of each row. The systemic velocity is 4770 $\pm$ 8 km s$^{-1}$. Note that the horizontal axis is slightly shifted between the three blocks composing the figure.}
     \label{fig:N973_models_channels}
\end{figure*}

\begin{figure*}[]
\centering
   \includegraphics[width=\textwidth]{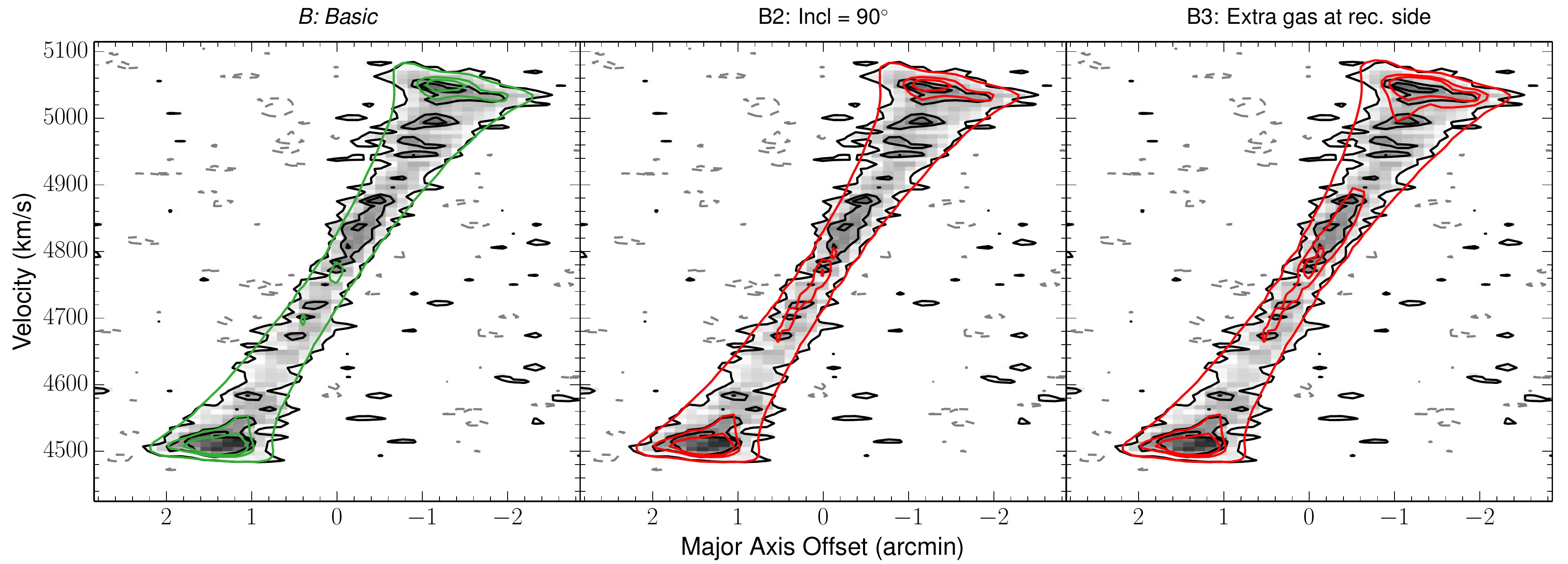}
     \caption{Observed major axis position-velocity diagram of NGC\,973 (black and gray contours) overlaid with different models. The green contours show the final (\textit{B}) model. Other models are shown as red contours. Contour levels are -2.3, 2.3 (1.5$\sigma$), 7.0 and 9.3 mJy beam$^{-1}$. The greyscale corresponds to the observations.}
     \label{fig:N973_pv_models}
\end{figure*}

For the modelling of the data cube we focus only on the H{\sc{i}} disk of NGC\,973. The satellite galaxy is not included in the models. The best fitting model in the basic geometry (which we will refer to as the basic or \emph{B} model) has a constant inclination of 88$\degr$ and scale height of 0.2 kpc, although the latter is very ill constrained, with values up to 1.1 kpc giving only marginally different models. A single, constant position angle of 228$\degr$ was found to be sufficient to reproduce the orientation of the disk on the plane of the sky. If we compare the zeroth moment map of this model to the observations (Fig. \ref{fig:N973_models_mom0}) we see that the agreement is generally very good, apart from the observed central density peak on the receding side which is not present in the model. The main trend in the channel maps is also well reproduced by the model (Fig. \ref{fig:N973_models_channels}), although (as expected) it often does not match the exact shape of the observed emission. Additionally, at intermediate rotational velocities on the receding side the model often underestimates the surface densities and hence lacks the high surface brightness contours. This effect can, for example, be seen in the channel maps at 4565, 4703, 4840 and 4909 km s$^{-1}$ and is also visible in the major axis position-velocity (XV) diagram (Fig. \ref{fig:N973_pv_models}). The latter is well modelled at the terminal sides (i.e. at the maximal velocities) and in the very central region, but lacks high surface brightness emission at intermediate velocities, especially on the receding side. In a regularly rotating, axisymmetric disk that is seen edge-on, the emission that is seen at these intermediate velocities is simply the projection along the major axis of the emission at the terminal velocities. Hence, the only ways to significantly enhance this emission are either to increase the inclination of the disk or to increase the surface brightness in part or all of the rings making up the model.
\par
The first option was tested by constructing a second basic model, the \emph{B2} model, with a constant inclination of 90$\degr$. The corresponding optimal scale height is 0.4 kpc. With the surface brightnesses optimised to match the terminal sides of the XV-diagram, this model offers little improvement on the receding side (Fig. \ref{fig:N973_pv_models}). On the approaching side the problem is almost solved, but at the cost of now overestimating the emission close to the systemic velocity.
\par
In a second attempt to improve our basic model we manually scaled up the surface brightness distribution on the receding half of the \emph{B2} model until we could reproduce the high contours at intermediate velocities in Fig. \ref{fig:N973_pv_models}. With an increase of 35 percent (\emph{B3} model) this is almost achieved, but the XV-diagram is now clearly too wide on the receding side. This effect can also be seen in the channel maps at 4909, 4977 and 5046 km s$^{-1}$ in Fig. \ref{fig:N973_models_channels}, where the left and right edges of the model extend well beyond those of the observed emission. In addition, the emission at the terminal and systemic velocities in the XV-diagram is strongly overestimated (note that the \emph{B3} model includes a 6$\sigma$ contour at the systemic velocity in Fig. \ref{fig:N973_pv_models}, while the observed XV-diagram does not). In the total H{\sc{i}} map (Fig. \ref{fig:N973_models_mom0}) the \emph{B3} model overestimates the peak density on the receding side and shows one big peak instead of two separate peaks.
\par
From the \emph{B2} and \emph{B3} models it is clear that the H{\sc{i}} disk of NGC\,973 cannot simply be modelled as an axisymmetric disk. Either we underestimate the amount of gas moving at intermediate velocities or we overestimate the amount of gas moving at the terminal velocities. The reason for this is most probably that the disk of NGC\,973 is far from smooth and has a clumpy structure with high-density regions alternated by regions with a lower density. Alternatively, a projected spiral arm might also cause this behaviour. In any case, although TiRiFiC allows to model different areas of a gas disk individually and to include spiral arms, the problem at hand, in an edge-on disk, is much too degenerate to attempt such a thing. In addition, apart from the arbitrary jumps of the observed emission in the channel maps, we do not see a clear trend indicating that the shape of the model emission is systematically wrong. Hence, we stick to the basic geometry and use the \emph{B} model as our final model for NGC\,973. We stress that this uncertainty concerning the exact morphology of the disk does not strongly affect our results and is primarily restricted to the geometric parameters, i.e. the inclination and the scale height. Slight adjustments of the surface density profile can be necessary when adding a flare or a line of sight warp to a model, but these are typically only of the order of 10$\%$ or less. In principle a LOS warp also affects the rotational velocities, but for small warps of only a few degrees (as would be the case here), these changes are negligible. 
\par
Fig. \ref{fig:N973_final_channels} of the appendix compares the model data cube to the observations. The satellite galaxy is indicated by arrows in the relevant channel maps. The main parameters describing the final model are given in Table \ref{tab:final_params} and Fig. \ref{fig:final_params}. The inclination and the position angle have errors of 2.0$\degr$ and 3.0$\degr$ respectively. The scale height has a value of 0.2$^{+0.9}_{-0.2}$ kpc. We note that this result seems to be consistent with a scale height of 0 kpc, which is of course unphysical. Unfortunately, due to the large observational beam it is not possible to determine a meaningful lower limit for the scale height of NGC\,973.

\subsection{UGC\,4277}

With a distance of 76.5 Mpc, UGC\,4277 is the most distant of the HEROES galaxies. It is classified as an Sc or Scd galaxy and, like NGC\,973, the details of its atomic gas content have not been investigated before.

\subsubsection{The data}

\begin{figure}[h]
  \resizebox{\hsize}{!}{\includegraphics{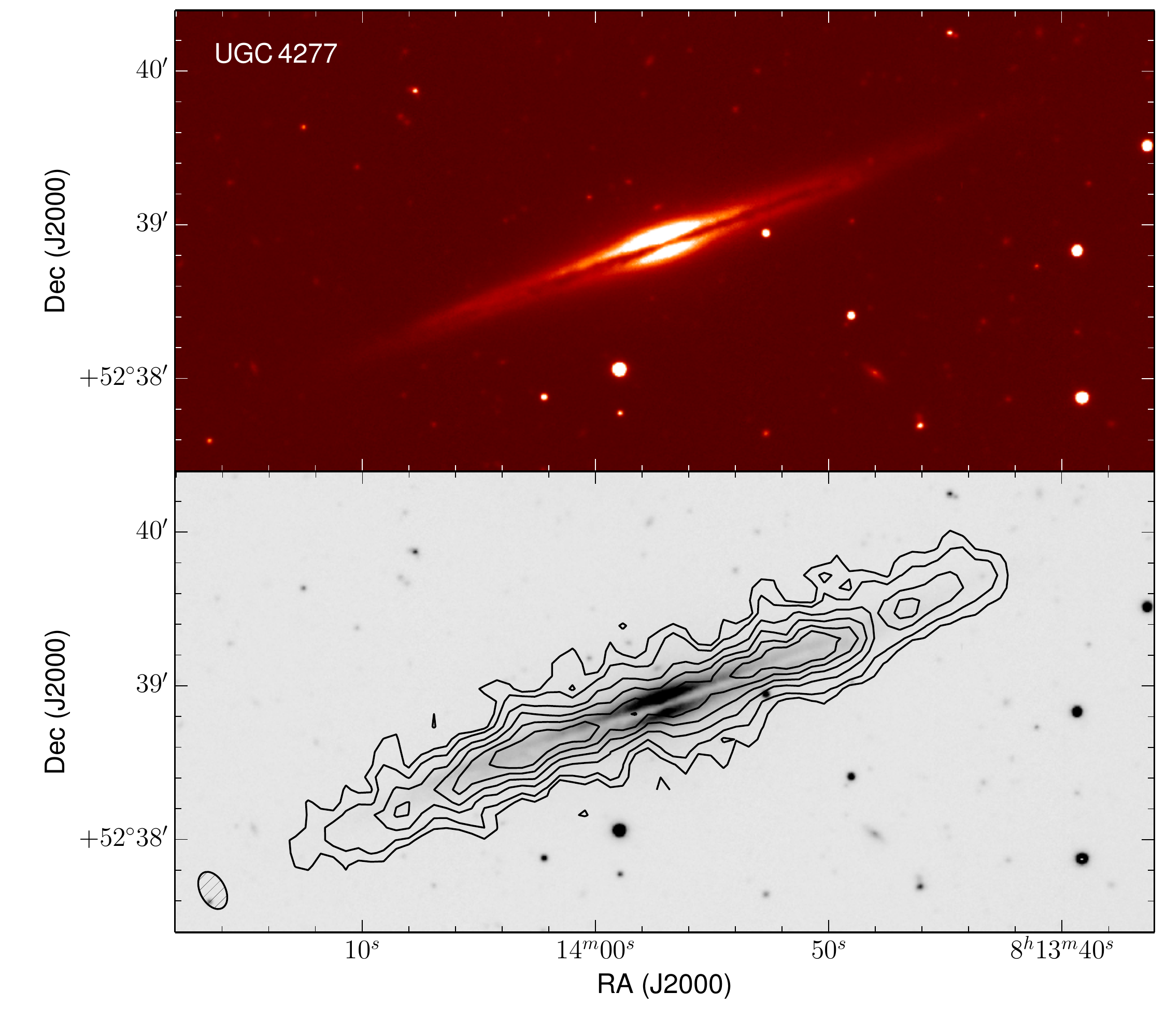}}
  \caption{Top: False colour TNG V-band image of UGC\,4277 from \cite{bianchi07}. Bottom: H{\sc{i}} contours overlaid on the same V-band image. Contours start at $3.61 \times 10^{20}$ atoms cm$^{-2}$ and increase as 2, 4, 6, 8 and 10 times this value. North is up, East is to the left.}
  \label{fig:U4277_V_mom0}
\end{figure}

As can be seen in Fig. \ref{fig:U4277_V_mom0}, both the stellar disk and the atomic gas disk of UGC\,4277 are slightly warped. Similarly to the case of NGC\,973, the limited observing time results in a rather low signal-to-noise ratio in our data cube, making the individual channel maps quite noisy and the shape of the emission quite irregular (see Fig. \ref{fig:U4277_final_channels} of the appendix). This is also reflected in the major axis position-velocity diagram (see Fig. \ref{fig:U4277_pv_models} later), where chunks of emission seem to be missing. However, the higher brightness contours, which should be less affected by the noise, are also very irregular. This suggests that UGC\,4277 has a rather clumpy H{\sc{i}} disk. Similarly, the FIR morphology, especially in the PACS bands, suggests that the dust disk of UGC\,4277 is also rather irregular \citep{verstappen13}.
\par 
From our zeroth moment map we measure a total H{\sc{i}} flux of 14.7 Jy km s$^{-1}$, corresponding to an atomic hydrogen mass of $2.03 \times 10^{10}$ M$_{\odot}$. \cite{huchtmeier89} report a single dish flux of 17.9 Jy km s$^{-1}$, which was corrected to 18.9 Jy km s$^{-1}$ by \cite{springob05}.

\subsubsection{Models}
\label{sec:U4277_models}

\begin{figure}[]
  \resizebox{\hsize}{!}{\includegraphics{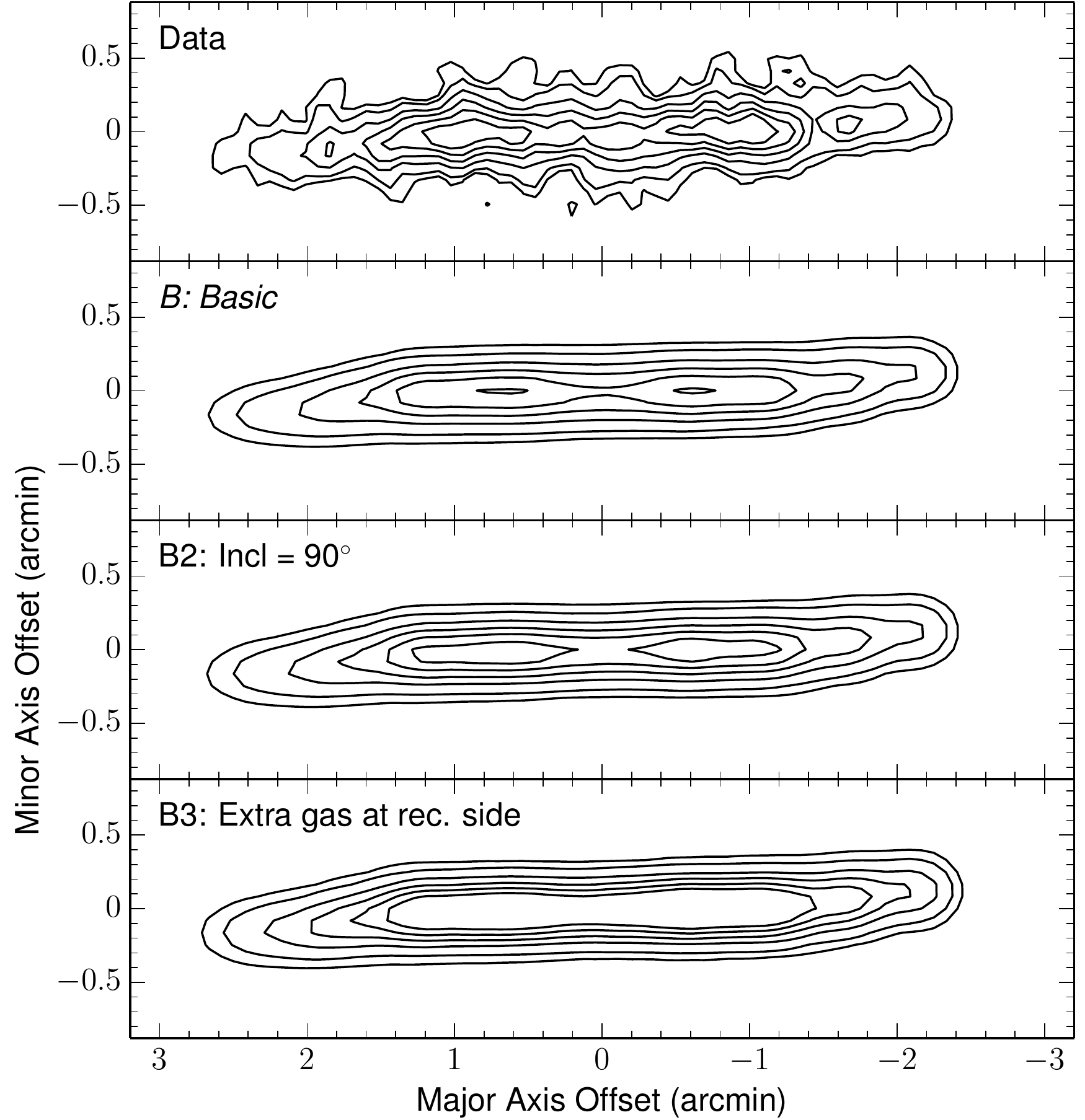}}
  \caption{Total H{\sc{i}} maps of the various models discussed here as compared to the observed total H{\sc{i}} map of UGC\,4277. The \textit{B} model is our best model. Contour levels are the same as in Fig. \ref{fig:U4277_V_mom0}.}
  \label{fig:U4277_models_mom0}
\end{figure}

\begin{figure*}[]
\centering
   \includegraphics[width=0.825\textwidth]{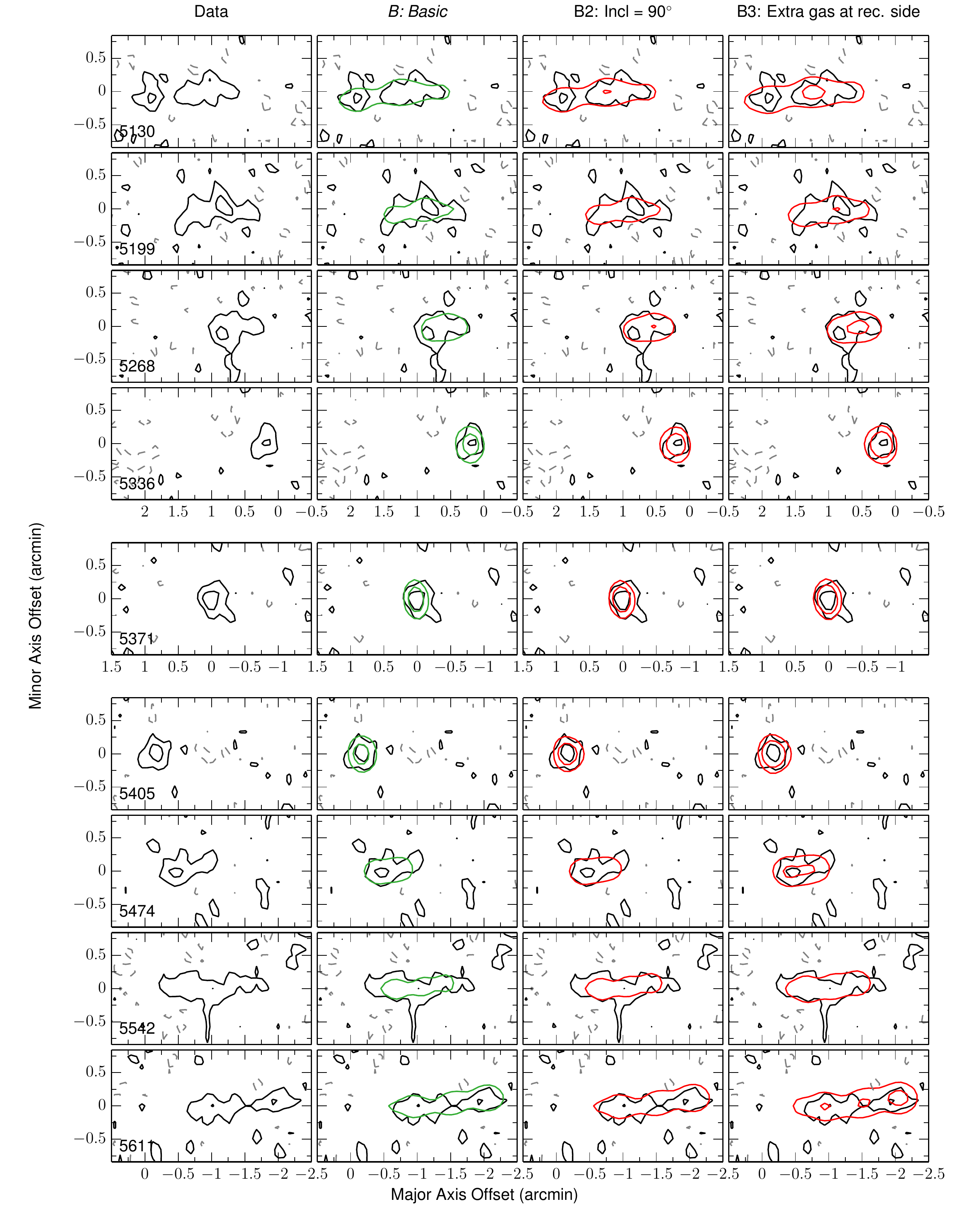}
     \caption{Representative channel maps from the observed data cube of UGC\,4277 and the various models. Contour levels are -1.5, 1.5 (1.5$\sigma$) and 4.0 mJy beam$^{-1}$. The black contours show the observations, with negative contours as dashed grey. The green contours represent the final (\textit{B}) model. Other models are shown as red contours. The systemic velocity is 5370 $\pm$ 6 km s$^{-1}$.}
     \label{fig:U4277_models_channels}
\end{figure*}

\begin{figure*}[]
\centering
   \includegraphics[width=\textwidth]{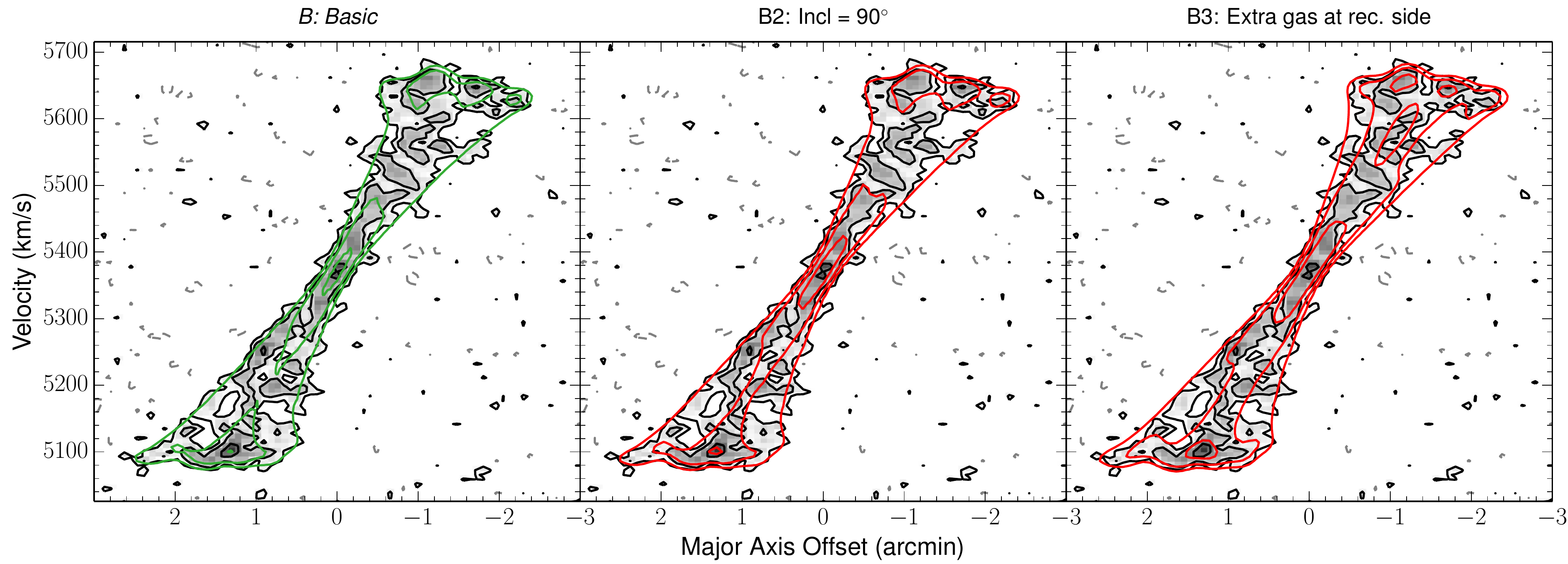}
     \caption{Observed major axis position-velocity diagram of UGC\,4277 (black and gray contours) overlaid with different models. The green contours represent the final (\textit{B}) model. Other models are shown as red contours. Contour levels are -1.5, 1.5 (1.5$\sigma$), 3.0 and 6.0 mJy beam$^{-1}$. The greyscale corresponds to the observations.}
     \label{fig:U4277_pv_models}
\end{figure*}

Following the standard procedure we first constructed a basic model consisting of a single disk with a constant inclination and scale height. This \emph{B} model has an inclination of 87.3$\degr$ and a rather large scale height of 1.8 kpc (but see last paragraph of this section). In the slightly warped outer edges of the disk the position angle deviates at most 3.6$\degr$ from the value in the inner disk. This model matches the observed total H{\sc{i}} map rather well (Fig. \ref{fig:U4277_models_mom0}), although it does not reproduce the highest brightness contours. If we compare the model channel maps (Fig. \ref{fig:U4277_models_channels}) and major axis XV-diagram (Fig. \ref{fig:U4277_pv_models}) to the observations, we see the same problems as we encountered for NGC\,973. The model often does not match the exact shape of the observed emission in the channel maps and lacks the observed high surface brightness contours at intermediate velocities. 
\par
We therefore followed the same modelling strategy and constructed a \emph{B2} model by imposing a constant inclination of 90$\degr$. TiRiFiC found the optimal corresponding scale height to be 2.2 kpc. As can be seen in Fig. \ref{fig:U4277_pv_models} this model does not solve the problems mentioned in the previous paragraph. The higher contours at intermediate velocities are still not reproduced, and additionally the contours in the central region of the XV-diagram are now overestimated. In the channel maps in Fig. \ref{fig:U4277_models_channels} this model also brings no improvement. Finally we note that the zeroth moment map of the \emph{B2} model does now include the highest contours from the observed moment-0 map (Fig. \ref{fig:U4277_models_mom0}), but they are located too far towards the center of the galaxy. This is actually the direct consequence of underestimating the emission at intermediate velocities and overestimating the emission close to the systemic velocity, as we described for the XV-diagram.
\par
Next, we again manually scaled up the surface brightness distribution of the \emph{B2} model to match the high contours at intermediate velocities. In the end an increase of 25 percent on the approaching side and 35 percent on the receding side was needed to more or less achieve this (\emph{B3} model). However, from Fig. \ref{fig:U4277_pv_models} it is clear that the shape of the model contours is quite different from the observed contours, which are clumpy and follow an irregular pattern. Additionally, the model now strongly overestimates the emission near the systemic velocity and at the terminal edges. We note that the innermost contour of the \emph{B3} model between 5540 and 5620 km s$^{-1}$ is actually a hole in the 3$\sigma$-contour and not a 6$\sigma$-contour. The agreement with the observations in the total H{\sc{i}} map (Fig. \ref{fig:U4277_models_mom0}) and in the channel maps (Fig. \ref{fig:U4277_models_channels}) is also clearly a lot worse than for the \emph{B} model. 
\par
Similarly to NGC\,973, the combination of a low signal-to-noise ratio and an irregular gas disk makes that little can be done to improve the model. Since, again, the shape of the model emission in the channel maps does not differ from that of the observed emission in a systematic way (on top of the irregular jumps of the latter), there is no reason to alter the geometry of the model and we take the \emph{B} model as our final model. The observed data cube is compared to the model cube in Fig. \ref{fig:U4277_final_channels} of the appendix and the main parameters of the model are shown in Fig. \ref{fig:final_params} and listed in Table \ref{tab:final_params}. The values of the inclination, position angle and scale height have an uncertainty of 1.5$\degr$, 2.0$\degr$ and 1.0 kpc respectively.
\par
The scale height of the final model is remarkably high. However, the large uncertainty on this value indicates that it should not be trusted blindly and still allows a lower, more common value of the scale height as well. In addition, a flaring gas disk, with a large scale height in the outer regions but a significantly lower scale height in the inner disk, is also not in disagreement with the data. The low signal-to-noise ratio and irregular shape of the observed emission in the channel maps unfortunately make it impossible to distinguish between these options and put more stringent constraints on the scale height.

\subsection{IC\,2531}

\begin{figure}[t]
  \resizebox{\hsize}{!}{\includegraphics{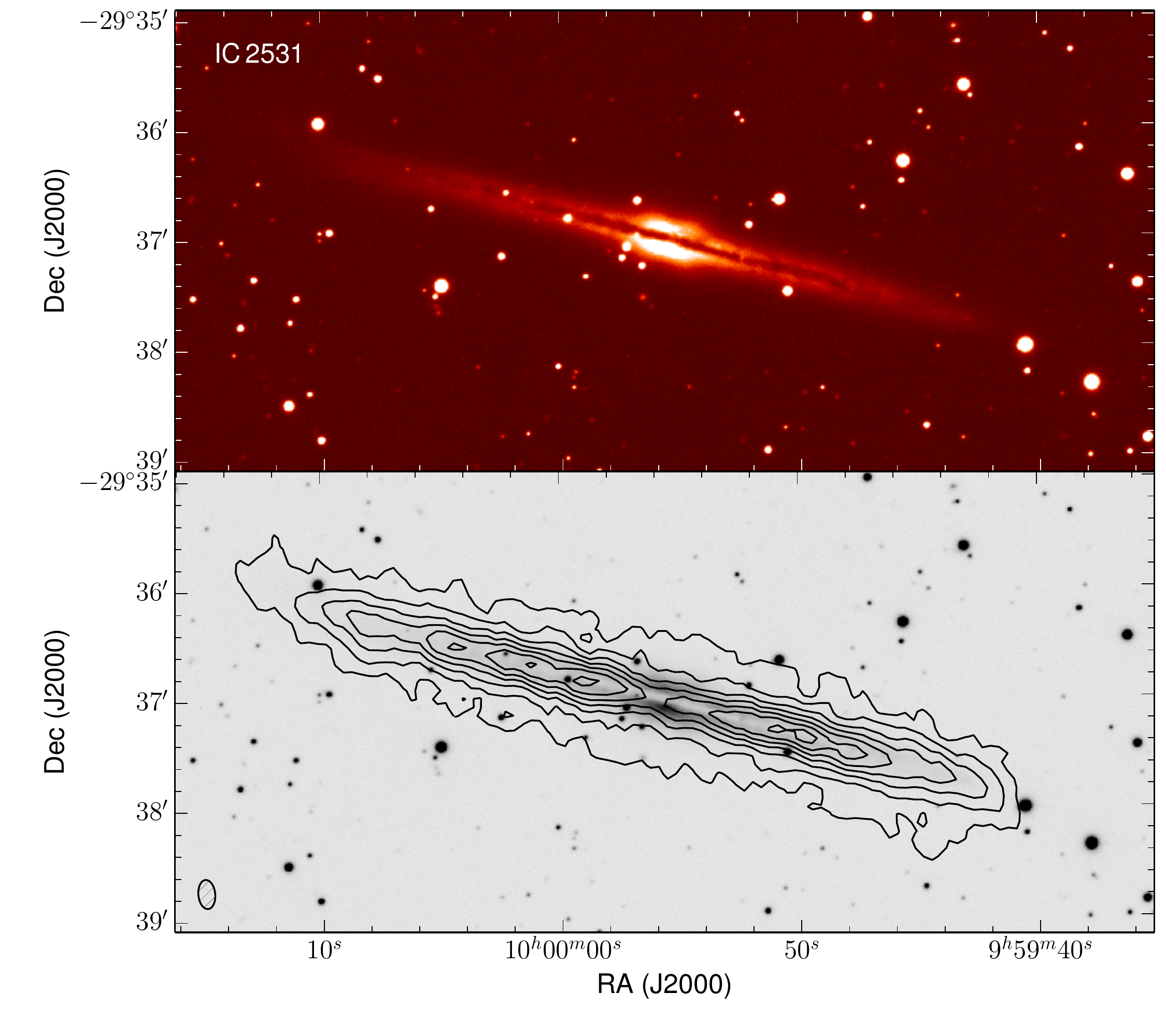}}
  \caption{Top: False colour V-band image of IC2531 taken with the Faulkes Telescope South. The data were cleaned and combined using standard data reduction techniques. Bottom: H{\sc{i}} contours overlaid on the same V-band image. Contours start at $2.98 \times 10^{20}$ atoms cm$^{-2}$ and increase as 4, 8, 12, 16, 20 and 24 times this value. North is up, East is to the left.}
  \label{fig:I2531_V_mom0}
\end{figure}

IC\,2531 is the only Southern galaxy in our sample and is located at a distance of 36.8 Mpc. It is classified as an Sb or Sc type galaxy and has a prominent peanut-shaped bulge. \cite{patsis06} ascribe this to a barred potential, although optical emission line observations do not show the characteristic line splitting in the bulge region \citep{bureau97,bureau99,kregel04b}. A false colour V-band image taken with the Faulkes Telescope South is shown in the top panel of Fig. \ref{fig:I2531_V_mom0}. The stellar disk shows a slight warp and is more extended on the approaching (NE) side, where it is also slightly misaligned with the dust lane. On the receding (SW) side, the dust lane shows a weak bump. The H{\sc{i}} content of IC\,2531 was already studied in the series of papers by Kregel et al. \citep{kregel04,kregel04b}. They focused mainly on deriving the H{\sc{i}} rotation curves, with the ultimate goal to analyse the contribution of the stellar disk to the (inner) rotation curve of a spiral galaxy. IC\,2531 was also part of the sample of \cite{obrien10a, obrien10b, obrien10c, obrien10d} and later \cite{peters13}, who studied the rotation and flaring of the H{\sc{i}} disks of a number of edge-on galaxies to derive the three-dimensional shape of their dark matter halos. These studies, however, did not yet include the possibility of a radially varying inclination. \cite{kregel04} discovered two companion galaxies at projected distances of 12$\arcmin$ (128 kpc) and 14$\arcmin$ (150 kpc) and with velocity separations of 200 km s$^{-1}$ and 182 km s$^{-1}$ in their study of the H{\sc{i}} content of IC\,2531.

\subsubsection{The data}

A contour plot of our total H{\sc{i}} map is shown in the bottom panel of Fig. \ref{fig:I2531_V_mom0}, overlaid on the V-band image from the top panel. Like the stellar disk, the H{\sc{i}} disk is also slightly warped and more extended towards the NE side. This was already observed by \cite{kregel04}. The individual channel maps show that the behaviour of the H{\sc{i}} disk somewhat deviates from that of a perfect regularly rotating disk (see Fig. \ref{fig:IC2531_final_channels} in the appendix). The approaching and receding side behave quite differently, for example regarding the shape of the channel maps and the (projected) peak intensities. In addition, (projected) extraplanar features seem to appear and disappear irregularly throughout the cube.
\par
We measure a total H{\sc{i}} flux of 42.7 Jy km s$^{-1}$, which corresponds to a total H{\sc{i}} mass of 1.37 $\times 10^{10}$ M$_{\odot}$ at the assumed distance of 36.8 Mpc. Our total H{\sc{i}} flux is remarkably higher than the flux reported by \cite{obrien10a} (13.1 Jy km s$^{-1}$), but in relatively good agreement with the values from \cite{kregel04} (44.1 Jy km s$^{-1}$) and \cite{peters13} (41.7 Jy km s$^{-1}$). These studies are based on the same data set as the current analysis with the exception of \cite{kregel04}, who did not include the observations from January 2002 taken in the 750A configuration (see our Table \ref{table:observations} and their Table 1). The single dish 21-cm fluxes reported on NED range from 33.1 Jy km s$^{-1}$ \citep{koribalski04} to 52.3 Jy km s$^{-1}$ \citep{mathewson96}.

\subsubsection{Models}
\label{sec:I2531_models}

\begin{figure}[]
  \resizebox{\hsize}{!}{\includegraphics{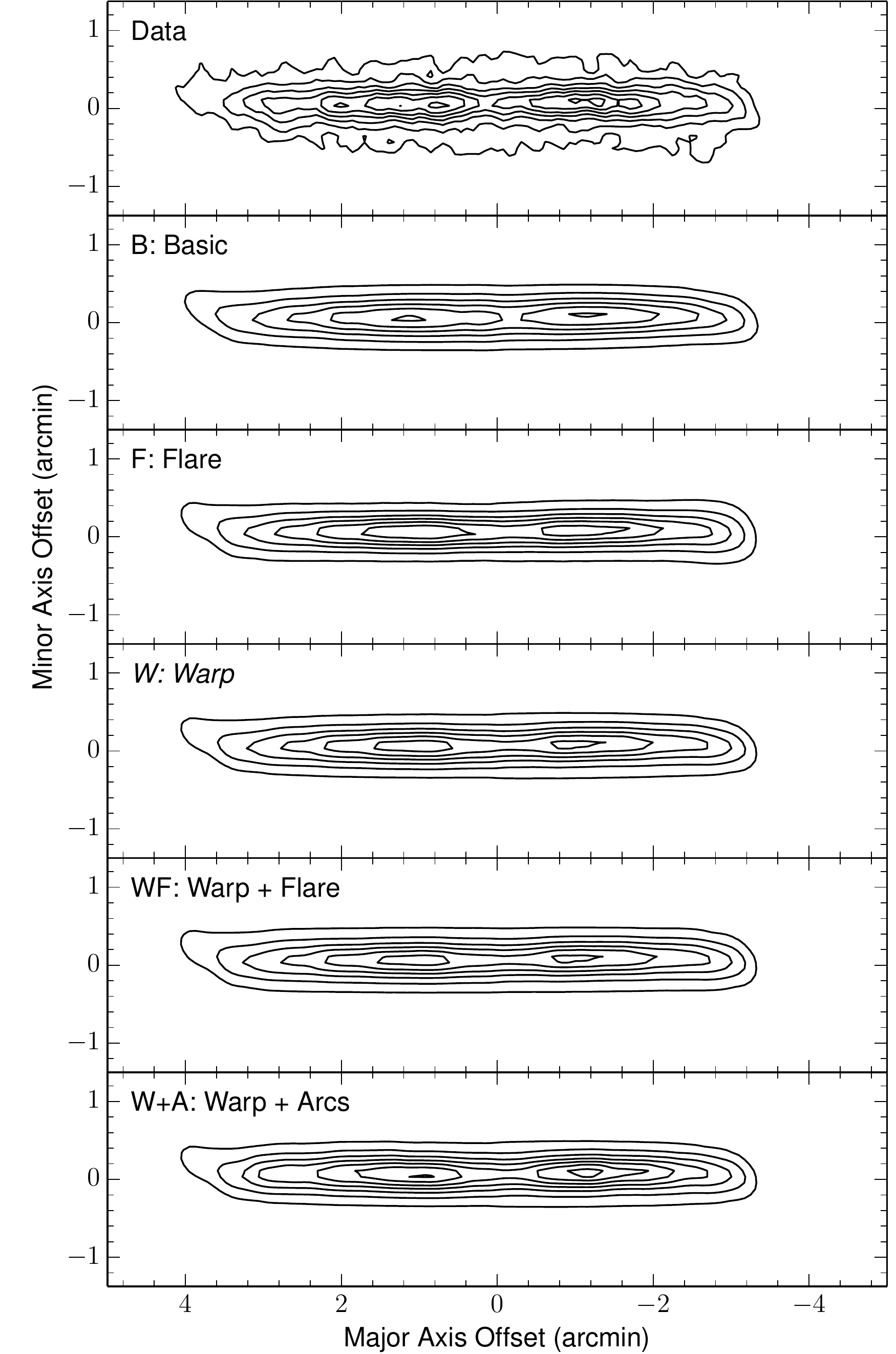}}
  \caption{Total H{\sc{i}} maps of the various models discussed here as compared to the observed total H{\sc{i}} map of IC\,2531. The \textit{W} model is our best model. Contour levels are the same as in Fig. \ref{fig:I2531_V_mom0}.}
  \label{fig:IC2531_models_mom0}
\end{figure}

\begin{figure*}[]
\centering
   \includegraphics[width=\textwidth]{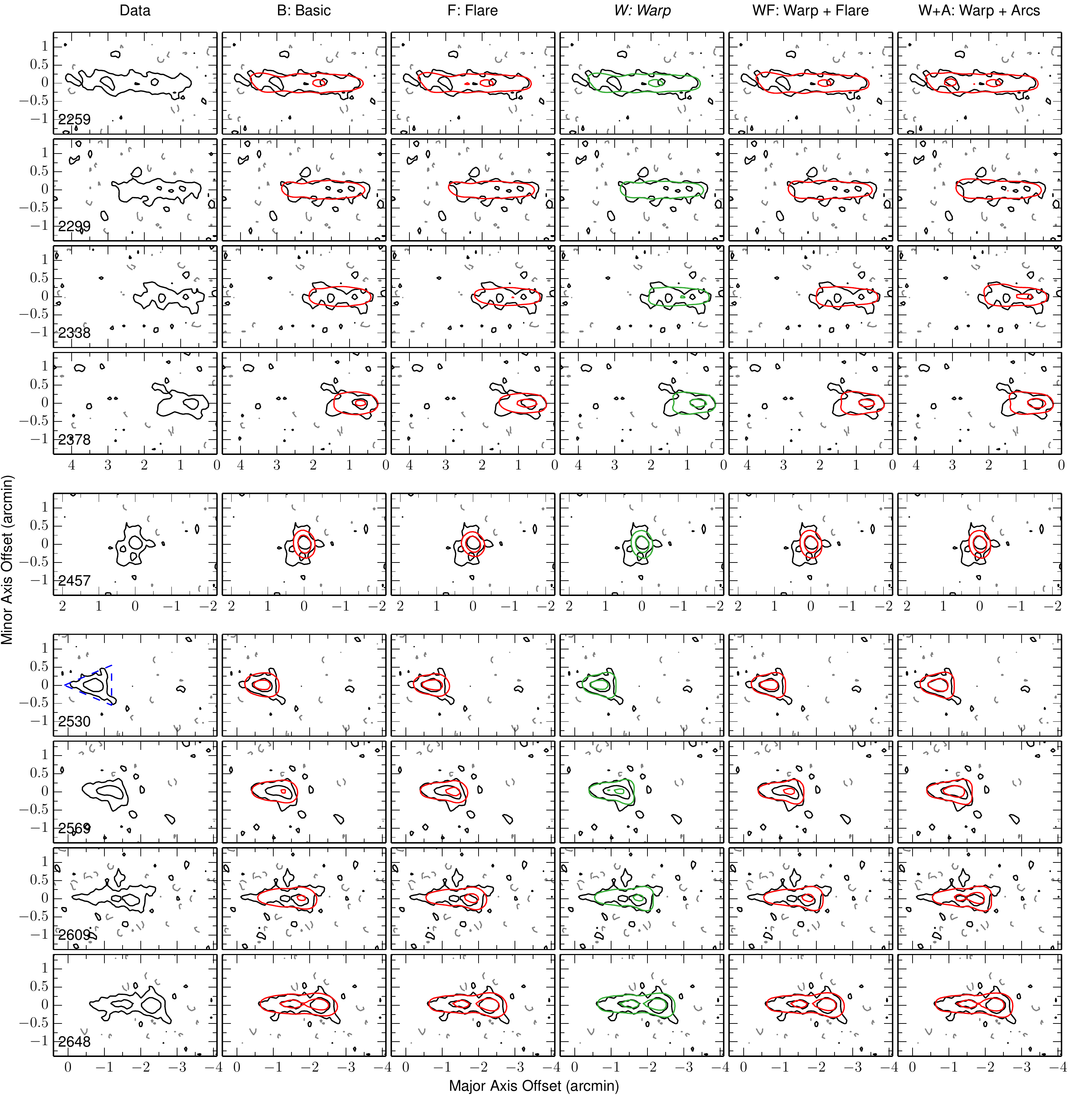}
     \caption{Representative channel maps from the observed data cube of IC\,2531 and the different models. Contour levels are -1.6, 1.6 (1.5$\sigma$) and 6.5 mJy beam$^{-1}$. The black contours show the observations, with negative contours as dashed grey. The green contours represent the final (\textit{W}) model. Other models are shown as red contours. The dashed blue lines illustrate the triangular-like shape of the observed emission. The systemic velocity is 2455 $\pm$ 6 km s$^{-1}$.}
     \label{fig:IC2531_models_channels}
\end{figure*}

Imposing the basic geometry, the best agreement of the \emph{B} model with the data was achieved with an inclination of 87.5$\degr$ and a scale height of 1.1 kpc. The warp deviates at most 4$\degr$ from the inner disk. Comparing the total H{\sc{i}} map of the basic model of IC\,2531 to the observed map (Fig. \ref{fig:IC2531_models_mom0}), it is clear that the low-density contours of the model disk are too thin. An obvious explanation for this would be that the scale height is underestimated or the inclination is overestimated, or a combination of both. In that case, the model emission (or at least a part of it) in the individual channel maps would be systematically thinner than the observed emission. Fig. \ref{fig:IC2531_models_channels} shows that this is not the case. The apparent thickness of the disk in the observed moment-0 map is in fact caused by the extraplanar features indicated in the previous section. At the sensitivity of our data these do not seem to be part of a regular structure, and it is not possible to model them in the frame of an axisymmetric disk. We therefore excluded these features in the modelling and focused on the underlying disk. As a consequence, all models appear slightly too thin in the total H{\sc{i}} map. Deeper observations might reveal the true nature of these irregular features and whether they are part of an extended, low column density structure such as a thick disk or a large-scale line of sight warp. 
\begin{figure*}[]
\centering
   \includegraphics[width=\textwidth]{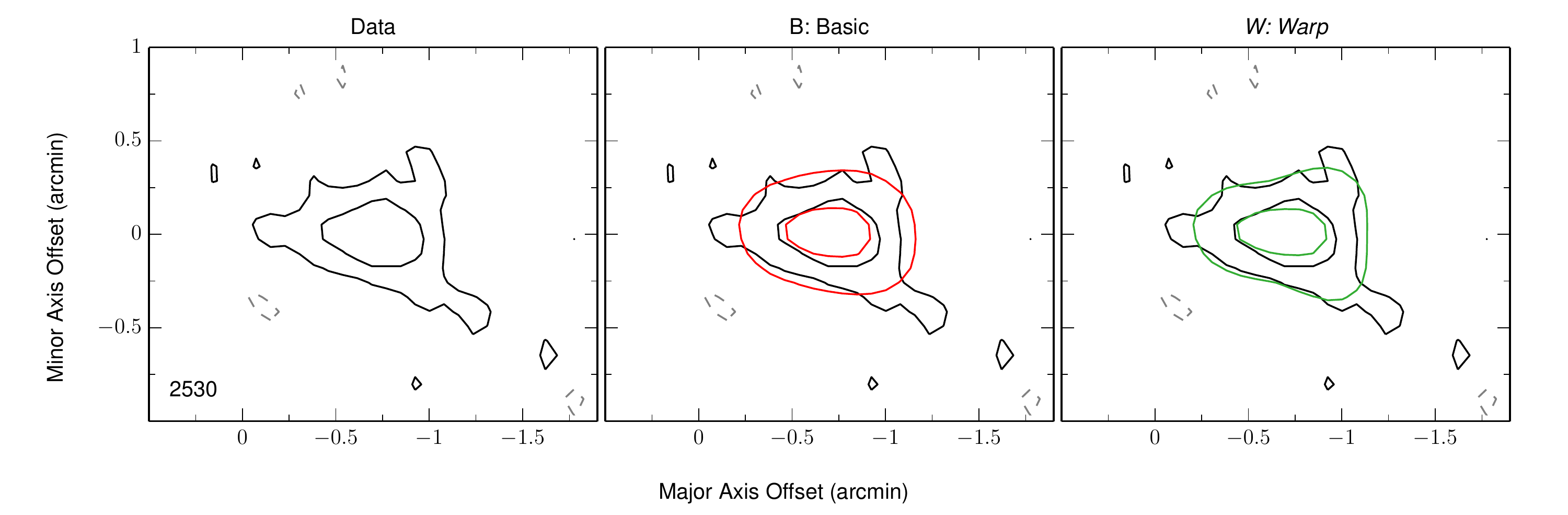}
     \caption{Channel map at 2530 km s$^{-1}$ from the observed cube of IC\,2531 (black and grey contours), the basic (\textit{B}) model (red contours) and the final (\textit{W}) model (green contours). Contour levels are -1.6, 1.6 (1.5$\sigma$) and 6.5 mJy beam$^{-1}$. The \textit{B} model is cleary too round. The \textit{W} model has a more triangular shape and provides a better fit to the data.}
     \label{fig:IC2531_models_channels_zoom}
\end{figure*}
\par
Ignoring the extraplanar emission, the comparison of some key channel maps in Fig. \ref{fig:IC2531_models_channels}, particularly on the receding side, shows that the geometry of the basic model is wrong. Indeed, in the observed channel maps we see that the emission has a triangular-like shape, with the inner part (i.e. the part closest to the center of the galaxy) being a bit thinner than the outer part. In addition, the outer tip of the observed emission generally appears quite flat, i.e. it looks more like a straight vertical line than a round arc. This is illustrated in Fig. \ref{fig:IC2531_models_channels} by the blue dashed lines in the map at 2530 km s$^{-1}$, where the effect is most obvious. The \textit{B} model, on the other hand, has a round shape. The difference between both might be difficult to see in Fig. \ref{fig:IC2531_models_channels}, but is obvious in Fig. \ref{fig:IC2531_models_channels_zoom}. The triangular shape of the observed emission indicates the presence of a line of sight warp or a flare. We therefore continued by investigating these two geometries.
\par
The best agreement with the inner parts of the observed emission in the channel maps was obtained with an inclination of 89.5$\degr$ and a scale height of 1.0 kpc for the inner disk. A flaring (\emph{F}) model was constructed by gradually increasing the scale height in the outer rings while fixing the inclination to 89.5$\degr$. The flare starts at a radius of 25 kpc on both sides of the disk and grows to a maximum scale height of 1.3 kpc on the approaching side and 1.8 kpc on the receding side. Fig. \ref{fig:IC2531_models_channels} shows that the flaring model provides a better fit to the data on the inner side of the emission, but the outer edge of the model is still systematically too round on the receding side. This rounding is inherent to a flaring geometry and can not be removed by changing the details of the flare.
\par
A model with a warp along the line of sight naturally explains these issues. The best fitting model (\emph{W}) has a constant scale height of 1.0 kpc, a central inclination of 89.5$\degr$ and a line of sight warp that reaches an inclination of 85$\degr$ on the receding side and 87$\degr$ on the approaching side. Figures \ref{fig:IC2531_models_channels} and \ref{fig:IC2531_models_channels_zoom} show that the triangular shape of the emission on the receding side is now nicely matched by the model. The high-density contours are a bit underestimated in some channels (e.g. at 2569 and 2609 km s$^{-1}$ in Fig. \ref{fig:IC2531_models_channels}), but this is most likely due to the presence of a spiral arm and will be discussed below. The agreement with the data on the approaching side seems less good in Fig. \ref{fig:IC2531_models_channels}, although it is still better than for the previous models. However, the misalignment between the stellar disk and the dust lane, the larger extent seen in the optical and in HI, and the irregular channel maps all suggest that this side of the galaxy is somewhat disrupted and not simply a smooth disk. The aim of this modelling is therefore not to obtain an exact match in every channel map, but to constrain the main properties of the H{\sc{i}} disk.
\par
Rather than just one specific type of geometry, real galaxy disks usually contain a combination of different features. Therefore we also investigated the combination of a line of sight warp with a flare. A substantial line of sight warp is in any case required to reproduce the flattening of the outer edge of the emission in the channel maps. Adding to this a flare of the same order as in the \emph{F} model makes the emission too thick in the outer channels. Adding a more modest flare does not cause significant changes compared to the \emph{W} model. This is shown for a flare with a maximum scale height of 1.3 kpc on both sides as the \emph{WF} model in Fig. \ref{fig:IC2531_models_channels}. Since the addition of a flare does not result in a siginificant improvement of the model, but does introduce extra free parameters, we discard this additional feature and stick to the simple line of sight warp geometry (\textit{W} model). 
\begin{figure*}[]
  \includegraphics[width=\textwidth]{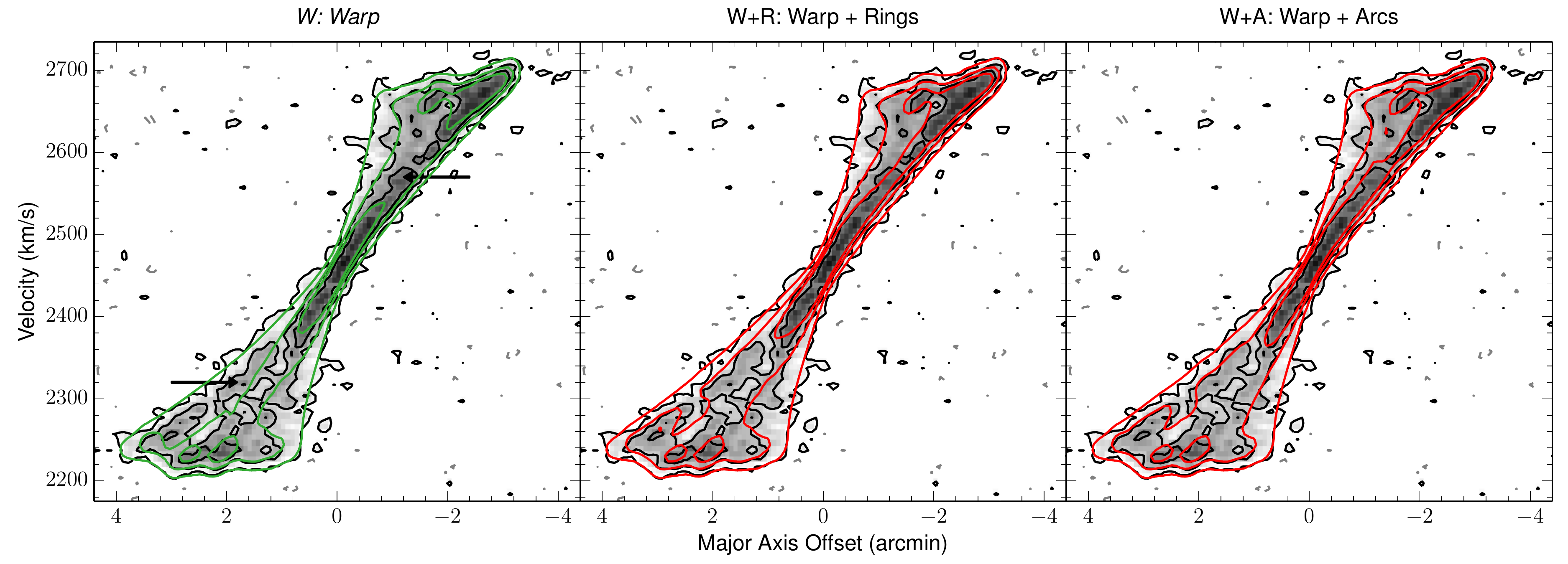}
  \caption{Observed major axis position-velocity diagram of IC\,2531 (black and gray contours) overlaid with different models. The green contours represent the final (\emph{W}) model. Other models are shown as red contours. Contour levels are -1.6, 1.6 (1.5$\sigma$), 4.9 and 8.1 mJy beam$^{-1}$ and the greyscale corresponds to the observations. The high-density ridge along the low-velocity edge on both sides (indicated by the arrows) is not reproduced by the \textit{W} model.}
  \label{fig:IC2531_final_pv}
\end{figure*}

\subsubsection{Spiral arms}
\label{sec:I2531_spiralarms}

The terminal side of the major axis position-velocity diagram (Fig. \ref{fig:IC2531_final_pv}) is nicely reproduced by the \textit{W} model. This indicates that we recover the rotation curve and the global H{\sc{i}} distribution of the disk quite well. Not reproduced by our model is the high-density ridge that is clearly visible along the low-velocity edge of the diagram. This ridge is present on both sides of the galaxy, albeit at lower contour levels on the approaching side, and was already noticed by \cite{kregel04} on the receding side. After inspecting their full-resolution data cube, they conclude that the ridge is a superposition of multiple rings. On the other hand, in the next paper of this series \cite{kregel04b} argue that the coincidence of HII emission with the H{\sc{i}} ridges in many galaxies could indicate the presence of spiral arms.
\par
We tested the first hypothesis by modelling the ridge as a set of rings in different ways. A first possibility is that the ridge is in fact a projection along the major axis of some high-density rings at large radii, rotating at the normal terminal velocities. In practice this approach requires increasing the densities of the outer rings. On the approaching side, we were unable to construct a ridge in this way, no matter how much the densities were increased. On the receding side, this strategy led to a huge overestimation of the intensities in the tip of the XV-diagram for the \textit{W} model. For a flaring model with a constant inclination of 88$\degr$, the outer rings are more strongly projected along the major axis. Hence a smaller increase of the outer densities suffices to create a ridge in the XV-diagram. But even in this case the intensities in the tip of the XV-diagram were overestimated.
\par
\begin{figure}[]
  \resizebox{\hsize}{!}{\includegraphics{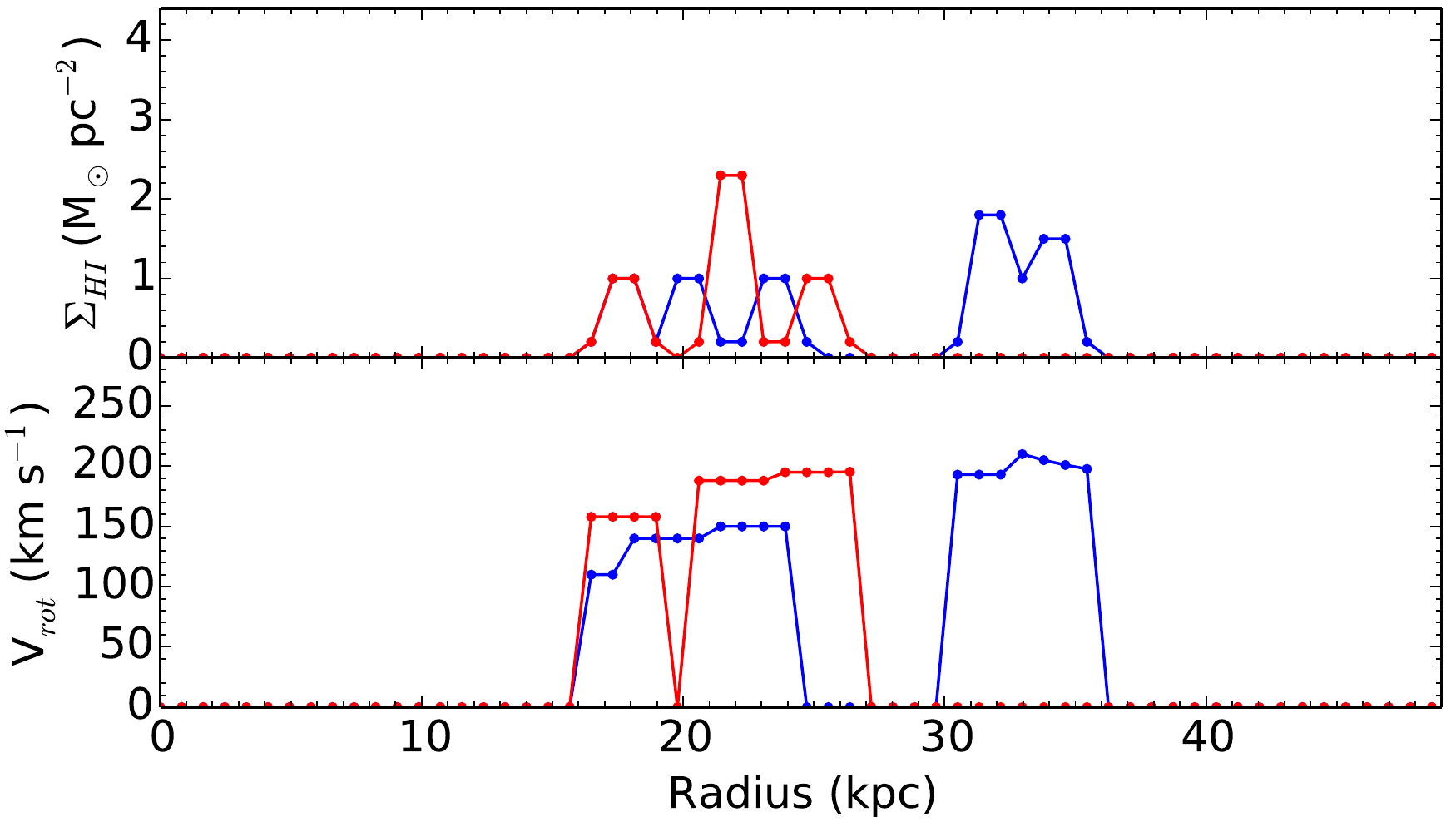}}
  \caption{Surface densities (top) and rotational velocities (bottom) of the rings added to the \emph{W} model of IC\,2531 in order to reproduce the high-density ridge in the major axis XV-diagram. Blue indicates the approaching side, red the receding side.}
  \label{fig:IC2531_ridgemodel_graphs}
\end{figure}

A second, more artificial possibility would be that the ridge is a superposition of multiple rings at different radii, rotating at velocities considerably lower than the terminal velocities. Such a model, based on the \textit{W} model, is shown in Fig.  \ref{fig:IC2531_final_pv}. Although it is indeed possible to reconstruct the observed ridge, this approach requires adding a large number of low-velocity rings on top of the main disk (see Fig. \ref{fig:IC2531_ridgemodel_graphs} for the details of these rings) and is highly unphysical.
\par
Based on our modelling we thus exclude the interpretation of the high-density ridge as a superposition of rings. A prominent spiral arm provides a much more natural explanation. Unfortunately, the edge-on orientation offers little constraints and modelling the spiral arms would be highly speculative at best. We therefore do not attempt to include spiral arms in our model. Because of this, the peak densities in the total H{\sc{i}} map and in some channel maps are slightly underestimated by the \textit{W} model. To prove that the latter is not in fact caused by a structural mistake in the model, we constructed a toy model (\textit{W+A}) that reproduces the H{\sc{i}} ridge in the XV-diagram by locally enhancing the surface brightness in several arcs within the disk (Fig. \ref{fig:IC2531_final_pv}). Figures \ref{fig:IC2531_models_mom0} and \ref{fig:IC2531_models_channels} show that with this addition the model indeed reproduces the observed peak densities.
\par
Despite the improved agreement with the data at high column densities, adding arcs to a model without a clear physical motivation is a dangerous practice and not unambiguous. The \textit{W+A} model therefore remains a toy model and we take the \textit{W} model as our final model for the atomic gas disk of IC\,2531. The total H{\sc{i}} map of the \textit{W} model is compared to the observed map in Fig. \ref{fig:IC2531_models_mom0}. Our final model data cube is compared to the observed cube in Fig. \ref{fig:IC2531_final_channels} of the appendix. The main parameters of the final model are given in Fig. \ref{fig:final_params} and Table \ref{tab:final_params}. The errors on the inclination, position angle and scale height are 1.3$\degr$, 1.8$\degr$ and 0.5 kpc respectively.

\subsection{NGC\,4217}

NGC\,4217 is located at a distance of 19.6 Mpc and is classified as an SAb galaxy. Its atomic gas content has previously been studied by \cite{verheijen01,verheijen01b}, who used the kinematics of the atomic gas in a sample of galaxies in the Ursa Major cluster to investigate the statistical properties of the Tully-Fisher relations. However, their sample consisted largely of moderately inclined galaxies and their method to derive the surface density distribution of the H{\sc{i}} is actually not suitable for an edge-on geometry, with Verheijen and Sancisi stating that the ``method for extracting the surface density profiles from integrated H{\sc{i}} maps breaks down for nearly edge-on systems.''

\subsubsection{The data}

\begin{figure}[t]
  \resizebox{\hsize}{!}{\includegraphics{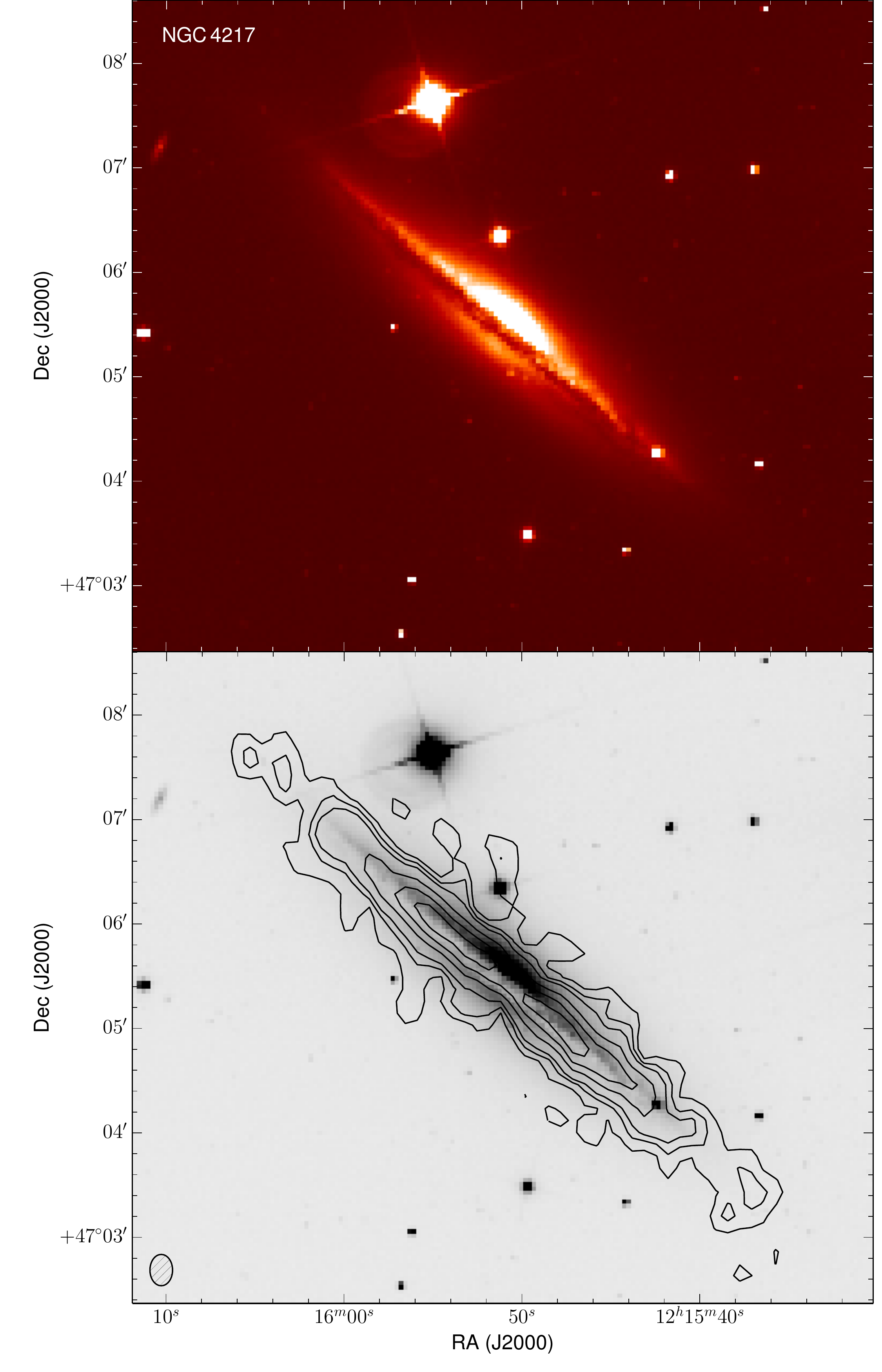}}
  \caption{Top: False color r-band image of NGC\,4217 (SDSS J121551.51+470538.1) from the EFIGI catalogue \citep{baillard11}, based on data from the SDSS DR4. Bottom: H{\sc{i}} contours overlaid on the same r-band image. Contours start at $3.76 \times 10^{20}$ atoms cm$^{-2}$ and increase as 2.5, 4, 8, 12 and 16 times this value. North is up, East is to the left.}
  \label{fig:N4217_r_mom0}
\end{figure}

\begin{figure}[]
  \resizebox{\hsize}{!}{\includegraphics{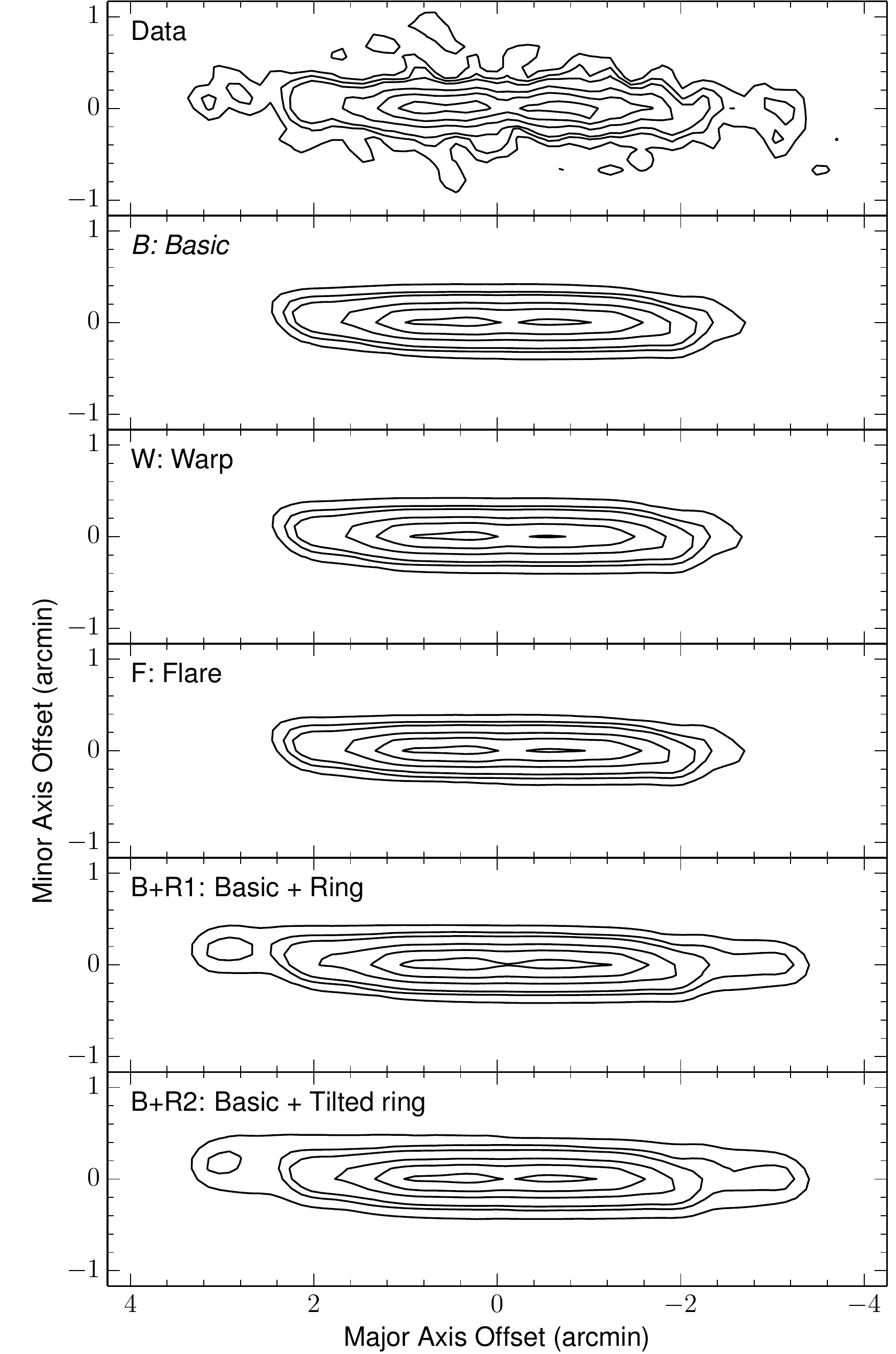}}
  \caption{Total H{\sc{i}} maps of the various models discussed here as compared to the observed total H{\sc{i}} map of NGC\,4217. The \textit{B} model is our best model. Contour levels are the same as in Fig. \ref{fig:N4217_r_mom0}.}
  \label{fig:N4217_models_mom0}
\end{figure}

In the total H{\sc{i}} map of NGC\,4217 (Figs. \ref{fig:N4217_r_mom0}, \ref{fig:N4217_models_mom0}) two blobs of gas on either side of the main disk immediately catch the eye. They suggest the presence of a ring, which, strikingly, has no optical counterpart in the r-band image in Fig. \ref{fig:N4217_r_mom0}. In what follows we will refer to this feature as the outer ring. Some filaments can also be seen extending out of the plane of the galaxy. The H{\sc{i}} disk is not significantly warped, apart from a slight upward bend (i.e. towards the NW in Fig. \ref{fig:N4217_r_mom0}) where it meets the outer ring. This upward bend is also visible in the stellar and dust disks.
\par
If we separate the outer ring from the main disk in our data (see section \ref{sec:N4217_models} for a discussion on the nature of the outer ring), we measure a total H{\sc{i}} flux of 27.6 Jy km s$^{-1}$ for the main disk and 4.2 Jy km s$^{-1}$ for the outer ring. At a distance of 19.6 Mpc, this translates to atomic hydrogen masses of 2.50$\times 10^{9}$ M$_{\odot}$ and 3.81$\times 10^{8}$ M$_{\odot}$. It should be noted that in the central channels of the data cube the emission from the outer ring and the main disk are projected on top of each other, making it very difficult to separate both components. Our estimate of the mass of the outer ring only includes the emission that unambiguously belongs to this component, so it should be regarded as a lower limit. \cite{verheijen01} also analysed the H{\sc{i}} content of NGC\,4217, based on the same data as used here, but did not make a distinction between the main disk and the outer ring. They found a total H{\sc{i}} flux of 33.8 Jy km s$^{-1}$, which is slightly higher than our total flux of 31.8 Jy km s$^{-1}$. The reason for this is probably that we were more strict in separating the real emission from the noise. (Corrected) single dish fluxes of 25.8 Jy km s$^{-1}$ and 31.0 Jy km s$^{-1}$ are reported by \cite{springob05} and \cite{huchtmeier89} respectively.  

\subsubsection{Models}
\label{sec:N4217_models}

\begin{figure*}[]
\centering
   \includegraphics[width=\textwidth]{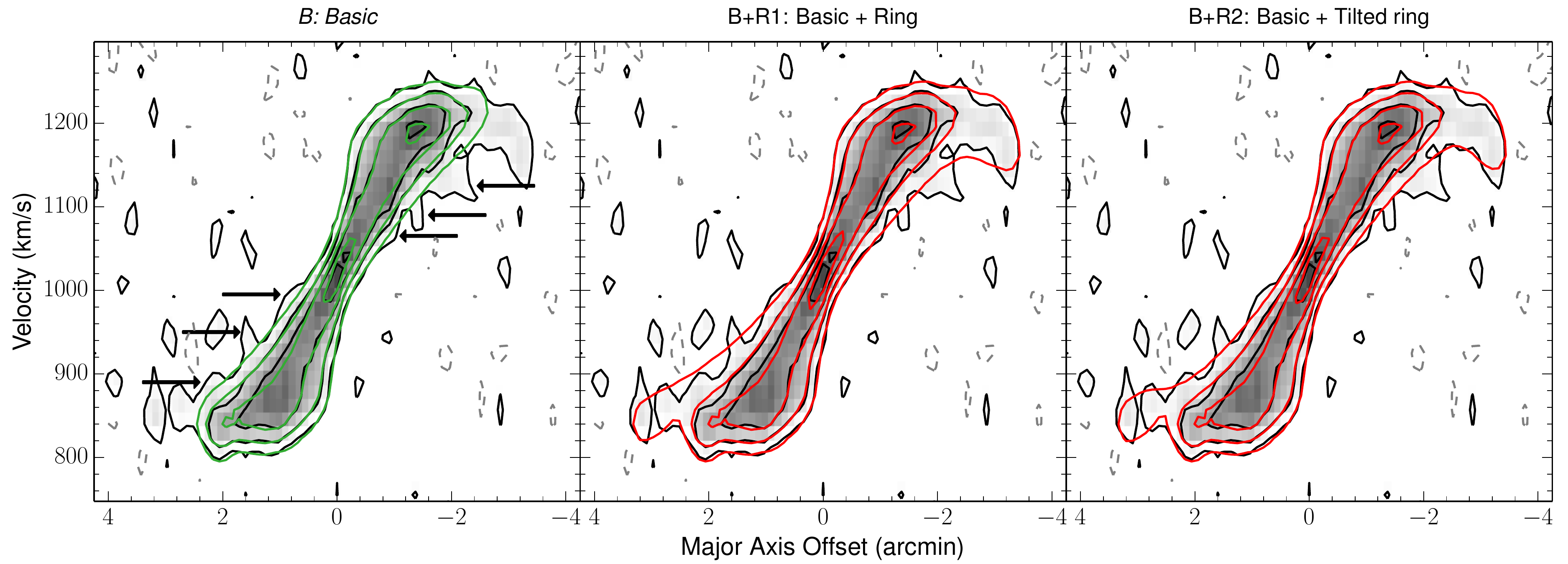}
     \caption{Observed major axis position-velocity diagram of NGC\,4217 (black and gray contours) overlaid with different models. The green contours show the final (\textit{B}) model. Other models are shown as red contours. Contour levels are -1.5, 1.5 (1.5$\sigma$), 4.5, 9.0 and 13.5 mJy beam$^{-1}$. The greyscale corresponds to the observations. We interpret the features indicated by the arrows as the projection of a clumpy ring along the major axis.}
     \label{fig:N4217_pv_models}
\end{figure*}

In the major axis position-velocity diagram (see Fig. \ref{fig:N4217_pv_models} later) the outer ring corresponds to the sudden drop in rotational velocities at major axis offsets of about $\pm$ 3 arcmin. The outer ring thus seems both spatially and kinematically separated from the main disk, indicating that it probably has an external origin. Because of this, we start by modelling only the main gas disk and add the outer ring as an extra feature in a later stage.
\par
\begin{figure*}[]
\centering
   \includegraphics[width=0.825\textwidth]{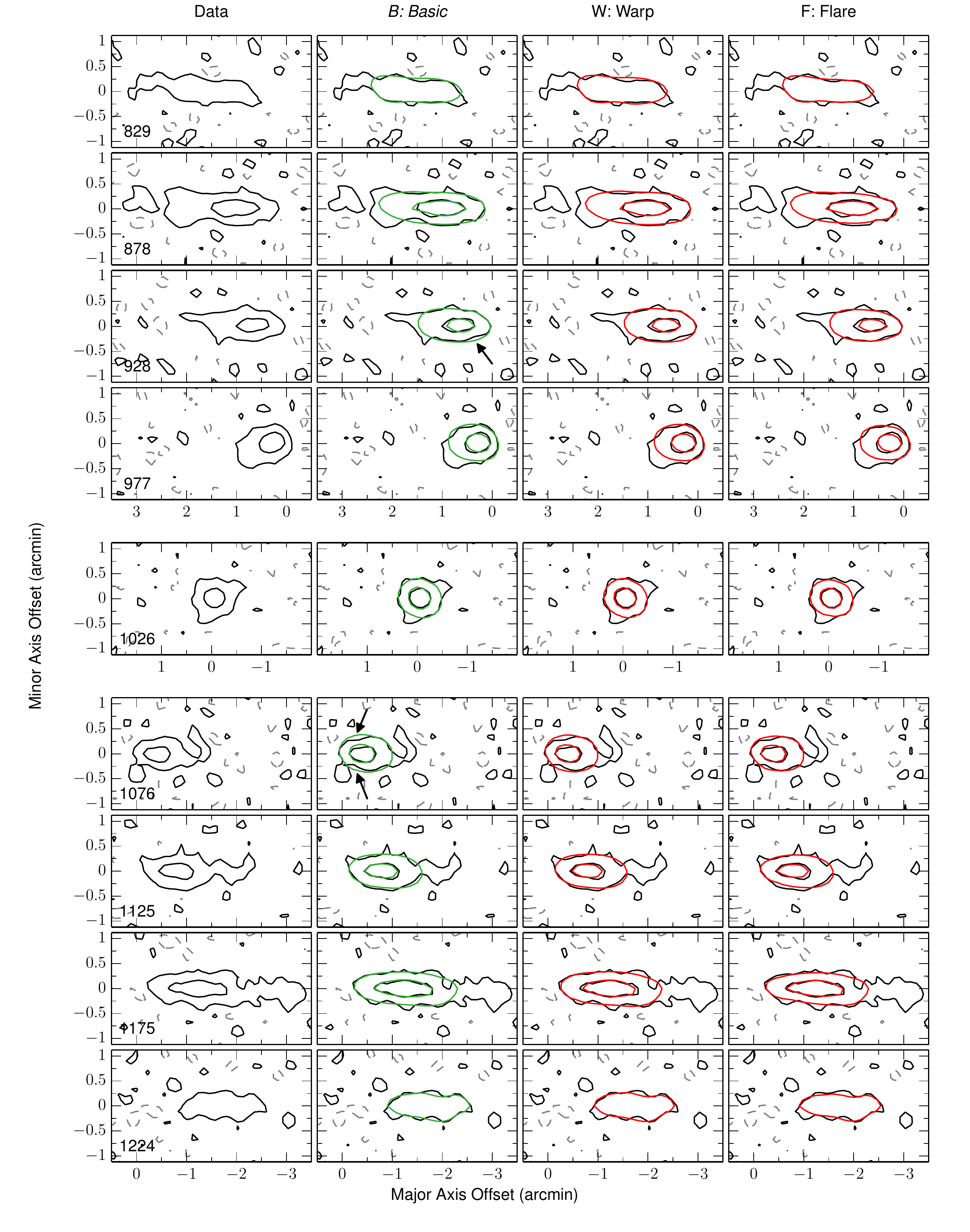}
     \caption{Representative channel maps from the observed data cube of NGC\,4217 and the different models of the main disk. Contour levels are -1.5, 1.5 (1.5$\sigma$) and 8.0 mJy beam$^{-1}$. The black contours show the observations, with negative contours as dashed grey. The green contours represent the final (\textit{B}) model. Other models are shown as red contours. The systemic velocity is 1022 $\pm$ 6 km s$^{-1}$. The arrows highlight regions where the \textit{B} model is significantly thicker than the observed emission, as explained in the text.}
     \label{fig:N4217_models_channels}
\end{figure*}

An initial fit (\emph{B}) imposing the basic geometry resulted in an inclination of 86.2$\degr$ and a scale height of 0.5 kpc. Fig. \ref{fig:N4217_models_channels} compares this basic model to the observations. Ignoring the outer ring, the model already provides a good fit to the data. The only exception to this are four channels on the receding side, between 1059 and 1109 km s$^{-1}$, and (to a lesser extent) two channels on the approaching side, between 928 and 944 km s$^{-1}$, where the model is significantly thicker than the observed emission in the inner regions. Two of these channels are shown in Fig. \ref{fig:N4217_models_channels}, at 928 and 1076 km s$^{-1}$, where the discrepant regions are indicated with arrows. This discrepancy could indicate a flaw in the model, but the fact that it only shows up in a few channels and that the affected channels are different on both sides (the central channel is at 1026 km s$^{-1}$) indicates that it is probably just caused by local irregularities in the disk and does not reflect a large-scale geometric feature. Nonetheless, we investigate whether a more complex geometry could provide a better fit.
\par
With the parameters of the basic model as inputs we conducted a new fit with TiRiFiC. The inclination and scale height were now allowed to vary in blocks of five rings in order to reveal a potential flare and/or line of sight warp, but at the same time smooth out the unphysical jumps that TiRiFiC usually finds. Fits in blocks of three and seven rings were also performed to verify the results. The inclination on the receding side was found to roughly increase towards the center, reaching a maximum of 89.5$\degr$ in the central rings. For the inclination on the approaching side and the scale height on both sides, the radial profile found by TiRiFiC was still dominated by unphysical jumps and did not display a clear trend. In the absence of strong constraints we constructed a model with a line of sight warp by applying the rising trend of the inclination that was found for the receding side to the approaching side as well and imposing a constant global scale height. A new TiRiFiC fit found a value of 0.6 kpc for the latter. As can be seen in Fig. \ref{fig:N4217_models_channels} the difference between this model (\emph{W}) and the \textit{B} model is very small and the discrepancy mentioned in the previous paragraph is still present. Moreover, looking at the r-band image in Fig. \ref{fig:N4217_r_mom0}, a central inclination of 89.5$\degr$ seems too high. Indeed, the bulge is clearly a lot stronger on the NW side of the galaxy (`above' the major axis) while the dust lane is located more towards the SE, suggesting that the galaxy is not seen perfectly edge-on. For these reasons we reject the \emph{W} model and continue with a constant inclination.
\par
In a final attempt to model the thinning of the disk we added a flare to the \emph{B} model. The \emph{F} model shown in Fig. \ref{fig:N4217_models_channels} has a constant scale height of 0.1 kpc out to a radius of 2.6 kpc, which then linearly increases to reach a maximum value of 0.6 kpc in the outer regions of the disk. The discrepancy in the six channels mentioned earlier is still not completely resolved, although it is (slightly) less severe. However, at the same level as this flaring model improves the fit in these six channels, it is now too thin in many other channels. Adding a flare to the model therefore introduces many free and poorly constrained parameters without significantly improving the quality of the fit. We thus decided not to include a flare and to stick with the basic geometry (\textit{B} model). It should be noted, though, that the difference between the various models we discussed here is very small and hence the geometry of the gas disk is not very well constrained by our modelling. However, this does not significantly affect the other parameters. Between the different geometries investigated here, only the surface brightness distribution needed slight adjustments and the changes were of the order of 10$\%$ at most.

\subsubsection{The outer ring}

\begin{figure*}[]
\centering
   \includegraphics[width=0.825\textwidth]{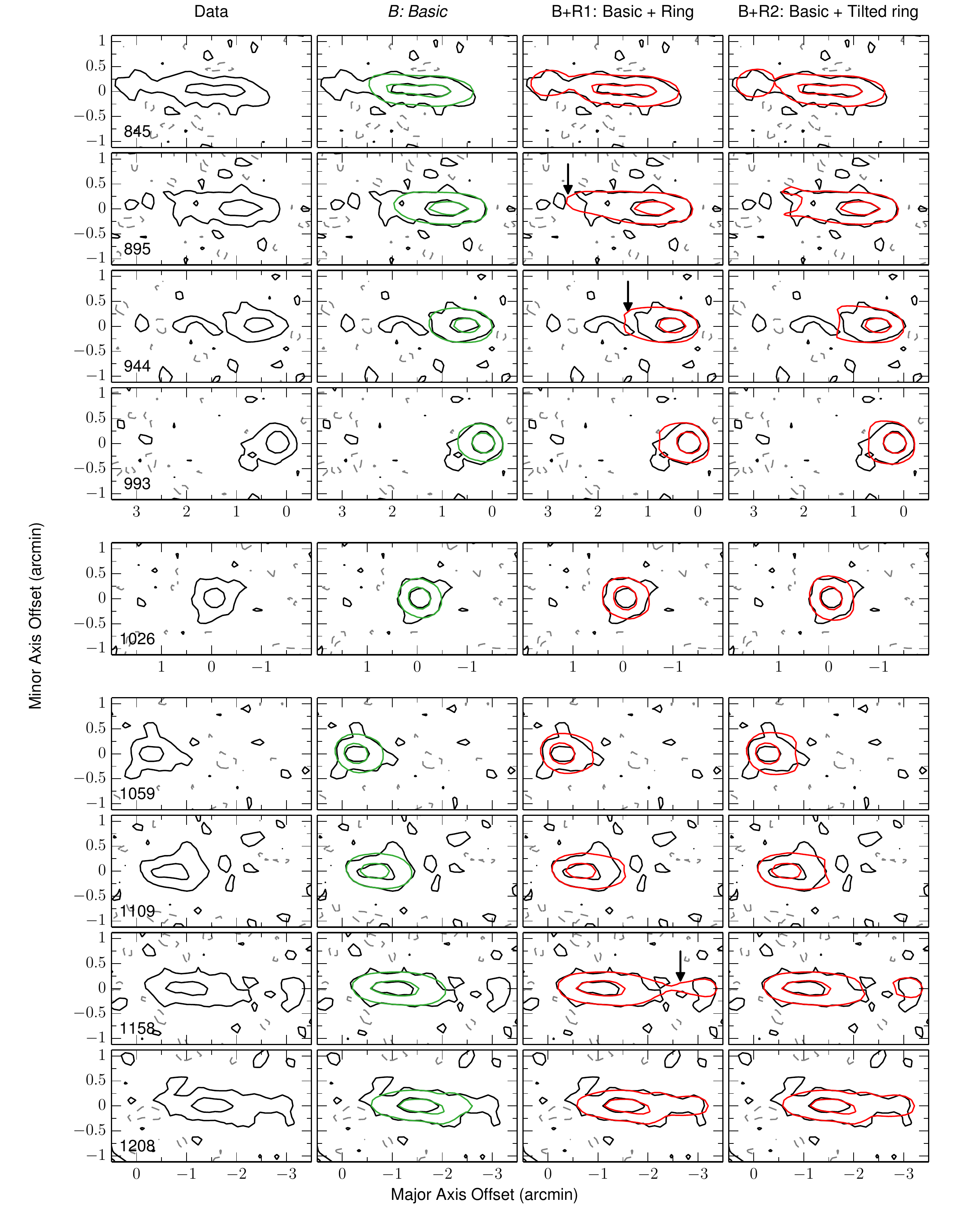}
     \caption{Representative channel maps from the observed data cube of NGC\,4217 and the different models that include the outer ring. Contour levels are -1.5, 1.5 (1.5$\sigma$) and 8.0 mJy beam$^{-1}$. The black contours show the observations, with negative contours as dashed grey. The green contours represent the final (\textit{B}) model. Other models are shown as red contours. Because we want to highlight different features, the channels shown here are different from those in Fig. \ref{fig:N4217_models_channels}. The arrows highlight the projection along the major axis of the outer ring in the \textit{B+R1} model, which is clearly in disagreement with the data.}
     \label{fig:N4217_models_channels_ring}
\end{figure*}

The next step in the modelling process is to determine the nature of the outer ring. We start by modelling it as a simple extension of the main disk, i.e. using the same inclination and scale height as in the disk. This model (model \emph{B+R1} in Fig. \ref{fig:N4217_models_channels_ring}) performs well in the very outer channels, but in the other channels the projection of the outer ring along the major axis is clearly in disagreement with the data, especially on the approaching side. Arrows highlight this effect in the channel maps at 895, 944 and 1158 km s$^{-1}$ in Fig. \ref{fig:N4217_models_channels_ring}. This is also visible in the XV-diagram (Fig. \ref{fig:N4217_pv_models}) and indicates that either the outer ring is not a complete ring, or that the inclination should be lower, placing less gas along the major axis. We test the latter option in the \emph{B+R2} model. Fig. \ref{fig:N4217_pv_models} shows that this model does indeed solve the aforementioned projection problem, but cannot account for the clumpy features along the low-velocity edges of the XV-diagram (indicated with arrows for the \textit{B} model). Moreover, the lower inclination makes this model too thick, as can be seen in the moment-0 map (Fig. \ref{fig:N4217_models_mom0}) and in the central channels in Fig. \ref{fig:N4217_models_channels_ring}. Finally, both these models also require a highly unphysical rotation curve on the receding side, with a rather abrupt decrease of about 60 km s$^{-1}$ (see Fig. \ref{fig:final_params}). All these arguments indicate that the outer ring is not simply an extension of the main disk. A much more natural interpretation would be that we are observing the accretion of a roughly coplanar satellite galaxy that has been gradually ripped apart, forming a clumpy arc or ring-like structure around NGC\,4217. This would explain the spatial and kinematical offset with respect to the main disk and the absence of a smooth projection of the ring along the major axis.
\par
Modelling the outer ring as a set of individual clumps is, in principle, possible with TiRiFiC, but this exercise would not contribute much to the overall quality of the model and is beyond the scope of this paper. We therefore take the \textit{B} model as our final model for the atomic gas disk of NGC\,4217 and note that a coplanar ring-like structure just outside of the disk is not included in this model. A comparison between the final model and the observed data cube is shown in Fig. \ref{fig:N4217_final_channels} of the appendix. The major axis position-velocity diagrams are compared in Fig. \ref{fig:N4217_pv_models} and the main parameters of the final model are shown in Fig. \ref{fig:final_params} and Table \ref{tab:final_params}. The inclination and position angle have respective uncertainties of 1.5$\degr$ and 2.0$\degr$ and the error on the scale height is 0.2 kpc.

\subsection{NGC\,5529}

NGC\,5529 is an Sc type galaxy located at a distance of 49.5 Mpc. Like IC\,2531, NGC\,5529 was also part of the sample of galaxies analysed by \cite{kregel04,kregel04b}. In the region of the peanut-shaped bulge, \cite{kregel04b} find a typical figure-of-eight pattern in their optical emission line position-velocity diagram, indicating the presence of a bar. In the outer regions of the galaxy, the optical emission line profiles are again double peaked. NGC\,5529 is part of a rich galaxy group with at least 16 other members \citep{irwin07}. Its stellar disk shows a clear warp and extends out to larger (projected) radii on the NW side. In their study of the atomic gas in this galaxy, \cite{kregel04} discovered that NGC\,5529 is connected to two of its companions, MGC +06-31-085a and NGC\,5529 B, through H{\sc{i}} bridges.
\begin{figure}[h]
  \resizebox{\hsize}{!}{\includegraphics{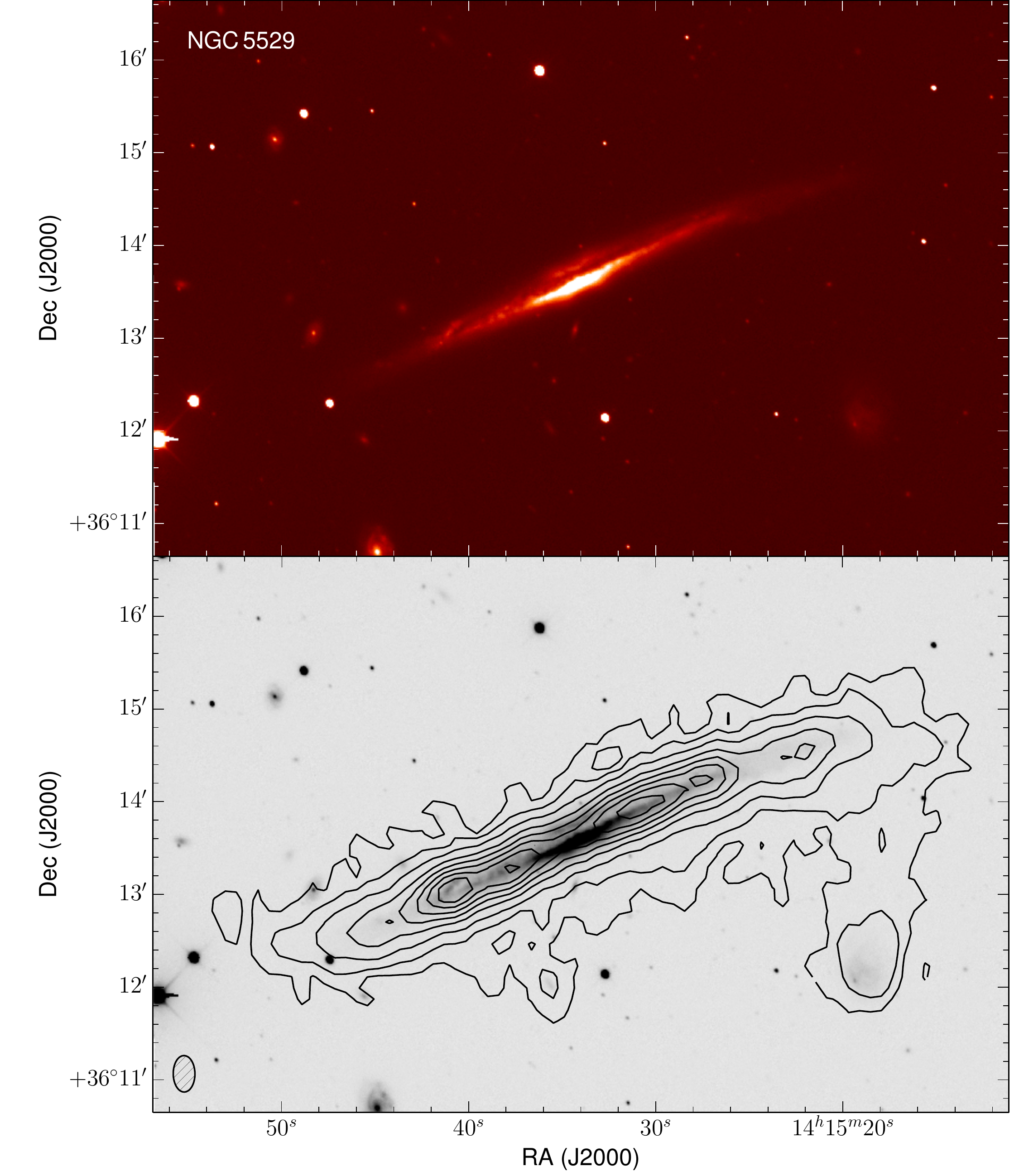}}
  \caption{Top: False colour V-band image of NGC\,5529 taken with the 1.3m telescope at the Skinakas Observatory in Crete. The data were cleaned and combined using standard data reduction techniques.. Bottom: H{\sc{i}} contours overlaid on the same V-band image. Contours start at $9.94 \times 10^{19}$ atoms cm$^{-2}$ and increase as 4, 10, 20, 30, 40, 50 and 60 times this value. North is up, East is to the left.}
  \label{fig:N5529_V_mom0}
\end{figure}

\subsubsection{The data}

Derived from the same WSRT data as used by \cite{kregel04}, our total H{\sc{i}} map (Fig. \ref{fig:N5529_V_mom0}) is very similar to theirs. On the approaching (NW) side the atomic gas follows the same moderate warp as the stars. On the receding (SE) side the H{\sc{i}} appears more strongly warped and shows a strong upturn beyond the stellar disk. The channel maps (see Fig. \ref{fig:N5529_final_channels} in the appendix) reveal that the H{\sc{i}} disk of NGC\,5529 is surrounded by a significant volume of extraplanar gas, especially on the approaching side, as a consequence of the interaction with the satellite galaxies. This will be discussed in further detail in the next section.
\par
It is not possible to unambiguously separate the actual H{\sc{i}} disk of NGC\,5529 from the extraplanar gas and the satellite galaxies in the channel maps. In order to determine the total H{\sc{i}} mass of the galaxy, we took a conservative approach and neglected only the emission that unambiguously comes from the satellite galaxies. From the resulting moment-0 map we measure a total H{\sc{i}} flux of 46.5 Jy km s$^{-1}$. This is slightly higher than the value of 43.2 Jy km s$^{-1}$ found by \cite{kregel04}. The small difference between both fluxes is most probably due to the rather subjective distinction between the emission from NGC\,5529 and that from the satellites. At a distance of 49.5 Mpc our H{\sc{i}} flux corresponds to a total H{\sc{i}} mass of $2.69 \times 10^{10}$ M$_{\odot}$. For NGC\,5529, \cite{huchtmeier89} report total HI fluxes of 26.8, 37.1 and 40.8 Jy km s$^{-1}$ obtained with different single dish telescopes.

\subsubsection{Models}
\label{sec:N5529_models}

\begin{figure*}[]
\centering
   \includegraphics[width=\textwidth]{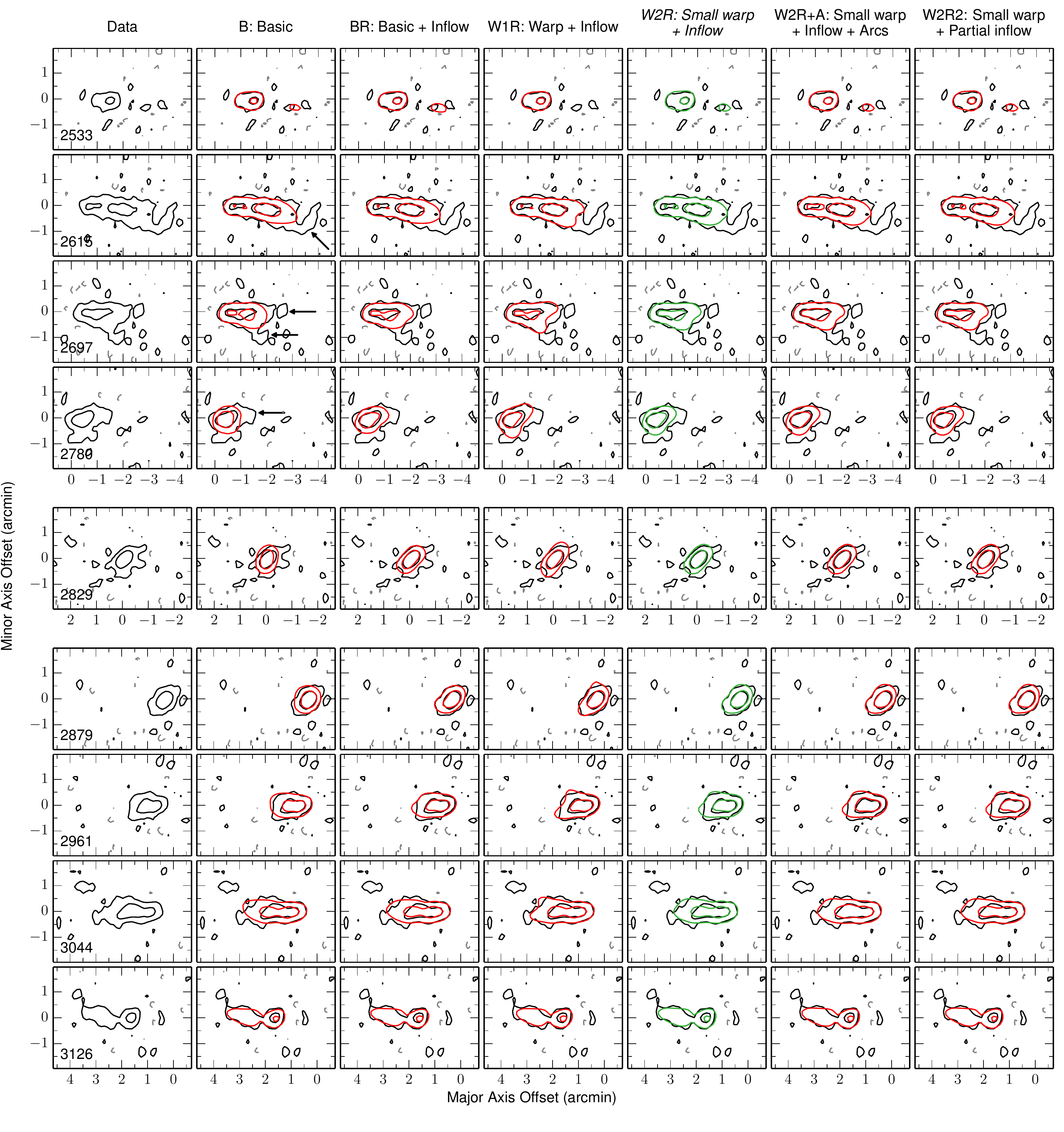}
     \caption{Representative channel maps from the observed data cube of NGC\,5529 and the various models. Contour levels are -0.8, 0.8 (1.5$\sigma$) and 5.6 mJy beam$^{-1}$. The black contours show the observations, with negative contours as dashed grey. The green contours represent the final (\textit{W2R}) model. Other models are shown as red contours. The systemic velocity is 2830 $\pm$ 6 km s$^{-1}$. The arrows highlight extraplanar gas that probably results from the interaction with the companions.}
     \label{fig:N5529_models_channels}
\end{figure*}
\begin{figure*}[]
\centering
   \includegraphics[width=\textwidth]{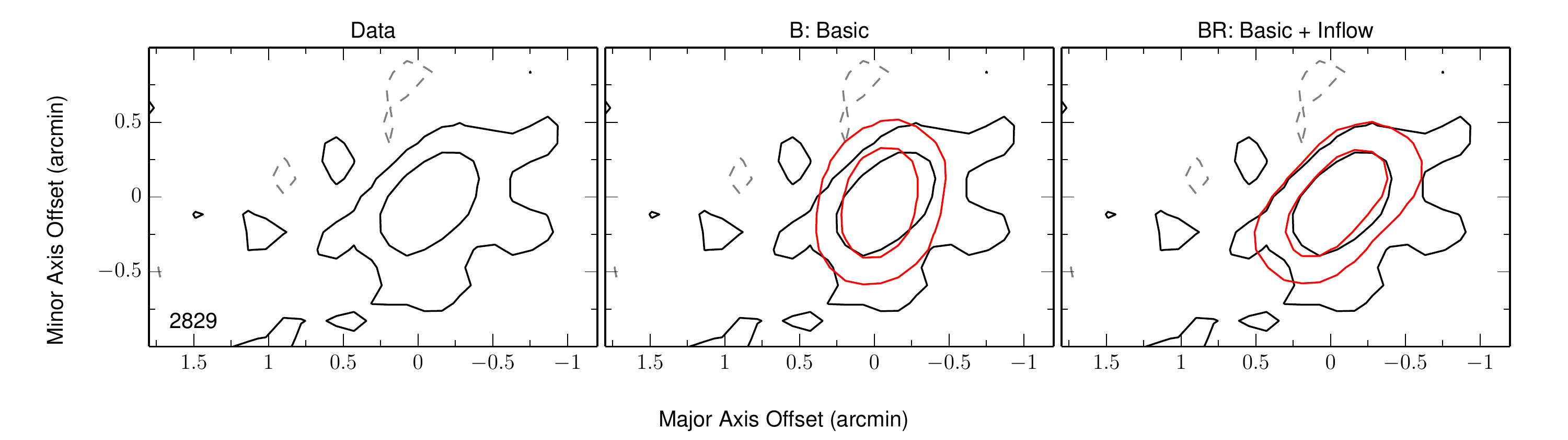}
     \caption{Channel map at 2829 km s$^{-1}$ from the observed data cube of NGC\,5529 (black and grey contours), the \textit{B} model and the \textit{BR} model. Contour levels are -0.8, 0.8 (1.5$\sigma$) and 5.6 mJy beam$^{-1}$. The orientation of the \textit{B} model is clearly too vertical. Adding a radial inflow significantly improves the agreement with the data.}
     \label{fig:N5529_models_channels_zoom}
\end{figure*}
\begin{figure}[]
  \resizebox{\hsize}{!}{\includegraphics{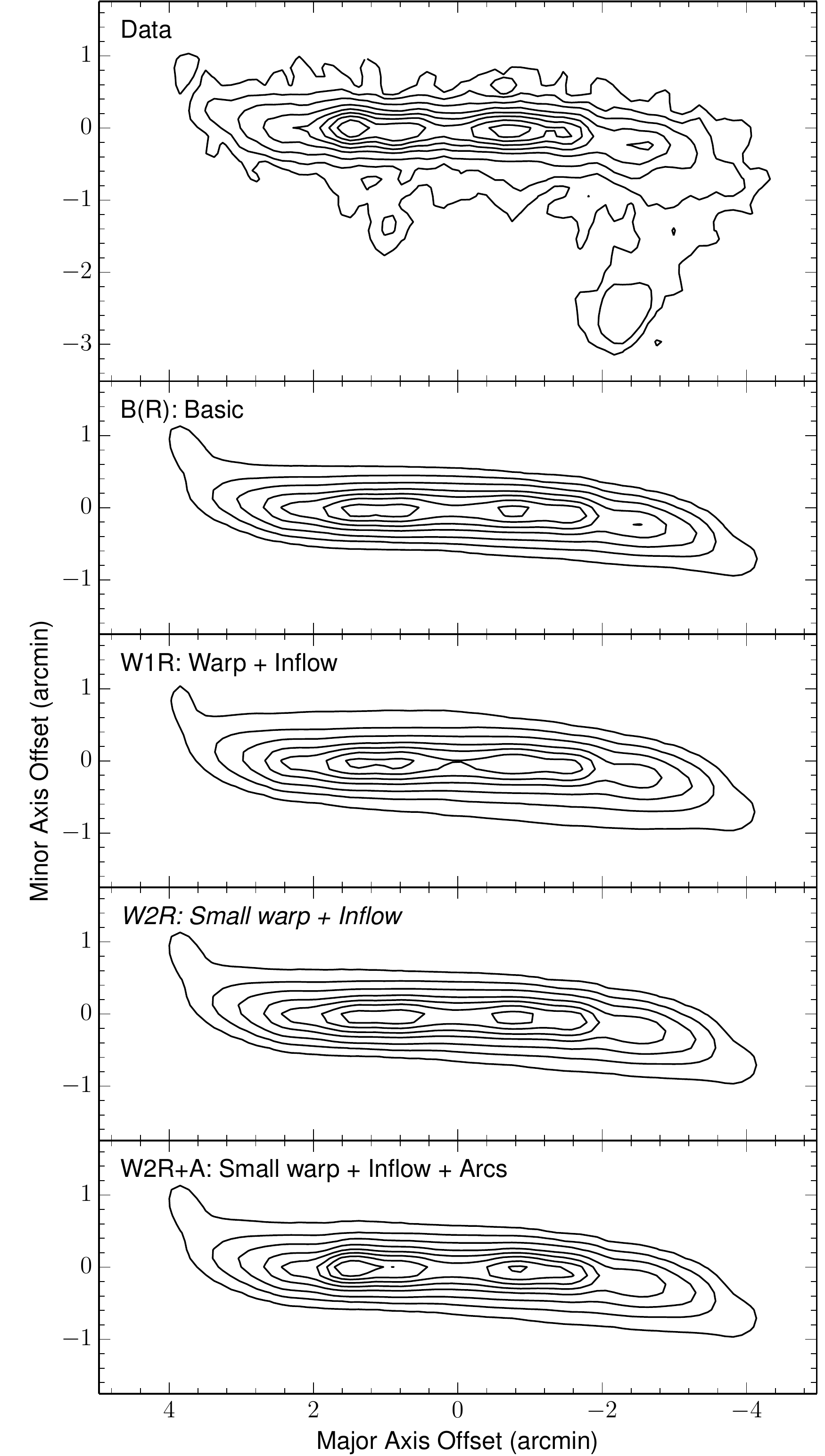}}
  \caption{Total H{\sc{i}} maps of the various models discussed here as compared to the observed total H{\sc{i}} map of NGC\,5529. The \textit{W2R} model is our best model. Contour levels are the same as in Fig. \ref{fig:N5529_V_mom0}.}
  \label{fig:N5529_models_mom0}
\end{figure}
Imposing a simple, one-component disk to start the modelling, the best agreement with the data was achieved with an inclination of 86.4$\degr$ and a scale height of 0.4 kpc. The approaching side of the resulting \emph{B} model is dominated by a gradual warp that starts early on in the disk, at a radius of 15 kpc, and reaches a maximum deviation of 10$\degr$ with respect to the central position angle. The receding side shows a more abrupt warp, starting at a radius of 37 kpc and rapidly increasing to a maximum deviation of 16$\degr$. Comparing the \textit{B} model to the data in the channel maps (Figs. \ref{fig:N5529_models_channels}, \ref{fig:N5529_models_channels_zoom}) and the moment-0 map (Fig. \ref{fig:N5529_models_mom0}), we see that the model already provides a reasonable approximation to the data, although there are still some issues to address. Firstly, part of the observed channel maps on the approaching side contain additional emission that is trailing behind the model. This emission is indicated with arrows at 2615, 2697 and 2780 km s$^{-1}$ in Fig. \ref{fig:N5529_models_channels} and is also visible in Fig. \ref{fig:N5529_final_channels} between 2598 and 2763 km s$^{-1}$. However, in the outermost channels the model reproduces all the observed emission very well. In addition, the extra emission never exceeds the outer edge of the (integrated) model disk. These two arguments indicate that the model disk extends out far enough radially and is rotating at the correct velocity and that the extra emission is not the projection of larger rings. We thus interpret this emission as extraplanar gas resulting from the interaction with MGC +06-31-085a and do not include it in our models of the main disk of NGC\,5529.
\par
A second, more subtle deficiency of the \emph{B} model is best visible in the innermost channel maps, for example at 2780 and 2879 km s$^{-1}$ in Fig. \ref{fig:N5529_models_channels}, and is highlighted for the 2829 km s$^{-1}$ channel in Fig. \ref{fig:N5529_models_channels_zoom}. While the model emission in these channels is oriented more or less perpendicular to the major axis (vertical in Figs. \ref{fig:N5529_models_channels}, \ref{fig:N5529_models_channels_zoom}), the observed emission shows a slant and is oriented more diagonally, from the top right to the bottom left. This discrepancy cannot be resolved by changing the geometry of the disk. Instead, we model the slant by introducing inward radial motions of -15 km s$^{-1}$ in the entire disk (\emph{BR} model). We note that from our H{\sc{i}} data alone it is in principle not possible to determine which half of the disk (north or south of the major axis in Fig. \ref{fig:N5529_V_mom0}) is the near side and which is the far side. In the optical, however, we clearly see that the dust lane is mainly located north of the major axis, indicating that this is the near side. The radial motions in our model are inward under this assumption. The addition of a radial inflow significantly improves the agreement with the data, especially in the inner channels. On the other hand, this purely kinematical adjustment has no influence on the moment-0 map.
\par
The total H{\sc{i}} map of the \emph{BR} model is therefore identical to that of the \emph{B} model. In general it is in good agreement with the observed moment-0 map (Fig. \ref{fig:N5529_models_mom0}), but at the lowest contour levels it is too thin. Adding a flare to the model can immediately be ruled out as potential solution to this problem. Indeed, in the outermost channel maps we see that the model emission is thick enough and matches the data very well. Adding a flare would make the model significantly too thick in these channels. A second alternative consists of adding a line of sight warp. We tested this option by gradually lowering the inclination from 86.4$\degr$ to 82.0$\degr$ between radii of 40 kpc (2.8$\arcmin$) and 47 kpc (3.3$\arcmin$) and keeping it fixed at 82.0$\degr$ beyond that. The moment-0 map of the resulting \emph{W1R} model now has the correct thickness, but in the channel maps this model does not provide a consistent improvement over the \emph{BR} model (Fig. \ref{fig:N5529_models_channels}). In some channels the observed emission broadens on the outer side and shows a triangular-like shape, which is nicely reproduced by the \emph{W1R} model (e.g. at 2615 km s$^{-1}$ in Fig. \ref{fig:N5529_models_channels}). In other channels, however, this broadening is less strong or even absent and the emission from the \emph{W1R} model is too thick on the outer side (e.g. at 2879 and 2961 km s$^{-1}$). The absence of a systematic trend in the data leads us to believe that the strong broadening of the observed emission in some of the channels is not a real feature of the disk, but is rather caused by the extraplanar gas from the interaction with the satellites. As a consequence, the lowest contours in the observed moment-0 map most probably also reflect the extraplanar gas and not the real thickness of the disk.
\par
\begin{figure*}[]
\centering
   \includegraphics[width=0.9\textwidth]{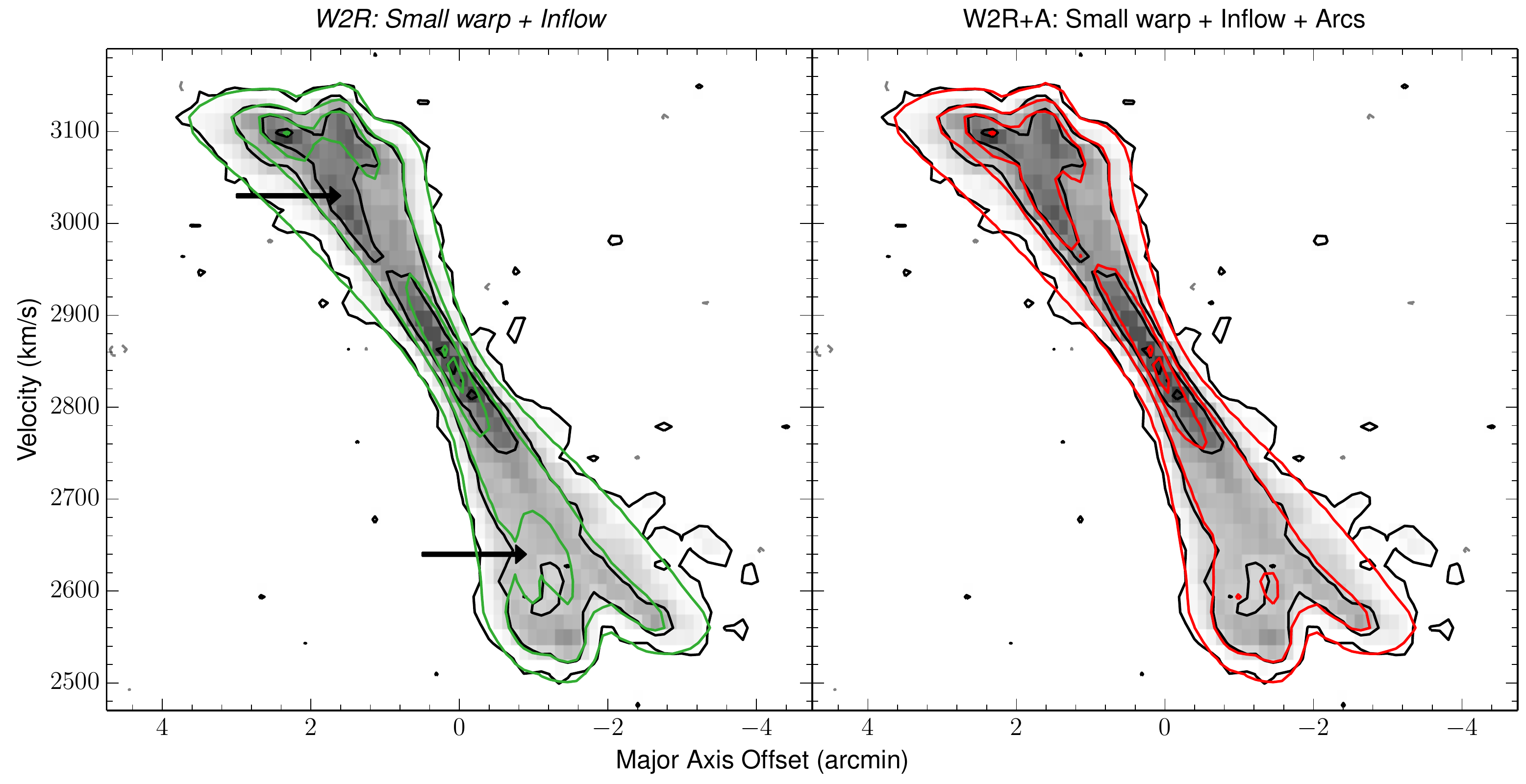}
   \caption{Observed major axis position-velocity diagram of NGC\,5529 (black and gray contours) overlaid with different models. The green contours show the final \textit{W2R} model, the red contours show the \textit{W2R+A} model. Contour levels are -0.8, 0.8 (1.5$\sigma$), 5.6, 11.2 and 16.8 mJy beam$^{-1}$. The greyscale corresponds to the observations. The arrows indicate a H{\sc{i}} ridge that is not reproduced by the \textit{W2R} model.}
     \label{fig:N5529_pv_models}
\end{figure*}
Notwithstanding the arguments from the previous paragraph, the data do show indications of a small line of sight warp. Indeed, comparing the \emph{BR} model to the data in the channel maps we see that the outer tip of the model emission is generally a bit too extended and sharp. In order to correct for these effects without affecting the central channels we constructed the \emph{W2R} model. This has a central inclination of 87.5$\degr$ and a constant scale height of 1.1 kpc. The inclination then gradually decreases to 85.0$\degr$ between 40 kpc and 47 kpc from the center and stays constant beyond that. This model provides a subtle but significant improvement over the \emph{BR} model.

\subsubsection{Spiral arms}
\label{sec:N5529_spiralarms}

The terminal edges of the major axis XV-diagram (Fig. \ref{fig:N5529_pv_models}, left panel) are reproduced rather well by the \textit{W2R} model, both on the approaching and the receding half. Similarly to the case of IC\,2531, however, the XV-diagram of NGC\,5529 shows an additional H{\sc{i}} ridge that is not reproduced by our model of the gas disk (indicated by arrows in the left panel of Fig. \ref{fig:N5529_pv_models}). And similarly to the case of IC\,2531, \cite{kregel04b} found that this ridge coincides with a ridge of H$\alpha$ emission, suggesting a spiral arm. Given the many uncertainties that are involved, we made no attempt to include spiral arms in our model, but we did again construct a toy model (\emph{W2R+A}) that reproduces the H{\sc{i}} ridge by locally enhancing the surface brightness in various arcs in the disk. The major axis XV-diagram of this model is shown in the right panel of Fig. \ref{fig:N5529_pv_models} and the corresponding zeroth moment map and channel maps can be seen in Figs. \ref{fig:N5529_models_channels} and \ref{fig:N5529_models_mom0}. We see that with the addition of the arcs, the peak intensities of the model are indeed in agreement with the data.
\par
As for IC\,2531, we note that the addition of arcs to the \textit{W2R} model is not really physical and not unambiguous. We therefore consider the \textit{W2R+A} model as a toy model and take the \textit{W2R} model as our final model for the atomic gas disk of NGC\,5529. A comparison of this model to the full observed data cube is shown in Fig. \ref{fig:N5529_final_channels} and the major axis position-velocity diagrams are compared in Fig. \ref{fig:N5529_pv_models}. Table \ref{tab:final_params} and Fig. \ref{fig:final_params} show the main parameters of the final model. The inclination, position angle and scale height have errors of 2.0$\degr$, 3.0$\degr$ and 0.7 kpc respectively.

\subsubsection{About the radial motions}

The central channel maps of the observed cube show a clear slant that indicates the presence of radial motions in the disk. To model this effect we used a global inflow in the entire gas disk. We acknowledge that such a parametrisation is most probably too simplistic, but the data do not provide sufficient constraints to model the radial motions in more detail. In order to reproduce the observed slant, an inflow of -15 $\pm$ 5 km s$^{-1}$ is definitely required in the outer half of the disk. On the other hand, radial motions in the inner parts of the disk only cause very subtle changes in the channel maps. Unfortunately these subtle changes are completely dominated by the effects of the extraplanar gas in the observed cube. It is therefore not possible to determine from the data whether an inflow is also present in the inner disk or not. In fact, even a small outflow in the inner regions is not inconsistent with the data. As an example of this, Fig. \ref{fig:N5529_models_channels} compares our final model to a similar model (\textit{W2R2}) with no radial motions in the inner disk. The difference between both is indeed very small and it is not possible to determine which model is better. In order to minimise the number of free parameters we therefore opted to use a global inflow in the entire disk.
\par
The velocity of the radial motions derived in this work is comparable to the inflow of -20 km s$^{-1}$ found by \cite{deblok14} in the outer part of the gas disk of NGC\,4414. \cite{zschaechner12} report a strong, local inflow of -50 km s$^{-1}$ in the inner part and an outflow of 10 km s$^{-1}$ in the outer part of the gas disk of NGC\,4565.
\par
How should we interpret this radial inflow? It was already mentioned that NGC\,5529 shows strong evidence for the presence of a bar and prominent spiral arms. These features are indeed known to generate non-circular flows in the disk of a galaxy. However, theory and simulations show that bars and spiral arms induce an inflow of gas inside the corotation radius and an outflow outside the corotation radius \citep{kenney94,kormendy04}. Since corotation generally occurs at only several kpc from the center of a galaxy \citep{elmegreen92}, it is unlikely that the inflow we find for NGC\,5529 is driven by a bar or spiral arms. Instead it might be related to ongoing accretion of gas from the satellite galaxies. Indeed, when the accreted gas has a lower angular momentum than the gas in the disk, it can generate a radial inflow in the galaxy \citep{matteucci12}.

\subsection{NGC\,5907}

The closest member of our sample is NGC\,5907, located at a distance of 16.3 Mpc. It is a disk-dominated galaxy with no significant bulge visible. In contrast to its morphological classification as an SA(s)c type galaxy, \cite{garcia-burillo97} saw strong evidence for the presence of a bar in their CO(1-0) observations of the nucleus of NGC\,5907. Almost 40 years ago, \cite{sancisi76} already noticed that the H{\sc{i}} disk of NGC\,5907 is strongly warped. This warp is also visible in the optical, although less obvious. NGC\,5907 is a member of the NGC\,5866 group (or LGG\,396), but the three other members of this group are located at large distances. NGC\,5907 was therefore long considered as the prototype of an isolated galaxy with a warped disk, until \cite{shang98} discovered a small companion, PGC\,54419, in their H{\sc{i}} map and an extended stellar tidal stream around the galaxy in their optical observations. Deeper optical imaging by \cite{martinez08} later revealed even more loops of stellar debris. The latter also conducted N-body simulations showing that most of the stellar loops they observed can be explained by the accretion of a single satellite galaxy.
\begin{figure}[h]
  \resizebox{\hsize}{!}{\includegraphics{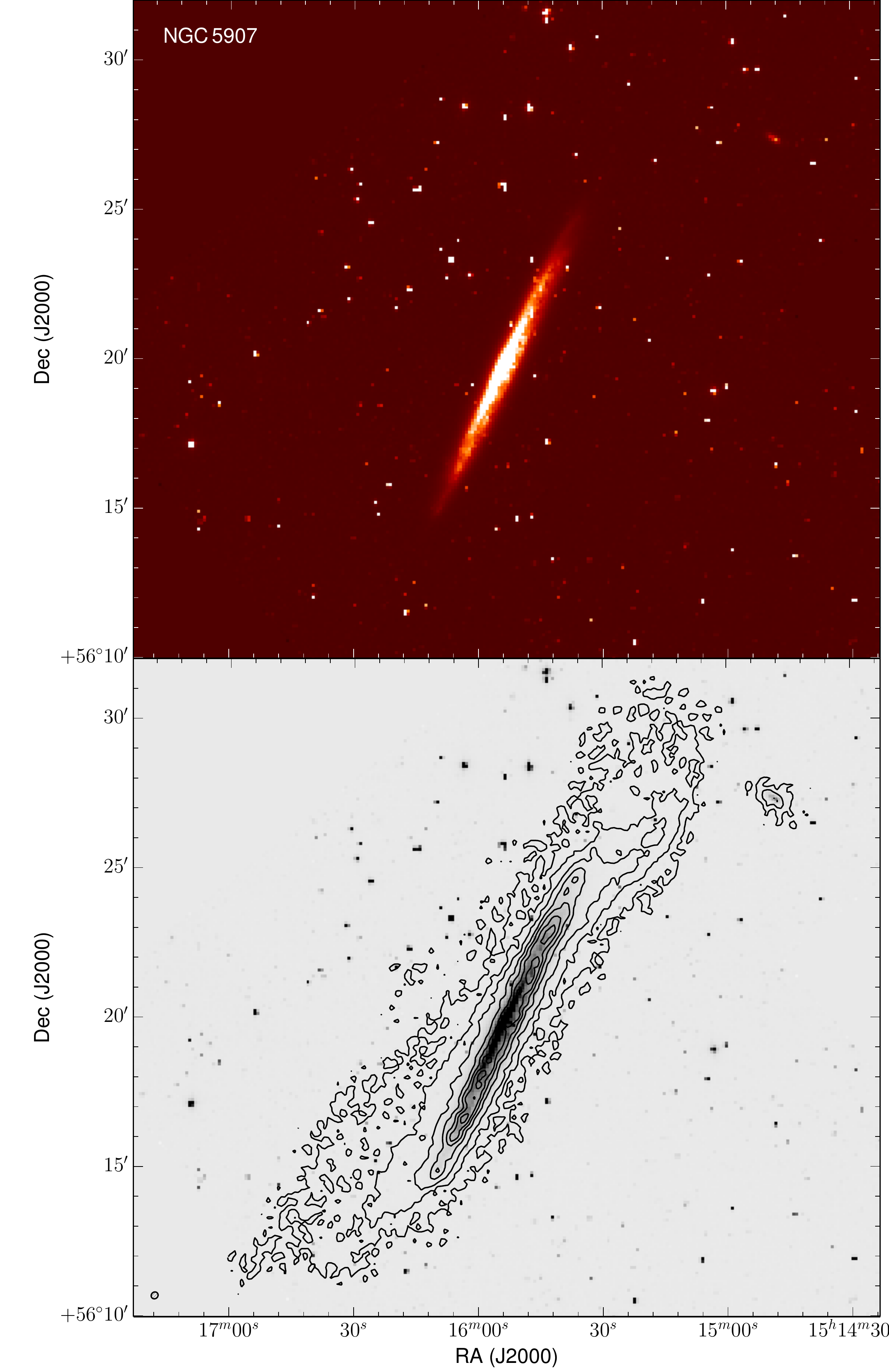}}
  \caption{Top: False color r-band image of NGC\,5907 (SDSS J151553.77+561943.5) from the EFIGI catalogue \citep{baillard11}, based on data from the SDSS DR4. Bottom: H{\sc{i}} contours overlaid on the same r-band image. Contours start at $1.01 \times 10^{20}$ atoms cm$^{-2}$ and increase as 5, 15, 30, 45, 60 and 75 times this value. North is up, East is to the left.}
  \label{fig:N5907_optical}
\end{figure}

\subsubsection{The data}

The total H{\sc{i}} map of NGC\,5907 as derived from our data is compared to an r-band image from the SDSS in Fig. \ref{fig:N5907_optical}. In the optical image we see that the dust lane is located slightly below or west of the major axis. We therefore interpret that side as the front half of the disk and the east side as the back half. The H{\sc{i}} reveals a more complex structure than would be expected from the optical image. Indeed, the main gas disk is strongly warped and slightly asymmetric, with the receding (NW) half showing a slight upturn in the outer regions. Additionally, an extended body of low-density gas seems to surround the main disk (in projection). Strikingly, this low-density gas is located almost entirely above (i.e., east of) the disk. Finally, the companion galaxy PGC\,54419 is visible toward the NW of NGC\,5907. We note that the atomic gas content of NGC\,5907 might actually extend even further than seen in our map since a significant amount of low-density gas is likely still hidden in the noise in our channel maps. Deeper WSRT observations with a total integration time of about $16 \times 12$ hours were recently conducted by G. J\'{o}zsa and a detailed modelling of these data is currently underway (Yim et al., in prep.).
\par
Excluding the companion, we measure a total H{\sc{i}} flux of 317.6 Jy km s$^{-1}$. This corresponds to an atomic hydrogen mass of $1.99 \times 10^{10}$ M$_{\odot}$ if we assume a distance of 16.3 Mpc. NED lists a variety of single dish fluxes, ranging from 36.6 Jy km s$^{-1}$ \citep{huchtmeier89} to 282.5 Jy km s$^{-1}$ \citep{springob05}. \cite{shang98} did not measure the total H{\sc{i}} flux of NGC\,5907. Somewhat surprisingly, our interferometric flux measurement is higher than all the single dish fluxes. Keeping in mind the large angular extent of NGC\,5907, the limited HPBW of many of the single dish observations is likely part of the reason for this. In addition, uncertainties in the flux calibration generally also cause differences.

\subsubsection{Models}
\label{sec:N5907_models}

\begin{figure}[]
  \resizebox{\hsize}{!}{\includegraphics{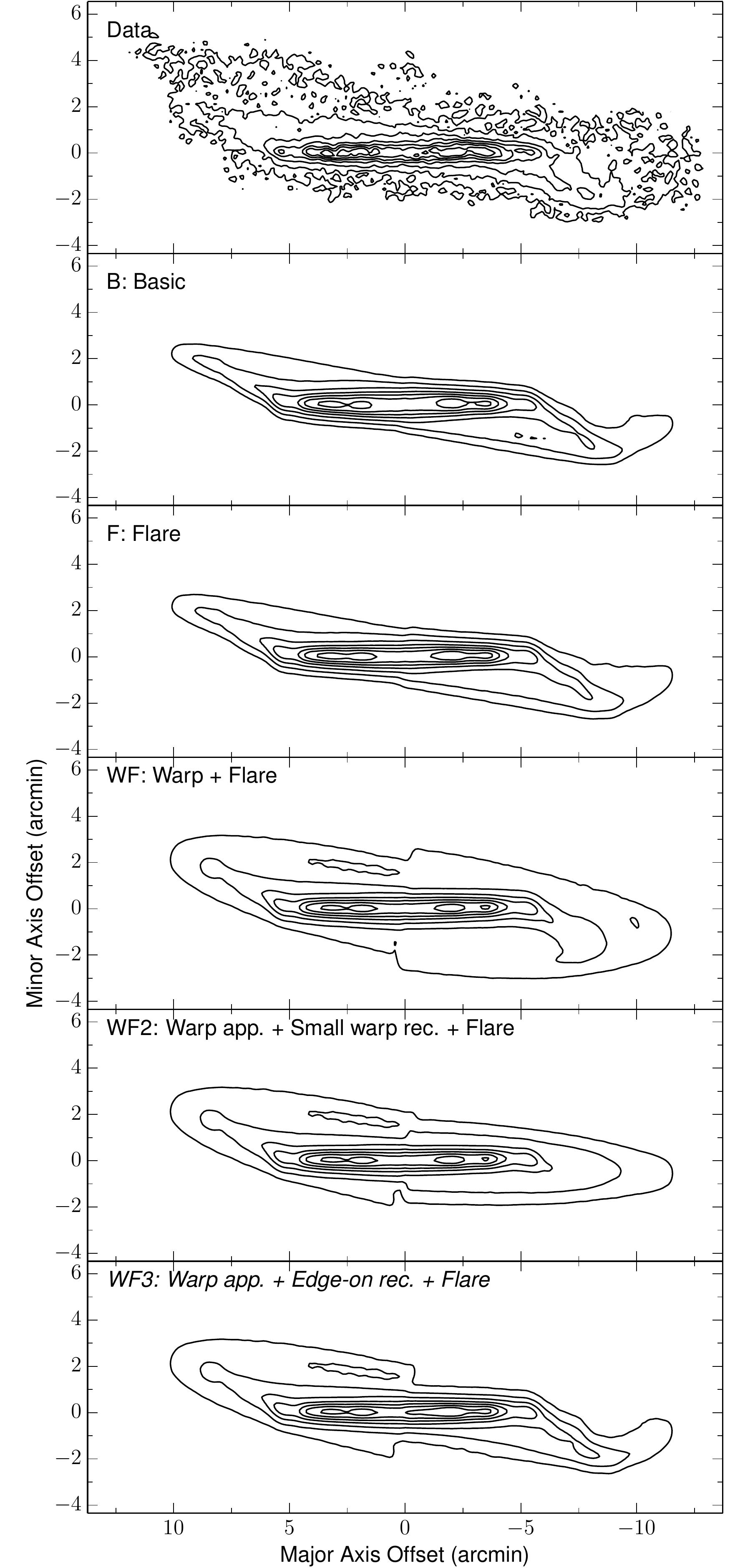}}
  \caption{Total H{\sc{i}} maps of the various models discussed here as compared to the observed total H{\sc{i}} map of NGC\,5907. The \textit{WF3} model is our best model. Contour levels are the same as in Fig. \ref{fig:N5907_optical}.}
  \label{fig:N5907_models_mom0}
\end{figure}
\begin{figure*}[]
\centering
   \includegraphics[width=\textwidth]{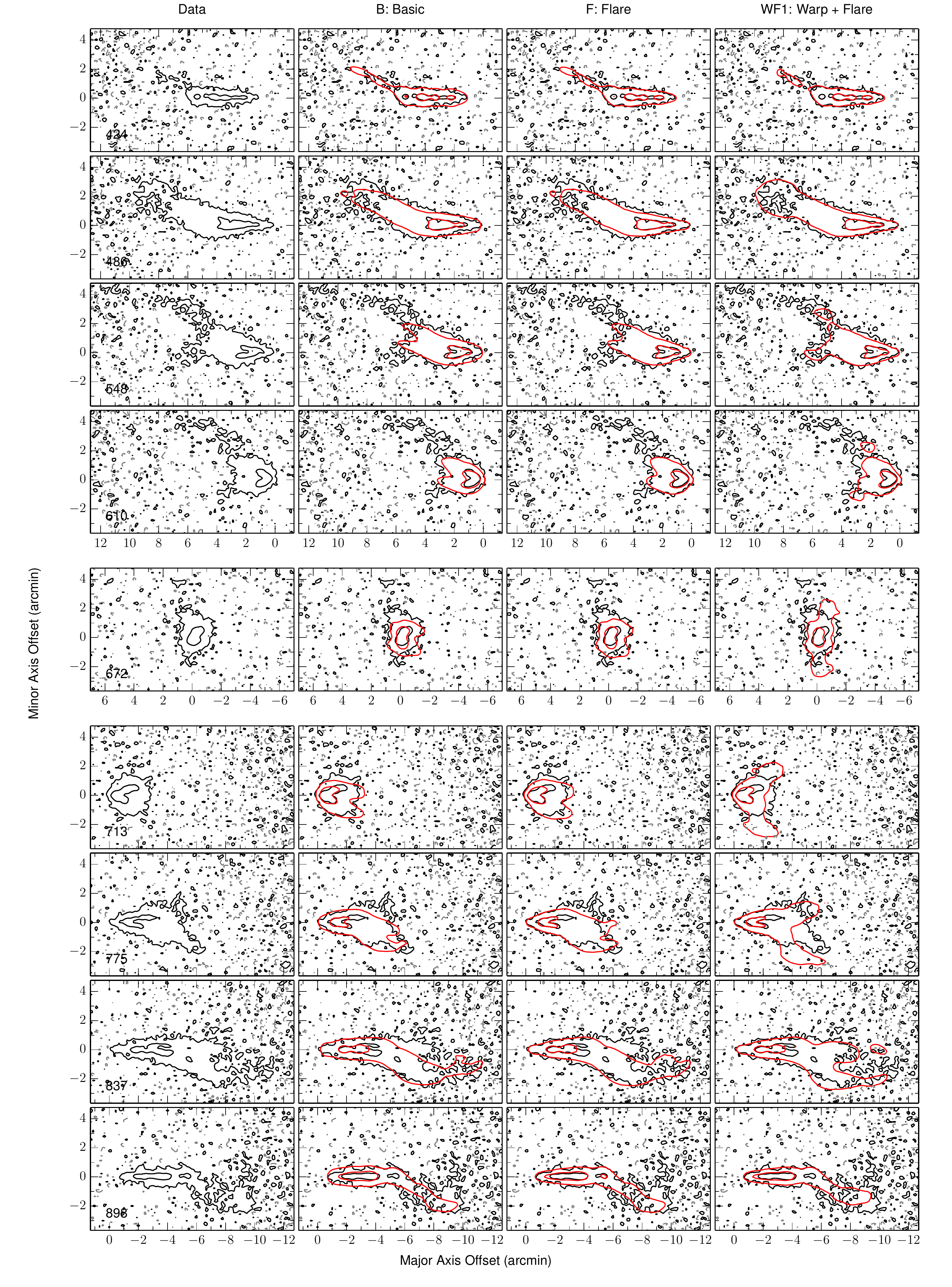}
     \caption{Representative channel maps from the observed data cube of NGC\,5907 and the various models. Contour levels are -0.5, 0.5 (1.5$\sigma$) and 6.4 mJy beam$^{-1}$. The black contours show the observations, with negative contours as dashed grey. The red contours represent the models. The systemic velocity is 664 $\pm$ 7 km s$^{-1}$.}
     \label{fig:N5907_models_channels}
\end{figure*}
\begin{figure*}[]
\ContinuedFloat
\centering
   \includegraphics[width=0.78125\textwidth]{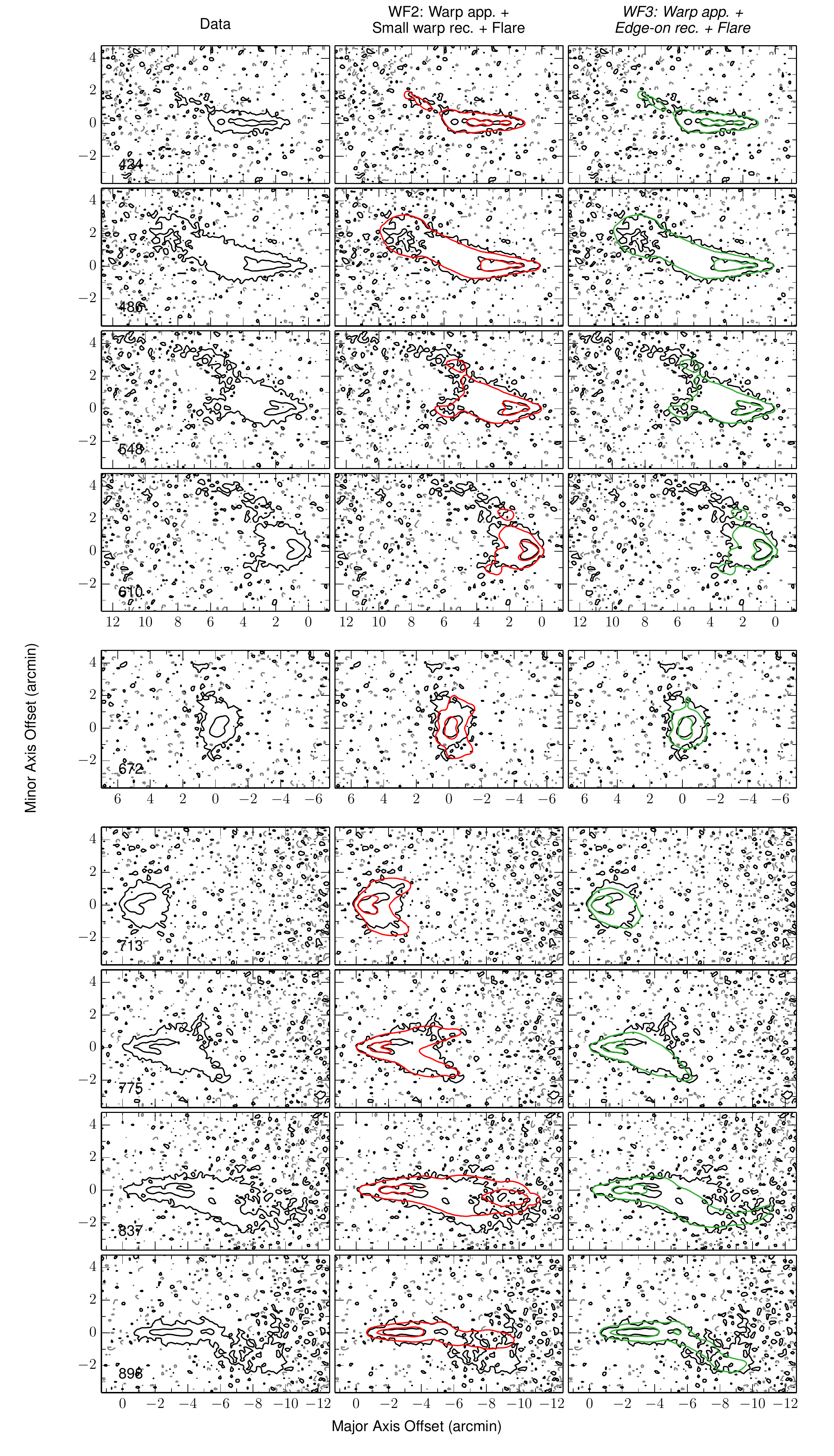}
     \caption{continued. The final (\textit{WF3}) models is shown as green contours.}
\end{figure*}
The best fitting single-component (\emph{B}) model has a constant inclination of 85.4$\degr$ and scale height of 1.2 kpc. The strong warp on the approaching side is reproduced with a sudden strong decrease of the position angle at a radius of 26 kpc (see e.g. Fig. \ref{fig:N5907_models_mom0}). This decrease then becomes more gradual and continues out to the edge of the disk, where the position angle reaches a maximum deviation of 12.9$\degr$ with respect to the center. On the receding side the position angle initially shows a similar behaviour with a maximum deviation of 12.9$\degr$ with respect to the center. However, at a radius of 45 kpc the position angle then starts to increase again, eventually deviating only 2.8$\degr$ from the central value in the outer regions of the disk (Fig. \ref{fig:N5907_models_mom0}). We also note that in the best fitting \emph{B} model the position angles in the inner, flat part of the disk differ by 1.8$\degr$ between the approaching and receding sides. Comparing this model to the observed data cube we see that in the outer channels (e.g. at 424, 486, 548, 837 and 898 km s$^{-1}$ in Fig. \ref{fig:N5907_models_channels}) the model is clearly too thick in the central regions of the disk. Additionally, the inclination in the outer regions of the model still seems too high, since the observed emission in the channel maps clearly `opens up' more than the model does. We address these issues one by one.
\par
Firstly we note that the central parts of the model disk appear too thick primarily in the outer channels, near the terminal velocities, where the inclination only has a negligible effect. This indicates that the problem is related to the scale height rather than the inclination. We therefore added a flare to the \emph{B} model by lowering the scale height in the central rings. The resulting \emph{F} model has a central scale height of 0.5 kpc out to a radius of 4.2 kpc. It then linearly increases to a maximum value of 1.6 kpc at a radius of 32.7 kpc and stays constant after that. As can be seen in Fig. \ref{fig:N5907_models_channels} the addition of a flare improves the agreement with the data in the inner regions of the disk.
\par
\begin{figure*}[]
\centering
   \includegraphics[angle=90,width=0.75\textwidth]{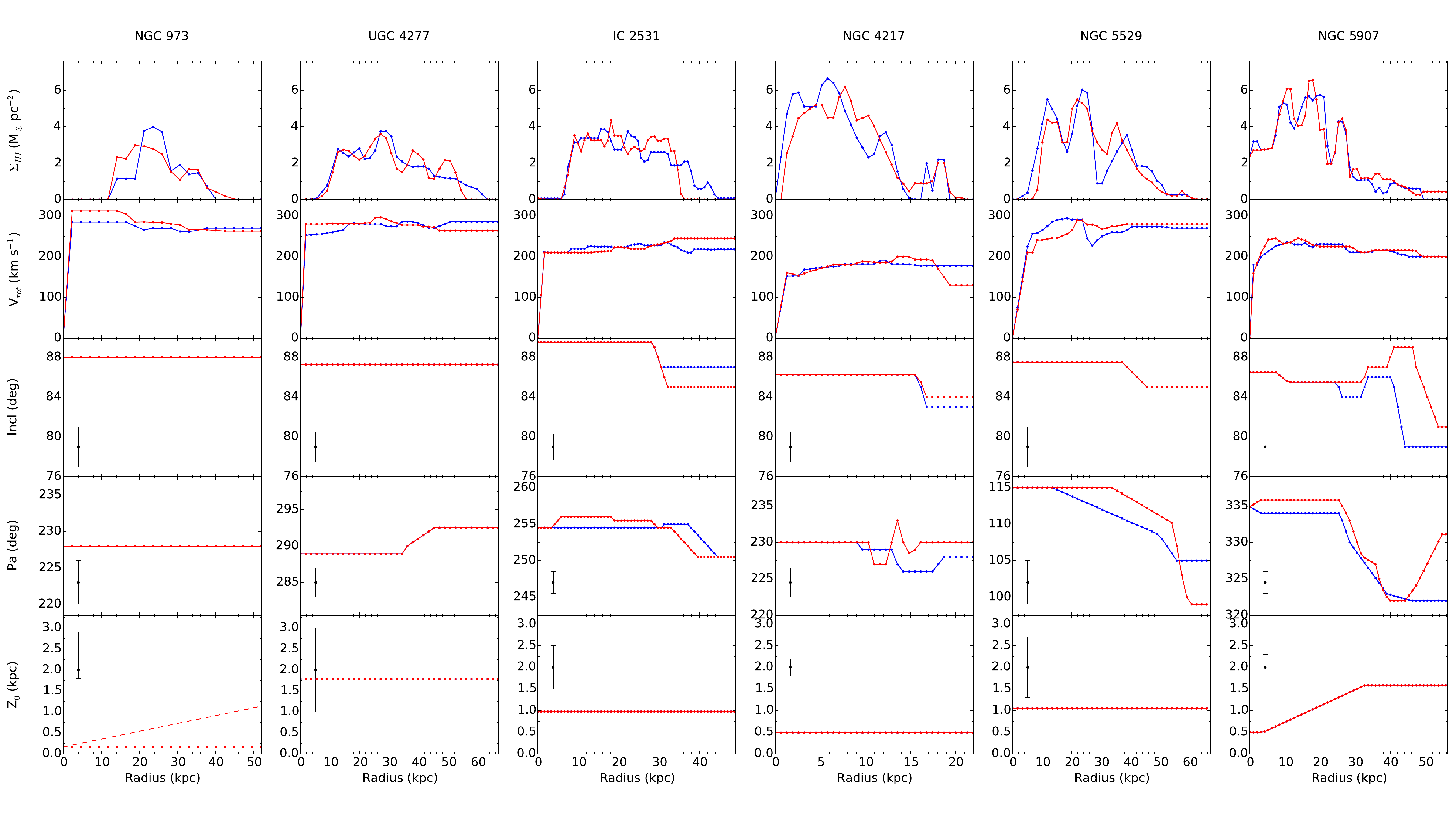}
     \caption{Variation of the radially dependent parameters of the final models. From top to bottom: the H{\sc{i}} surface density, the rotation velocity, the inclination, the position angle and the vertical scale height. Blue indicates the approaching side, red the receding side. For NGC\,4217, the parameters of the \textit{B+R2} model for the outer ring are also shown. The dashed vertical line shows (approximately) where the main disk meets the outer ring. For NGC\,973, the flare from the \textit{F} model (Section \ref{sec:flare}) is shown as the dashed red line. The uncertainties are represented by the error bars.}
     \label{fig:final_params}
\end{figure*}

On the other hand the \emph{F} model still does not reproduce the forked shape of the observed emission in the outer parts of the disk. This forked shape is present on both the approaching and the receding half, which seem to behave very similarly at first sight. In a first attempt to model this behaviour we included a strong and abrupt line of sight warp in the \emph{F} model. With an inclination of 79$\degr$ in the outer part of the warp, the \emph{WF1} model broadly matches the forked shape of the observed emission on the approaching half. Indeed, although the model is slightly offset from the data in the very central channels and it does not include the faint tail that is observed above or on the back side of the disk, it is able to reproduce the general trend very well. On the receding half this is not the case. If the position of the model matches that of the observed emission in the outer channels and the inclination is taken to reproduce the opening of the upper arm, the lower arm of the model opens up far too strongly. This is clearly visible in Fig. \ref{fig:N5907_models_channels}, e.g. at 775 km s$^{-1}$. A smaller line of sight warp could resolve this problem, but also requires a change of the position angles in the warp to avoid now underestimating the opening of upper arm. Although this causes a clear offset between the model and the data at terminal velocities (at 898 km s$^{-1}$ in Fig. \ref{fig:N5907_models_channels}) we investigated this option with the \emph{WF2} model. The agreement of this model with the data in the lower arm is still very poor. Indeed, it now underestimates the opening of the data in the outer channels while still opening up a bit too strongly in the central channels. In fact, the only way to match the observed position of the lower arm in all receding channels is with an almost edge-on orientation in the warp. This is demonstrated in Fig. \ref{fig:N5907_models_channels} by the \emph{WF3} model, with inclinations between 87$\degr$ and 89$\degr$ throughout most of the warp and dropping to 81$\degr$ only in the outer tip of the disk (as shown in Fig. \ref{fig:final_params}). 
\par
The two arms of the forked emission on the receding side thus seem to behave systematically different and cannot be modelled simultaneously with a simple, axisymmetric model. The lower arm behaves like an edge-on disk while the upper arm opens up and seems to suggest a significantly lower inclination. Unfortunately it is not possible to give a clear interpretation of this feature based on our data. Maybe only the back side of the receding half has a line of sight warp and the front side does not. On the other hand, the upper arm could also be caused by an extraplanar, arc-like feature related to the recent or ongoing interaction. Given the close companion and the stellar tidal streams that have been observed in the optical, it is indeed not unlikely that the gas disk of NGC\,5907 is somewhat disturbed. As further modelling based on our data would degrade into guessing and speculating, we take the \emph{WF3} model as our (preliminary) final model and refer the reader to the future work of Yim et al. (in prep.). Their much deeper WSRT data will reveal the full extent of the atomic gas disk of NGC\,5907 and provide additional constraints on its structure.
\par
\begin{figure}[]
   \resizebox{\hsize}{!}{\includegraphics{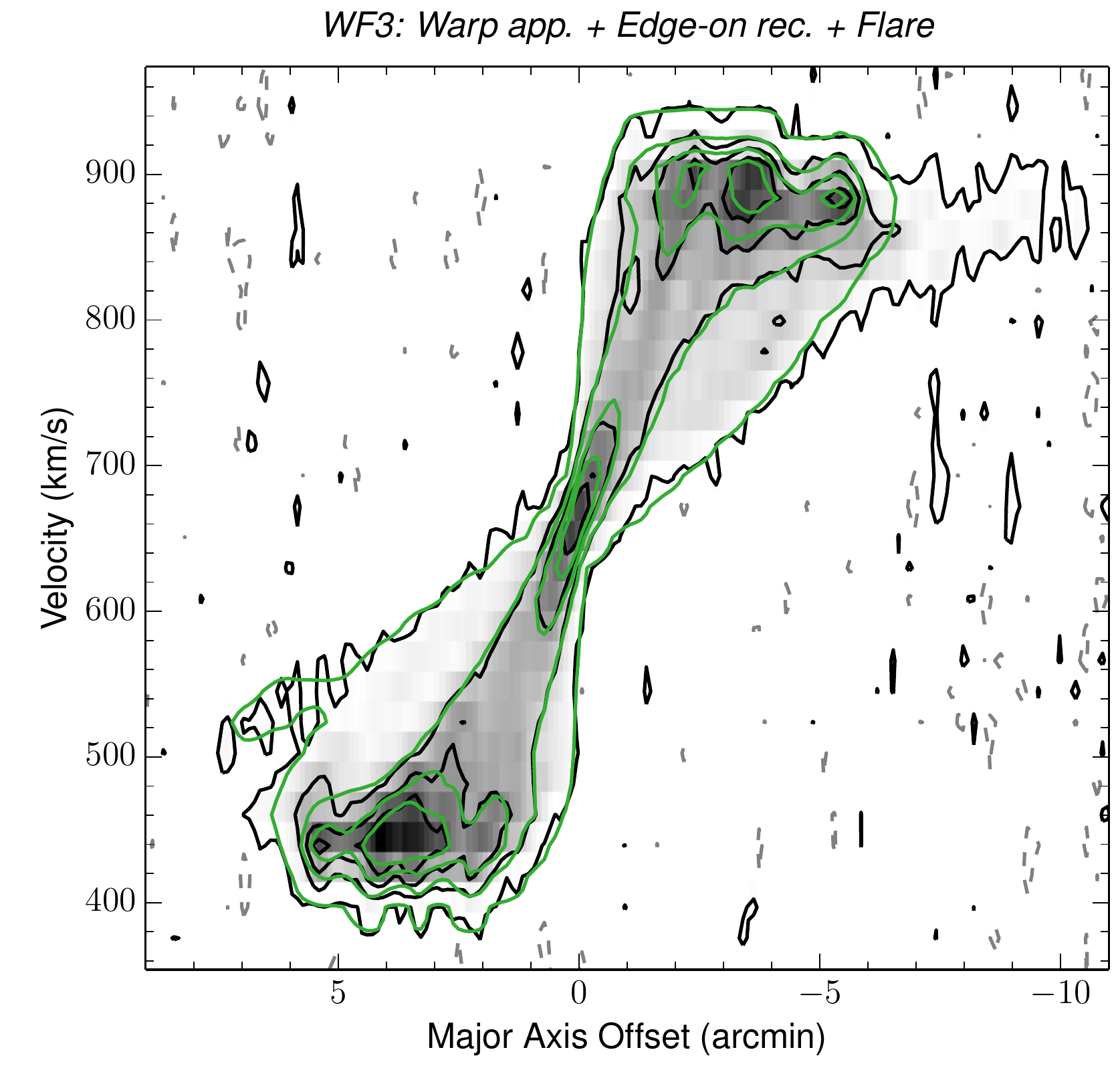}}
   \caption{Observed major axis position-velocity diagram of NGC\,5907 (black and gray contours) overlaid with the final (\emph{WF3}) model (green contours). Contour levels are -0.5, 0.5 (1.5$\sigma$), 4.8, 9.6 and 14.4 mJy beam$^{-1}$. The greyscale corresponds to the observations.}
     \label{fig:N5907_pv_models}
\end{figure}

The variation of the radially dependent parameters of the model is shown in Fig. \ref{fig:final_params}. In contrast to the warped outer regions, the inner disk of NGC\,5907 appears to be remarkably symmetric. The best fit values of the remaining parameters are listed in Table \ref{tab:final_params}. The inclination, position angle and scale height have errors of 1.0$\degr$, 1.5$\degr$ and 0.3 kpc respectively. A full comparison of the final model to the observed data cube is given in Fig. \ref{fig:N5907_final_channels} of the appendix and the major axis position-velocity diagrams are compared in Fig. \ref{fig:N5907_pv_models}. 
\begin{table*}[t]
\caption{Radially constant parameters of the final models.}
\label{tab:final_params}
\centering
\begin{tabular*}{\textwidth}{l @{\extracolsep{\fill}} c c c c c}
\hline \hline\noalign{\smallskip}
Galaxy & Center $\alpha$ & Center $\delta$ & $V_{sys}$ & Velocity dispersion & Radial motions \\
 & (J2000) & (J2000) & (km s$^{-1}$) & (km s$^{-1}$) & (km s$^{-1}$) \\
\hline\noalign{\smallskip}
NGC\,973 & 02\textsuperscript{h}34\textsuperscript{m}20.1\textsuperscript{s} $\pm$ 0.4\textsuperscript{s} & $+$32$\degr$30$\arcmin$19.9$\arcsec$ $\pm$ 4.0$\arcsec$ & 4770 $\pm$ 8 & 8.2 $\pm$ 2.0 & - \\
UGC\,4277 & 08\textsuperscript{h}13\textsuperscript{m}57.0\textsuperscript{s} $\pm$ 0.7\textsuperscript{s} & $+$52$\degr$38$\arcmin$53.9$\arcsec$ $\pm$ 5.0$\arcsec$ & 5370 $\pm$ 6 & 10.4 $\pm$ 3.0 & - \\
IC\,2531 & 09\textsuperscript{h}59\textsuperscript{m}55.5\textsuperscript{s} $\pm$ 0.5\textsuperscript{s} & $-$29$\degr$37$\arcmin$02.4$\arcsec$ $\pm$ 4.0$\arcsec$ & 2455 $\pm$ 6 & 11.2 $\pm$ 1.5 & - \\
NGC\,4217 & 12\textsuperscript{h}15\textsuperscript{m}51.1\textsuperscript{s} $\pm$ 0.6\textsuperscript{s} & $+$49$\degr$05$\arcmin$29.4$\arcsec$ $\pm$ 4.0$\arcsec$ & 1022 $\pm$ 6 & 19.1 $\pm$ 3.0 & - \\
NGC\,5529 & 14\textsuperscript{h}15\textsuperscript{m}34.0\textsuperscript{s} $\pm$ 1.0\textsuperscript{s} & $+$36$\degr$13$\arcmin$39.0$\arcsec$ $\pm$ 8.0$\arcsec$ & 2830 $\pm$ 6 & 12.3 $\pm$ 3.0 & $-$15 $\pm$ 5 \\
NGC\,5907 & 15\textsuperscript{h}15\textsuperscript{m}53.2\textsuperscript{s} $\pm$ 0.9\textsuperscript{s} & $+$56$\degr$19$\arcmin$42$\arcsec$ $\pm$ 15$\arcsec$ & 664 $\pm$ 7 & 13.3 $\pm$ 3.0 & - \\
\hline
\end{tabular*}
\end{table*}
\begin{figure}[]
  \resizebox{\hsize}{!}{\includegraphics{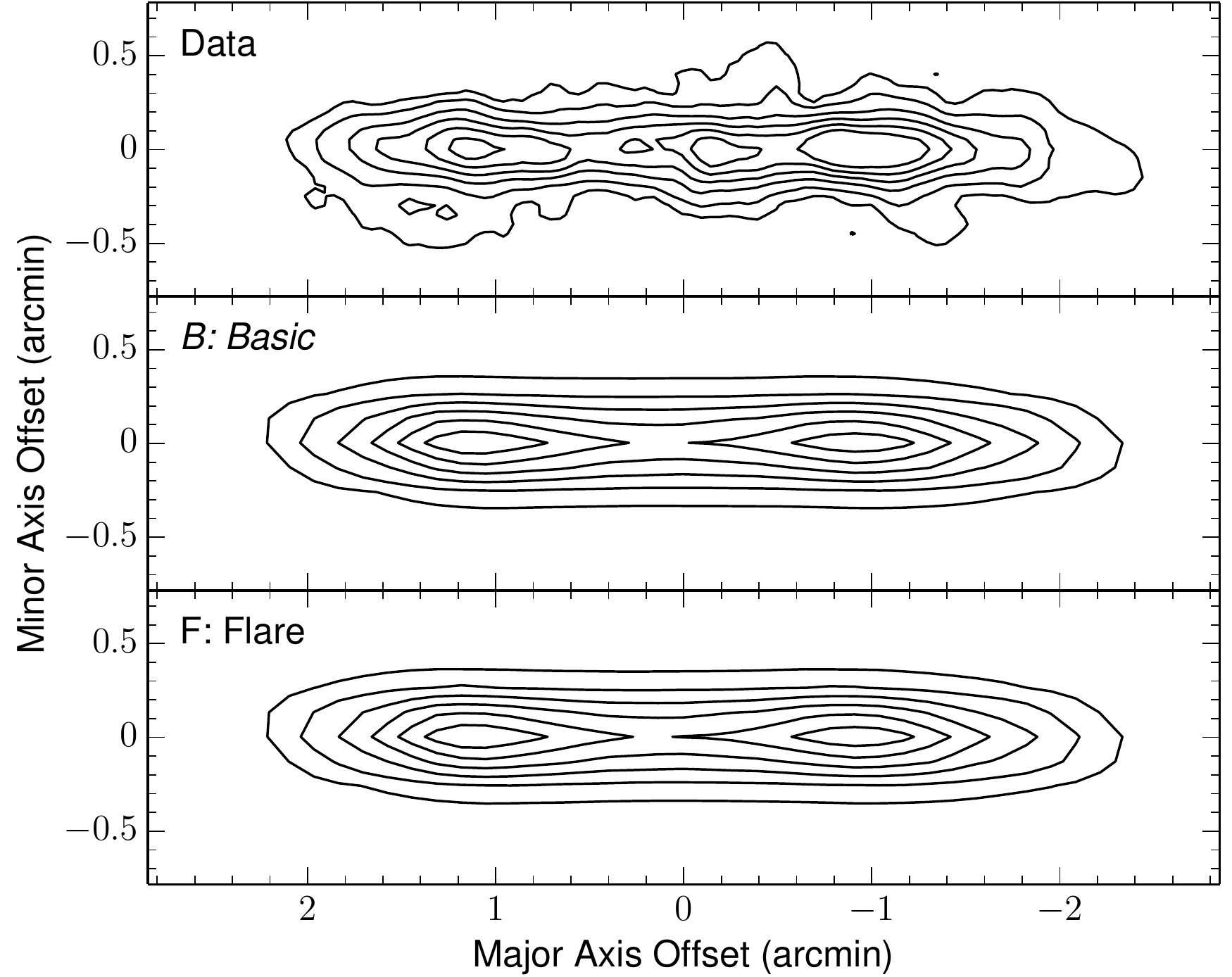}}
  \caption{Total H{\sc{i}} maps of the basic (\textit{B}) and flaring (\textit{F}) models of NGC\,973 as compared to the observed total H{\sc{i}} map. Contour levels are the same as in Fig. \ref{fig:N973_V_mom0}.}
  \label{fig:N973_models_mom0_flare}
\end{figure}
\begin{figure*}[]
\centering
   \includegraphics[width=0.65\textwidth]{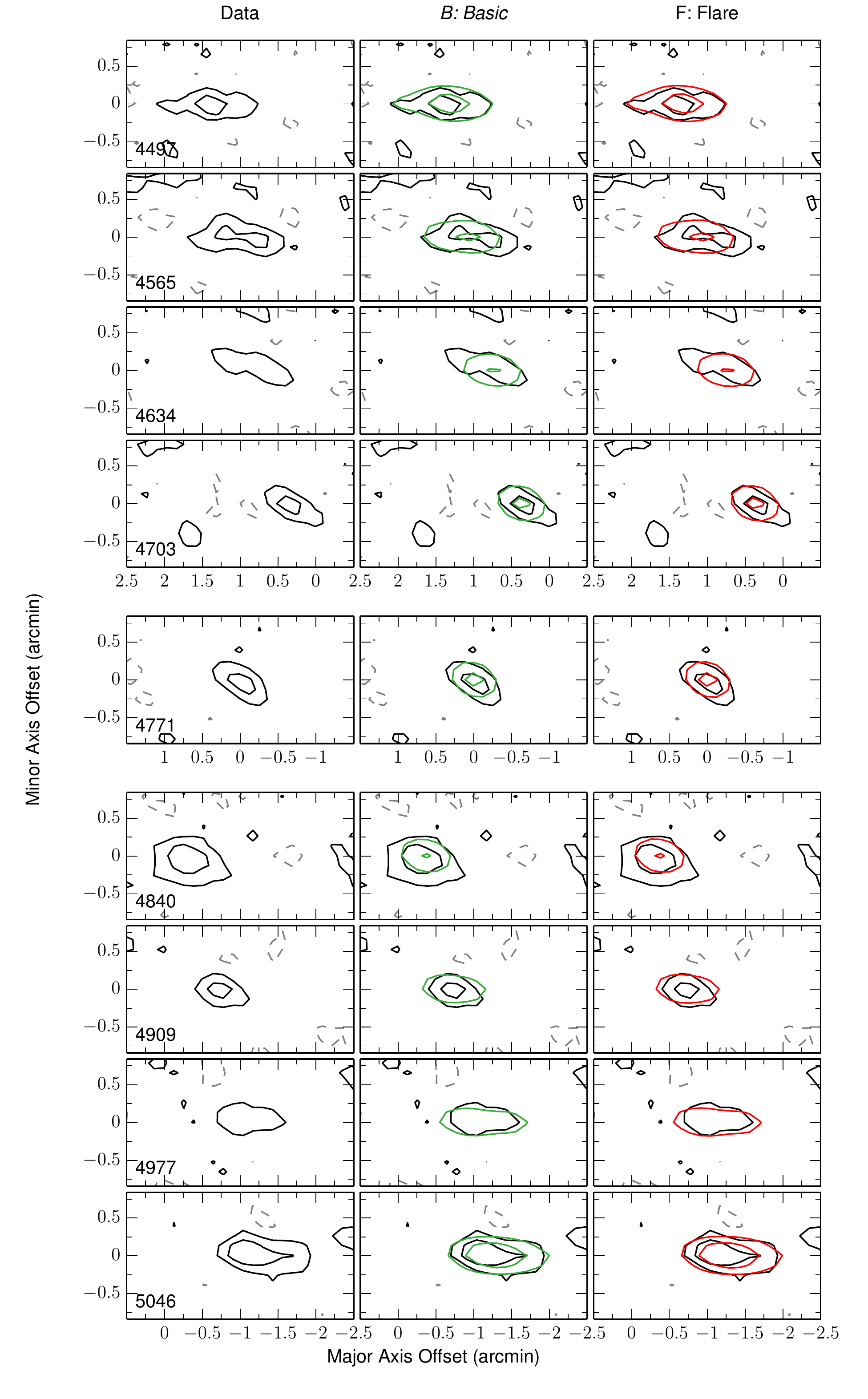}
     \caption{Representative channel maps from the observed data cube of NGC\,973 (black and grey contours), the basic (\textit{B}) model (green contours) and the flaring (\textit{F}) model (red contours). Contour levels are -2.3, 2.3 (1.5$\sigma$) and 6.2 mJy beam$^{-1}$. The systemic velocity is 4770 $\pm$ 8 km s$^{-1}$.}
     \label{fig:N973_models_channels_flare}
\end{figure*}

\section{Constant scale height or flare}
\label{sec:flare}

As was discussed in Section \ref{sec:results}, our final models generally have a constant vertical scale height throughout the entire gas disk. A flaring H{\sc{i}} layer is only required for NGC\,5907. Theoretically, however, it is expected that all spiral galaxies have a flared atomic gas disk. Indeed, in the inner regions, the gravitational potential is dominated by the stellar disk. If the vertical velocity dispersion of the gas (and hence the internal pressure in the gas disk) is independent of radius, the decreasing surface density of the stars with radius will result in an increasing scale height of the gas \citep[e.g.][]{vanderkruit81}. Beyond the stellar disk, the gravitational potential is dominated by the dark matter and the flaring of the gas disk is directly related to the shape of the DM halo \citep[e.g.][]{olling95,olling96,becquaert97,obrien10a,obrien10b,obrien10c,obrien10d,peters13}.
\par
We stress that these theoretical expectations are not inconsistent with our models. Indeed, as can be seen in Fig. \ref{fig:final_params}, the uncertainties on the scale height are generally significant and, in fact, leave room for a (moderate) flare. We illustrate this point for NGC\,973, by replacing the constant scale height of 0.2 kpc in the final (\textit{B}) model with a flare. In the resulting \textit{F} model, the scale height linearly increases from 0.2 kpc in the center to 1 kpc at a radius of 45 kpc (dashed red line in Fig. \ref{fig:final_params}). From Figs. \ref{fig:N973_models_mom0_flare} and \ref{fig:N973_models_channels_flare} it is clear that the \textit{B} and \textit{F} models are very similar. A flaring model is thus still fully consistent with the data and would, in fact, be more realistic from a theoretical point of view. The problem is that the details of the flare are generally ill constrained. Flares with different slopes or shapes would fit the data equally well and there is no way to tell which is best. For this reason we stick to a constant scale height in our final models, but note the reader that a flare is probably present within the uncertainties on the (constant) scale height. The range of scale heights allowed by our uncertainties is indeed generally of the same order as the amplitude of the flares found in other galaxies, like NGC\,4244 \citep{olling96} or the Milky Way \citep{kalberla08}.
\par
Finally we remark that from Sections \ref{sec:I2531_models} and \ref{sec:N4217_models} it might seem like we explicitely investigated the possibility of a flare for IC\,2531 and NGC\,4217 and discarded it. In fact, this is not entirely true. For both galaxies, only a certain type of flare was tested to model a specific feature in the data cube. In the case of IC\,2531, this flare did not significantly improve the model, but was not excluded by the data either. It was simply discarded to reduce the number of free parameters. For NGC\,4217, a flare was tested to reproduce the thinning of the emission in certain channel maps. When this was achieved, the resulting model now appeared slightly too thin in other channels and was therefore discarded. It should be noted, however, that the central scale height of this model was only 0.1 kpc, which is far beyond our error bars (0.5 $\pm$ 0.2 kpc). A flare within these uncertainties is not in disagreement with the data, but does not significantly improve the model either (and was therefore not discussed in Section \ref{sec:N4217_models}). Hence, the arguments from the previous paragraph are also valid for IC\,2531 and NGC\,4217. Within our uncertainties on the scale height, a flare can be added to the models without affecting the quality of the fit, but its parameters are ill constrained.

\section{Impact of the interactions}
\label{sec:discussion}

\begin{table*}[t!]
\caption{Main features of the final models}
\label{tab:main_features}
\centering
\begin{tabular*}{\textwidth}{l @{\extracolsep{\fill}} c c c c c}
\hline \hline\noalign{\smallskip}
Galaxy & Edge-on warp & LOS warp & Flare & Signs of spiral arms & Signs of interaction \\
\hline\noalign{\smallskip}
\smallskip
NGC\,973 & - & - & - & - & - \\
\smallskip
UGC\,4277 & \checkmark & - & - & - & - \\
IC\,2531 & \checkmark & \checkmark & - & \multicolumn{1}{l}{Bright ridge in} & - \\
\smallskip
 & & & & \multicolumn{1}{l}{XV-diagram} & \\
NGC\,4217 & \checkmark & - & - & - & \multicolumn{1}{l}{Coplanar ring outside} \\
\smallskip
 & & & & & \multicolumn{1}{l}{main disk} \\
NGC\,5529 & \checkmark & \checkmark & - & \multicolumn{1}{l}{Bright ridge in} & \multicolumn{1}{l}{H{\sc{i}} bridges to companions,} \\
 & & & & \multicolumn{1}{l}{XV-diagram} & \multicolumn{1}{l}{extraplanar gas,} \\
\smallskip
 & & & & & \multicolumn{1}{l}{radial motions} \\
NGC\,5907 & \checkmark & \checkmark & \checkmark & - & \multicolumn{1}{l}{Extended reservoir of} \\
 & & & & & \multicolumn{1}{l}{low-density gas,} \\
 & & & & & \multicolumn{1}{l}{asymmetric outer disk} \\
\hline
\end{tabular*}
\end{table*}
Figure \ref{fig:final_params} compares the radial variation of the main parameters of the final model for each galaxy. The final parameters that do not vary with radius are listed in Table \ref{tab:final_params} and an overview of the main features of each model is given in Table \ref{tab:main_features}. Although there are, of course, differences between the individual galaxies, they all show normal behaviour and there are no clear outliers in the sample. 
\par
As was discussed in Section \ref{sec:results}, the data cubes of NGC\,4217, NGC\,5529 and NGC\,5907 contain a number of morphologic and kinematic indications of recent or ongoing interactions with satellite galaxies. From Fig. \ref{fig:final_params} it appears that, apart from the radial inflow in the (outer) disk of NGC\,5529, these interactions have not strongly affected the main H{\sc{i}} disks of the galaxies in question and their influence is mostly limited to the position angles and inclinations in the outer regions. Indeed, the radial column density profiles and rotation curves of NGC\,4217, NGC\,5529 and NGC\,5907 do not behave systematically differently than those of the rest of the sample and there are no strong asymmetries between the approaching and the receding halves. On the other hand, NGC\,5529 and NGC\,5907 do have a significantly stronger edge-on warp than the other galaxies. For NGC\,4217 the edge-on warp has an unusual profile on the receding side, but its amplitude is similar to the warps of UGC\,4277 and IC\,2531. NGC\,5907 also has a stronger line of sight warp than the other galaxies, but, on the other hand, NGC\,4217 shows no line of sight warp in its H{\sc{i}} disk (only in the outer ring, as represented by the \textit{B+R2} model).
\par
\begin{figure}[]
  \resizebox{\hsize}{!}{\includegraphics{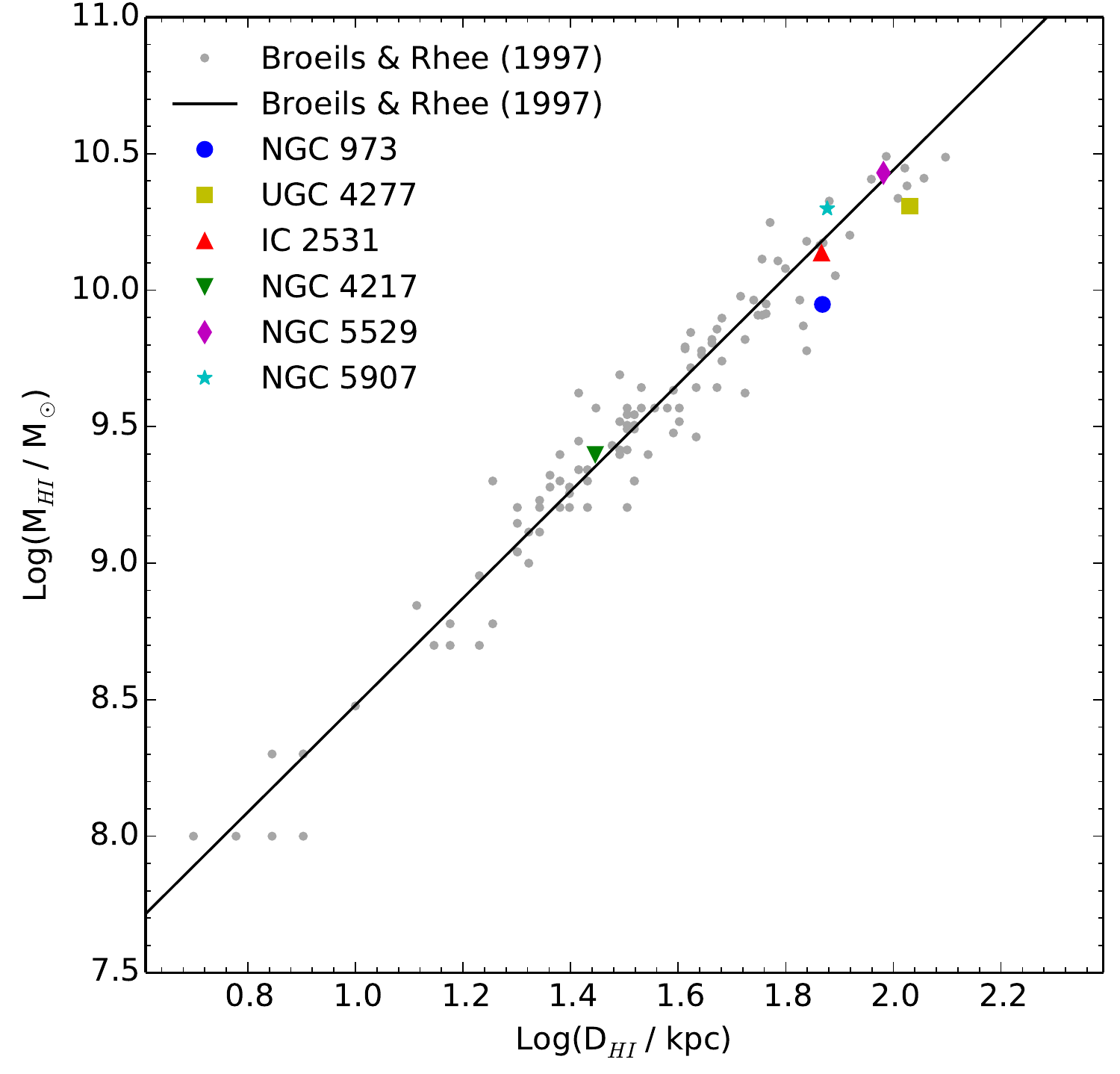}}
  \caption{The correlation between the total H{\sc{i}} mass and the diameter of the H{\sc{i}} disk for our sample (coloured symbols) and the sample of \cite{broeils97} (grey dots, based on their Table 1). The black line shows the best fit linear relation from \cite{broeils97}.}
  \label{fig:mass-size}
\end{figure}

In addition to these arguments, the galaxies in our sample are all in agreement with the H{\sc{i}} mass-size relation that was found by \cite{broeils97}. This can be seen in Fig. \ref{fig:mass-size} and further suggests that the atomic gas disks of the `interacting galaxies' have not been strongly disrupted by the interactions. Following \cite{broeils97}, we determined the H{\sc{i}} diameter $D_{HI}$ from our final models as the diameter where the H{\sc{i}} column density falls below 1 M$_{\odot}$ pc$^{-2}$. Since our models cannot adequately reproduce the outer ring of NGC\,4217, we excluded this structure and only used the main disk of NGC\,4217 to determine the data point in Fig. \ref{fig:mass-size}. We note that with the inclusion of the outer ring, based on the \textit{B+R2} model, NGC\,4217 is still in agreement with the relation from \cite{broeils97}.

\section{Conclusions}
\label{sec:conclusions}

In the frame of the HEROES project we present an analysis of the atomic gas content of 6 edge-on spiral galaxies. This work fits in a larger effort to derive the radial and vertical distribution and the properties of the gas, dust, stars and dark matter in a sample of 7 massive edge-on spiral galaxies. New GMRT H{\sc{i}} 21-cm observations of NGC\,973 and UGC\,4277 are combined with re-reduced archival data of IC\,2531, NGC\,4217, NGC\,5529 and NGC\,5907. The atomic gas content of the seventh HEROES galaxy, NGC\,4013, was recently studied according to the same strategy by \cite{zschaechner15}.
\par
The data cube of each galaxy is analysed through a set of detailed tilted-ring models. The primary goal of our modelling is to determine the main properties of the atomic gas disks such as the face-on column density distribution, the kinematics and the 3D geometry in an accurate and homogeneous way. Furthermore we identify and analyse peculiar features such as (potential) spiral arms and signs of recent interaction and separate them from the main gas disk. 
\begin{itemize}[topsep=2pt, itemsep=2pt]
\item[$\bullet$] Edge-on warps of varying strength are found in the outer regions of all galaxies except for NGC\,973. In addition we also identify moderate line of sight warps in IC\,2531, NGC\,5529 and NGC\,5907. Evidence for a flaring gas disk is only found in NGC\,5907, although, for the other galaxies, the uncertainties on the scale height leave room for a flare as well, in agreement with theoretical expectations. 
\item[$\bullet$] The major axis position-velocity diagrams of IC\,2531 and NGC\,5529 show a bright H{\sc{i}} ridge. Our models fail to account for these ridges, although they adequately reproduce the other features in the XV-diagrams of these two galaxies. This suggests that the ridges are not part of an axisymmetric disk and supports the interpretation by \cite{kregel04b} as prominent spiral arms.
\item[$\bullet$] Three galaxies in our sample show evidence of recent or ongoing interactions. We detect a (clumpy) coplanar ring of H{\sc{i}} gas just outside the main disk of NGC\,4217. This structure has not been reported before and seems to be both kinematically and spatially offset from the main disk. We suggest this could be the relic of a recent minor merger event. For NGC\,5907 \cite{shang98} already reported the detection of a companion galaxy close to and a giant stellar tidal stream around the galaxy. Our modelling further reveals that the outer H{\sc{i}} disk is disrupted, with a strong (edge-on and line of sight) warp and large asymmetries between the approaching and the receding side. Finally, as already noted by \cite{kregel04}, H{\sc{i}} bridges connect NGC\,5529 to two nearby companions. Apart from these H{\sc{i}} bridges, we find that the gas disk is further surrounded (in projection) by a significant reservoir of extraplanar gas, mainly on the approaching side. A radial inflow of 15 $\pm$ 5 km s$^{-1}$ in the gas disk is possibly also related to the ongoing gas accretion.
\item[$\bullet$] The effect of these interactions on the gas disks of the galaxies in question is mostly limited to the orientation (inclination and position angle) of the outer parts. The column density profiles and rotation curves behave normally and show no strong asymmetries between the approaching and the receding halves.
\item[$\bullet$] Finally, a small companion galaxy is detected in the data cube of NGC\,973, located at a projected distance of 36 kpc from the center of the main galaxy. We find no optical counterpart of this galaxy.
\end{itemize}
In addition to the atomic gas, we are currently collecting CO line data to constrain the molecular gas content as well. An analysis of the dust and stellar content of the HEROES galaxies by fitting radiative transfer models to optical/NIR images is also underway. The next step in the HEROES project will be to fit dynamical models to the rotation curves derived in this work to determine the distribution of the dark matter. Combining our analyses of all the different components, we will then ultimately investigate the correlations between them. The peculiar features identified in this paper will be an important factor in the interpretation of these results.

\begin{acknowledgements}

We thank the anonymous referee for the useful comments and suggestions that helped to improve this paper. We also wish to thank M. Verheijen and S. Peters for kindly sharing their reduced H{\sc{i}} data of NGC\,4217 and IC\,2531 respectively. F.A., M.B., G.G. and S.V. acknowledge the support of the Flemish Fund for Scientific Research (FWO-Vlaanderen). M.B. and T.M.H. acknowledge financial support from the Belgian Science Policy Office (BELSPO) as part of the PRODEX project C90370 (\textit{Herschel}-PACS Guaranteed Time and Open Time Programs: Science Exploitation). We thank the staff of the GMRT that made these observations possible. GMRT is run by the National Centre for Radio Astrophysics of the Tata Institute of Fundamental Research. The Faulkes Telescopes are maintained and operated by Las Cumbres Observatory Global Telescope Network. We also thank Peter Hill and the staff and students of College Le Monteil ASAM (France), The Thomas Aveling School (Rochester, England), Glebe School (Bromley, England) and St David's Catholic College (Cardiff, Wales). This research has made use of NASA's Astrophysics Data System and the NASA/IPAC Extragalactic Database (NED), which is operated by the Jet Propulsion Laboratory, California Institute of Technology, under contract with the National Aeronautics and Space Administration. Funding for the Sloan Digital Sky Survey (SDSS) has been provided by the Alfred P. Sloan Foundation, the Participating Institutions, the National Aeronautics and Space Administration, the National Science Foundation, the U.S. Department of Energy, the Japanese Monbukagakusho, and the Max Planck Society. The SDSS Web site is http://www.sdss.org/. The SDSS is managed by the Astrophysical Research Consortium (ARC) for the Participating Institutions. The Participating Institutions are The University of Chicago, Fermilab, the Institute for Advanced Study, the Japan Participation Group, The Johns Hopkins University, the Korean Scientist Group, Los Alamos National Laboratory, the Max-Planck-Institute for Astronomy (MPIA), the Max-Planck-Institute for Astrophysics (MPA), New Mexico State University, University of Pittsburgh, University of Portsmouth, Princeton University, the United States Naval Observatory, and the University of Washington.

\end{acknowledgements}

\bibliography{bibliography_HEROES_II}{}

\begin{thebibliography}{102}
\expandafter\ifx\csname natexlab\endcsname\relax\def\natexlab#1{#1}\fi

\bibitem[{{Alatalo} {et~al.}(2011){Alatalo}, {Blitz}, {Young}, {Davis},
  {Bureau}, {Lopez}, {Cappellari}, {Scott}, {Shapiro}, {Crocker},
  {Mart{\'{\i}}n}, {Bois}, {Bournaud}, {Davies}, {de Zeeuw}, {Duc}, {Emsellem},
  {Falc{\'o}n-Barroso}, {Khochfar}, {Krajnovi{\'c}}, {Kuntschner}, {Lablanche},
  {McDermid}, {Morganti}, {Naab}, {Oosterloo}, {Sarzi}, {Serra}, \&
  {Weijmans}}]{alatalo11}
{Alatalo}, K., {Blitz}, L., {Young}, L.~M., {et~al.} 2011, \apj, 735, 88

\bibitem[{{Baes} {et~al.}(2011){Baes}, {Verstappen}, {De Looze}, {Fritz},
  {Saftly}, {Vidal P{\'e}rez}, {Stalevski}, \& {Valcke}}]{baes11}
{Baes}, M., {Verstappen}, J., {De Looze}, I., {et~al.} 2011, \apjs, 196, 22

\bibitem[{{Baillard} {et~al.}(2011){Baillard}, {Bertin}, {de Lapparent},
  {Fouqu{\'e}}, {Arnouts}, {Mellier}, {Pell{\'o}}, {Leborgne}, {Prugniel},
  {Makarov}, {Makarova}, {McCracken}, {Bijaoui}, \& {Tasca}}]{baillard11}
{Baillard}, A., {Bertin}, E., {de Lapparent}, V., {et~al.} 2011, \aap, 532, A74

\bibitem[{{Becquaert} \& {Combes}(1997)}]{becquaert97}
{Becquaert}, J.-F. \& {Combes}, F. 1997, \aap, 325, 41

\bibitem[{{Bianchi}(2007)}]{bianchi07}
{Bianchi}, S. 2007, \aap, 471, 765

\bibitem[{{Bocchio} {et~al.}(2012){Bocchio}, {Micelotta}, {Gautier}, \&
  {Jones}}]{bocchio12}
{Bocchio}, M., {Micelotta}, E.~R., {Gautier}, A.-L., \& {Jones}, A.~P. 2012,
  \aap, 545, A124

\bibitem[{{Braun}(2012)}]{braun12}
{Braun}, R. 2012, \apj, 749, 87

\bibitem[{{Braun} {et~al.}(2009){Braun}, {Thilker}, {Walterbos}, \&
  {Corbelli}}]{braun09}
{Braun}, R., {Thilker}, D.~A., {Walterbos}, R.~A.~M., \& {Corbelli}, E. 2009,
  \apj, 695, 937

\bibitem[{{Broeils} \& {Rhee}(1997)}]{broeils97}
{Broeils}, A.~H. \& {Rhee}, M.-H. 1997, \aap, 324, 877

\bibitem[{{Bureau} \& {Freeman}(1997)}]{bureau97}
{Bureau}, M. \& {Freeman}, K.~C. 1997, \pasa, 14, 146

\bibitem[{{Bureau} \& {Freeman}(1999)}]{bureau99}
{Bureau}, M. \& {Freeman}, K.~C. 1999, \aj, 118, 126

\bibitem[{{Cazaux} \& {Tielens}(2002)}]{cazaux02}
{Cazaux}, S. \& {Tielens}, A.~G.~G.~M. 2002, \apjl, 575, L29

\bibitem[{{Clark}(1980)}]{clark80}
{Clark}, B.~G. 1980, \aap, 89, 377

\bibitem[{{Dalgarno} \& {McCray}(1972)}]{dalgarno72}
{Dalgarno}, A. \& {McCray}, R.~A. 1972, \araa, 10, 375

\bibitem[{{de Blok}(2010)}]{deblok10}
{de Blok}, W.~J.~G. 2010, Advances in Astronomy, 2010, 5

\bibitem[{{de Blok} {et~al.}(2014){de Blok}, {J{\'o}zsa}, {Patterson},
  {Gentile}, {Heald}, {J{\"u}tte}, {Kamphuis}, {Rand}, {Serra}, \&
  {Walterbos}}]{deblok14}
{de Blok}, W.~J.~G., {J{\'o}zsa}, G.~I.~G., {Patterson}, M., {et~al.} 2014,
  \aap, 566, A80

\bibitem[{{De Geyter} {et~al.}(2014){De Geyter}, {Baes}, {Camps}, {Fritz}, {De
  Looze}, {Hughes}, {Viaene}, \& {Gentile}}]{degeyter14}
{De Geyter}, G., {Baes}, M., {Camps}, P., {et~al.} 2014, \mnras, 441, 869

\bibitem[{{De Geyter} {et~al.}(2013){De Geyter}, {Baes}, {Fritz}, \&
  {Camps}}]{degeyter13}
{De Geyter}, G., {Baes}, M., {Fritz}, J., \& {Camps}, P. 2013, \aap, 550, A74

\bibitem[{{De Looze} {et~al.}(2012){De Looze}, {Baes}, {Bendo}, {Ciesla},
  {Cortese}, {de Geyter}, {Groves}, {Boquien}, {Boselli}, {Brondeel}, {Cooray},
  {Eales}, {Fritz}, {Galliano}, {Gentile}, {Gordon}, {Hony}, {Law}, {Madden},
  {Sauvage}, {Smith}, {Spinoglio}, \& {Verstappen}}]{delooze12}
{De Looze}, I., {Baes}, M., {Bendo}, G.~J., {et~al.} 2012, \mnras, 427, 2797

\bibitem[{{Di Teodoro} \& {Fraternali}(2014)}]{diteodoro14}
{Di Teodoro}, E.~M. \& {Fraternali}, F. 2014, \aap, 567, A68

\bibitem[{{Dwek}(1998)}]{dwek98}
{Dwek}, E. 1998, \apj, 501, 643

\bibitem[{{Dwek} \& {Scalo}(1980)}]{dwekscalo80}
{Dwek}, E. \& {Scalo}, J.~M. 1980, \apj, 239, 193

\bibitem[{{Elmegreen} {et~al.}(1992){Elmegreen}, {Elmegreen}, \&
  {Montenegro}}]{elmegreen92}
{Elmegreen}, B.~G., {Elmegreen}, D.~M., \& {Montenegro}, L. 1992, \apjs, 79, 37

\bibitem[{{Famaey} \& {McGaugh}(2012)}]{famaey12}
{Famaey}, B. \& {McGaugh}, S.~S. 2012, Living Reviews in Relativity, 15, 10

\bibitem[{{Feng}(2010)}]{feng10}
{Feng}, J.~L. 2010, \araa, 48, 495

\bibitem[{{Garcia-Burillo} {et~al.}(1997){Garcia-Burillo}, {Guelin}, \&
  {Neininger}}]{garcia-burillo97}
{Garcia-Burillo}, S., {Guelin}, M., \& {Neininger}, N. 1997, \aap, 319, 450

\bibitem[{{Gentile} {et~al.}(2013){Gentile}, {J{\'o}zsa}, {Serra}, {Heald}, {de
  Blok}, {Fraternali}, {Patterson}, {Walterbos}, \& {Oosterloo}}]{gentile13}
{Gentile}, G., {J{\'o}zsa}, G.~I.~G., {Serra}, P., {et~al.} 2013, \aap, 554,
  A125

\bibitem[{{Gibson} {et~al.}(2005){Gibson}, {Taylor}, {Higgs}, {Brunt}, \&
  {Dewdney}}]{gibson05}
{Gibson}, S.~J., {Taylor}, A.~R., {Higgs}, L.~A., {Brunt}, C.~M., \& {Dewdney},
  P.~E. 2005, \apj, 626, 195

\bibitem[{{Gould} \& {Salpeter}(1963)}]{gould63}
{Gould}, R.~J. \& {Salpeter}, E.~E. 1963, \apj, 138, 393

\bibitem[{{Greisen}(2003)}]{greisen03}
{Greisen}, E.~W. 2003, Information Handling in Astronomy - Historical Vistas,
  285, 109

\bibitem[{{Greisen} {et~al.}(2009){Greisen}, {Spekkens}, \& {van
  Moorsel}}]{greisen09}
{Greisen}, E.~W., {Spekkens}, K., \& {van Moorsel}, G.~A. 2009, \aj, 137, 4718

\bibitem[{{Hagen} \& {McClain}(1954)}]{hagen54}
{Hagen}, J.~P. \& {McClain}, E.~F. 1954, \apj, 120, 368

\bibitem[{{H{\"o}gbom}(1974)}]{hogbom74}
{H{\"o}gbom}, J.~A. 1974, \aaps, 15, 417

\bibitem[{{Hollenbach} \& {Salpeter}(1971)}]{hollenbach71}
{Hollenbach}, D. \& {Salpeter}, E.~E. 1971, \apj, 163, 155

\bibitem[{{Holwerda} {et~al.}(2011){Holwerda}, {Pirzkal}, {de Blok},
  {Bouchard}, {Blyth}, \& {van der Heyden}}]{holwerda11}
{Holwerda}, B.~W., {Pirzkal}, N., {de Blok}, W.~J.~G., {et~al.} 2011, \mnras,
  416, 2437

\bibitem[{{Huchtmeier} \& {Richter}(1989)}]{huchtmeier89}
{Huchtmeier}, W.~K. \& {Richter}, O.-G. 1989, {A General Catalog of HI
  Observations of Galaxies. The Reference Catalog.}

\bibitem[{{Hughes} {et~al.}(2014){Hughes}, {Baes}, {Fritz}, {Smith}, {Parkin},
  {Gentile}, {Bendo}, {Wilson}, {Allaert}, {Bianchi}, {De Looze}, {Verstappen},
  {Viaene}, {Boquien}, {Boselli}, {Clements}, {Davies}, {Galametz}, {Madden},
  {R{\'e}my-Ruyer}, \& {Spinoglio}}]{hughes14}
{Hughes}, T.~M., {Baes}, M., {Fritz}, J., {et~al.} 2014, \aap, 565, A4

\bibitem[{{Irwin} {et~al.}(2007){Irwin}, {Kennedy}, {Parkin}, \&
  {Madden}}]{irwin07}
{Irwin}, J.~A., {Kennedy}, H., {Parkin}, T., \& {Madden}, S. 2007, \aap, 474,
  461

\bibitem[{{Jones}(2004)}]{jones04}
{Jones}, A.~P. 2004, in Astronomical Society of the Pacific Conference Series,
  Vol. 309, Astrophysics of Dust, ed. A.~N. {Witt}, G.~C. {Clayton}, \& B.~T.
  {Draine}, 347

\bibitem[{{Jones} \& {Nuth}(2011)}]{jones11}
{Jones}, A.~P. \& {Nuth}, J.~A. 2011, \aap, 530, A44

\bibitem[{{J{\'o}zsa} {et~al.}(2007){J{\'o}zsa}, {Kenn}, {Klein}, \&
  {Oosterloo}}]{jozsa07}
{J{\'o}zsa}, G.~I.~G., {Kenn}, F., {Klein}, U., \& {Oosterloo}, T.~A. 2007,
  \aap, 468, 731

\bibitem[{{Kalberla} \& {Dedes}(2008)}]{kalberla08}
{Kalberla}, P.~M.~W. \& {Dedes}, L. 2008, \aap, 487, 951

\bibitem[{{Kamphuis} \& {Briggs}(1992)}]{kamphuis92}
{Kamphuis}, J. \& {Briggs}, F. 1992, \aap, 253, 335

\bibitem[{{Kamphuis} {et~al.}(2013){Kamphuis}, {Rand}, {J{\'o}zsa},
  {Zschaechner}, {Heald}, {Patterson}, {Gentile}, {Walterbos}, {Serra}, \& {de
  Blok}}]{kamphuis13}
{Kamphuis}, P., {Rand}, R.~J., {J{\'o}zsa}, G.~I.~G., {et~al.} 2013, \mnras,
  434, 2069

\bibitem[{{Kenney}(1994)}]{kenney94}
{Kenney}, J.~D.~P. 1994, in Mass-Transfer Induced Activity in Galaxies, ed.
  I.~{Shlosman}, 78

\bibitem[{{Kere{\v s}} {et~al.}(2009{\natexlab{a}}){Kere{\v s}}, {Katz},
  {Dav{\'e}}, {Fardal}, \& {Weinberg}}]{keres09b}
{Kere{\v s}}, D., {Katz}, N., {Dav{\'e}}, R., {Fardal}, M., \& {Weinberg},
  D.~H. 2009{\natexlab{a}}, \mnras, 396, 2332

\bibitem[{{Kere{\v s}} {et~al.}(2009{\natexlab{b}}){Kere{\v s}}, {Katz},
  {Fardal}, {Dav{\'e}}, \& {Weinberg}}]{keres09a}
{Kere{\v s}}, D., {Katz}, N., {Fardal}, M., {Dav{\'e}}, R., \& {Weinberg},
  D.~H. 2009{\natexlab{b}}, \mnras, 395, 160

\bibitem[{{Koribalski} {et~al.}(2004){Koribalski}, {Staveley-Smith}, {Kilborn},
  {Ryder}, {Kraan-Korteweg}, {Ryan-Weber}, {Ekers}, {Jerjen}, {Henning},
  {Putman}, {Zwaan}, {de Blok}, {Calabretta}, {Disney}, {Minchin}, {Bhathal},
  {Boyce}, {Drinkwater}, {Freeman}, {Gibson}, {Green}, {Haynes}, {Juraszek},
  {Kesteven}, {Knezek}, {Mader}, {Marquarding}, {Meyer}, {Mould}, {Oosterloo},
  {O'Brien}, {Price}, {Sadler}, {Schr{\"o}der}, {Stewart}, {Stootman}, {Waugh},
  {Warren}, {Webster}, \& {Wright}}]{koribalski04}
{Koribalski}, B.~S., {Staveley-Smith}, L., {Kilborn}, V.~A., {et~al.} 2004,
  \aj, 128, 16

\bibitem[{{Kormendy} \& {Kennicutt}(2004)}]{kormendy04}
{Kormendy}, J. \& {Kennicutt}, Jr., R.~C. 2004, \araa, 42, 603

\bibitem[{{Kozasa} {et~al.}(1991){Kozasa}, {Hasegawa}, \& {Nomoto}}]{kozasa91}
{Kozasa}, T., {Hasegawa}, H., \& {Nomoto}, K. 1991, \aap, 249, 474

\bibitem[{{Kregel} \& {van der Kruit}(2004)}]{kregel04b}
{Kregel}, M. \& {van der Kruit}, P.~C. 2004, \mnras, 352, 787

\bibitem[{{Kregel} {et~al.}(2004){Kregel}, {van der Kruit}, \& {de
  Blok}}]{kregel04}
{Kregel}, M., {van der Kruit}, P.~C., \& {de Blok}, W.~J.~G. 2004, \mnras, 352,
  768

\bibitem[{{Krumholz}(2012)}]{krumholz12}
{Krumholz}, M.~R. 2012, \apj, 759, 9

\bibitem[{{Lagos} {et~al.}(2013){Lagos}, {Lacey}, \& {Baugh}}]{lagos13}
{Lagos}, C.~d.~P., {Lacey}, C.~G., \& {Baugh}, C.~M. 2013, \mnras, 436, 1787

\bibitem[{{Leroy} {et~al.}(2008){Leroy}, {Walter}, {Brinks}, {Bigiel}, {de
  Blok}, {Madore}, \& {Thornley}}]{leroy08}
{Leroy}, A.~K., {Walter}, F., {Brinks}, E., {et~al.} 2008, \aj, 136, 2782

\bibitem[{{Lucy}(1974)}]{lucy74}
{Lucy}, L.~B. 1974, \aj, 79, 745

\bibitem[{{L{\"u}tticke} {et~al.}(2000){L{\"u}tticke}, {Dettmar}, \&
  {Pohlen}}]{lutticke00}
{L{\"u}tticke}, R., {Dettmar}, R.-J., \& {Pohlen}, M. 2000, \aaps, 145, 405

\bibitem[{{Mart{\'{\i}}nez-Delgado} {et~al.}(2008){Mart{\'{\i}}nez-Delgado},
  {Pe{\~n}arrubia}, {Gabany}, {Trujillo}, {Majewski}, \& {Pohlen}}]{martinez08}
{Mart{\'{\i}}nez-Delgado}, D., {Pe{\~n}arrubia}, J., {Gabany}, R.~J., {et~al.}
  2008, \apj, 689, 184

\bibitem[{{Mathewson} \& {Ford}(1996)}]{mathewson96}
{Mathewson}, D.~S. \& {Ford}, V.~L. 1996, \apjs, 107, 97

\bibitem[{{Matteucci}(2012)}]{matteucci12}
{Matteucci}, F. 2012, {Chemical Evolution of Galaxies}

\bibitem[{{Micelotta} {et~al.}(2010){Micelotta}, {Jones}, \&
  {Tielens}}]{micelotti10}
{Micelotta}, E.~R., {Jones}, A.~P., \& {Tielens}, A.~G.~G.~M. 2010, \aap, 510,
  A37

\bibitem[{{Milgrom}(1983)}]{milgrom83}
{Milgrom}, M. 1983, \apj, 270, 365

\bibitem[{{Moster} {et~al.}(2010){Moster}, {Macci{\`o}}, {Somerville},
  {Johansson}, \& {Naab}}]{moster10}
{Moster}, B.~P., {Macci{\`o}}, A.~V., {Somerville}, R.~S., {Johansson}, P.~H.,
  \& {Naab}, T. 2010, \mnras, 403, 1009

\bibitem[{{O'Brien} {et~al.}(2010{\natexlab{a}}){O'Brien}, {Freeman}, \& {van
  der Kruit}}]{obrien10c}
{O'Brien}, J.~C., {Freeman}, K.~C., \& {van der Kruit}, P.~C.
  2010{\natexlab{a}}, \aap, 515, A62

\bibitem[{{O'Brien} {et~al.}(2010{\natexlab{b}}){O'Brien}, {Freeman}, \& {van
  der Kruit}}]{obrien10b}
{O'Brien}, J.~C., {Freeman}, K.~C., \& {van der Kruit}, P.~C.
  2010{\natexlab{b}}, \aap, 515, A61

\bibitem[{{O'Brien} {et~al.}(2010{\natexlab{c}}){O'Brien}, {Freeman}, \& {van
  der Kruit}}]{obrien10d}
{O'Brien}, J.~C., {Freeman}, K.~C., \& {van der Kruit}, P.~C.
  2010{\natexlab{c}}, \aap, 515, A63

\bibitem[{{O'Brien} {et~al.}(2010{\natexlab{d}}){O'Brien}, {Freeman}, {van der
  Kruit}, \& {Bosma}}]{obrien10a}
{O'Brien}, J.~C., {Freeman}, K.~C., {van der Kruit}, P.~C., \& {Bosma}, A.
  2010{\natexlab{d}}, \aap, 515, A60

\bibitem[{{Ocvirk} {et~al.}(2008){Ocvirk}, {Pichon}, \& {Teyssier}}]{ocvirk08}
{Ocvirk}, P., {Pichon}, C., \& {Teyssier}, R. 2008, \mnras, 390, 1326

\bibitem[{{Olling}(1995)}]{olling95}
{Olling}, R.~P. 1995, \aj, 110, 591

\bibitem[{{Olling}(1996)}]{olling96}
{Olling}, R.~P. 1996, \aj, 112, 457

\bibitem[{{Ostriker} \& {Silk}(1973)}]{ostriker73}
{Ostriker}, J. \& {Silk}, J. 1973, \apjl, 184, L113

\bibitem[{{Patsis} \& {Xilouris}(2006)}]{patsis06}
{Patsis}, P.~A. \& {Xilouris}, E.~M. 2006, \mnras, 366, 1121

\bibitem[{{Peters} {et~al.}(2013){Peters}, {van der Kruit}, {Allen}, \&
  {Freeman}}]{peters13}
{Peters}, S.~P.~C., {van der Kruit}, P.~C., {Allen}, R.~J., \& {Freeman}, K.~C.
  2013, ArXiv e-prints

\bibitem[{{Pilbratt} {et~al.}(2010){Pilbratt}, {Riedinger}, {Passvogel},
  {Crone}, {Doyle}, {Gageur}, {Heras}, {Jewell}, {Metcalfe}, {Ott}, \&
  {Schmidt}}]{pilbratt10}
{Pilbratt}, G.~L., {Riedinger}, J.~R., {Passvogel}, T., {et~al.} 2010, \aap,
  518, L1

\bibitem[{{Radhakrishnan}(1960)}]{radhak60}
{Radhakrishnan}, V. 1960, \pasp, 72, 296

\bibitem[{{Sahni}(2005)}]{sahni05}
{Sahni}, V. 2005, in Lecture Notes in Physics, Berlin Springer Verlag, Vol.
  653, The Physics of the Early Universe, ed. K.~{Tamvakis}, 141

\bibitem[{{Sancisi}(1976)}]{sancisi76}
{Sancisi}, R. 1976, \aap, 53, 159

\bibitem[{{Sancisi}(1999)}]{sancisi99}
{Sancisi}, R. 1999, in IAU Symposium, Vol. 186, Galaxy Interactions at Low and
  High Redshift, ed. J.~E. {Barnes} \& D.~B. {Sanders}, 71

\bibitem[{{Sancisi} {et~al.}(2008){Sancisi}, {Fraternali}, {Oosterloo}, \& {van
  der Hulst}}]{sancisi08}
{Sancisi}, R., {Fraternali}, F., {Oosterloo}, T., \& {van der Hulst}, T. 2008,
  \aapr, 15, 189

\bibitem[{{Sault} {et~al.}(1995){Sault}, {Teuben}, \& {Wright}}]{sault95}
{Sault}, R.~J., {Teuben}, P.~J., \& {Wright}, M.~C.~H. 1995, in Astronomical
  Society of the Pacific Conference Series, Vol.~77, Astronomical Data Analysis
  Software and Systems IV, ed. R.~A. {Shaw}, H.~E. {Payne}, \& J.~J.~E.
  {Hayes}, 433

\bibitem[{{Schechtman-Rook} \& {Bershady}(2013)}]{schechtman13}
{Schechtman-Rook}, A. \& {Bershady}, M.~A. 2013, \apj, 773, 45

\bibitem[{{Schwab}(1984)}]{schwab84}
{Schwab}, F.~R. 1984, \aj, 89, 1076

\bibitem[{{Shang} {et~al.}(1998){Shang}, {Zheng}, {Brinks}, {Chen}, {Burstein},
  {Su}, {Byun}, {Deng}, {Deng}, {Fan}, {Jiang}, {Li}, {Lin}, {Ma}, {Sun},
  {Wills}, {Windhorst}, {Wu}, {Xia}, {Xu}, {Xue}, {Yan}, {Zhou}, {Zhu}, \&
  {Zou}}]{shang98}
{Shang}, Z., {Zheng}, Z., {Brinks}, E., {et~al.} 1998, \apjl, 504, L23

\bibitem[{{Sharma} \& {Nath}(2013)}]{sharma13}
{Sharma}, M. \& {Nath}, B.~B. 2013, \apj, 763, 17

\bibitem[{{Spitzer}(1942)}]{spitzer42}
{Spitzer}, Jr., L. 1942, \apj, 95, 329

\bibitem[{{Springob} {et~al.}(2005){Springob}, {Haynes}, {Giovanelli}, \&
  {Kent}}]{springob05}
{Springob}, C.~M., {Haynes}, M.~P., {Giovanelli}, R., \& {Kent}, B.~R. 2005,
  \apjs, 160, 149

\bibitem[{{Steer} {et~al.}(1984){Steer}, {Dewdney}, \& {Ito}}]{steer84}
{Steer}, D.~G., {Dewdney}, P.~E., \& {Ito}, M.~R. 1984, \aap, 137, 159

\bibitem[{{Trimble}(1987)}]{trimble87}
{Trimble}, V. 1987, \araa, 25, 425

\bibitem[{{van der Hulst} {et~al.}(1992){van der Hulst}, {Terlouw}, {Begeman},
  {Zwitser}, \& {Roelfsema}}]{vanderhulst92}
{van der Hulst}, J.~M., {Terlouw}, J.~P., {Begeman}, K.~G., {Zwitser}, W., \&
  {Roelfsema}, P.~R. 1992, in Astronomical Society of the Pacific Conference
  Series, Vol.~25, Astronomical Data Analysis Software and Systems I, ed. D.~M.
  {Worrall}, C.~{Biemesderfer}, \& J.~{Barnes}, 131

\bibitem[{{van der Kruit}(1981)}]{vanderkruit81}
{van der Kruit}, P.~C. 1981, \aap, 99, 298

\bibitem[{{Verheijen}(2001)}]{verheijen01b}
{Verheijen}, M.~A.~W. 2001, \apj, 563, 694

\bibitem[{{Verheijen} \& {Sancisi}(2001)}]{verheijen01}
{Verheijen}, M.~A.~W. \& {Sancisi}, R. 2001, \aap, 370, 765

\bibitem[{{Verstappen} {et~al.}(2013){Verstappen}, {Fritz}, {Baes}, {Smith},
  {Allaert}, {Bianchi}, {Blommaert}, {De Geyter}, {De Looze}, {Gentile},
  {Gordon}, {Holwerda}, {Viaene}, \& {Xilouris}}]{verstappen13}
{Verstappen}, J., {Fritz}, J., {Baes}, M., {et~al.} 2013, \aap, 556, A54

\bibitem[{{Warmels}(1988)}]{warmels88}
{Warmels}, R.~H. 1988, \aaps, 72, 427

\bibitem[{{Xilouris} {et~al.}(1999){Xilouris}, {Byun}, {Kylafis}, {Paleologou},
  \& {Papamastorakis}}]{xilouris99}
{Xilouris}, E.~M., {Byun}, Y.~I., {Kylafis}, N.~D., {Paleologou}, E.~V., \&
  {Papamastorakis}, J. 1999, \aap, 344, 868

\bibitem[{{Xilouris} {et~al.}(1997){Xilouris}, {Kylafis}, {Papamastorakis},
  {Paleologou}, \& {Haerendel}}]{xilouris97}
{Xilouris}, E.~M., {Kylafis}, N.~D., {Papamastorakis}, J., {Paleologou}, E.~V.,
  \& {Haerendel}, G. 1997, \aap, 325, 135

\bibitem[{{Yun} {et~al.}(1994){Yun}, {Ho}, \& {Lo}}]{yun94}
{Yun}, M.~S., {Ho}, P.~T.~P., \& {Lo}, K.~Y. 1994, \nat, 372, 530

\bibitem[{{Zhukovska}(2014)}]{zhukovska14}
{Zhukovska}, S. 2014, \aap, 562, A76

\bibitem[{{Zhukovska} \& {Henning}(2013)}]{zhukovska13}
{Zhukovska}, S. \& {Henning}, T. 2013, \aap, 555, A99

\bibitem[{{Zschaechner} \& {Rand}(2015)}]{zschaechner15}
{Zschaechner}, L.~K. \& {Rand}, R.~J. 2015, ArXiv e-prints

\bibitem[{{Zschaechner} {et~al.}(2012){Zschaechner}, {Rand}, {Heald},
  {Gentile}, \& {J{\'o}zsa}}]{zschaechner12}
{Zschaechner}, L.~K., {Rand}, R.~J., {Heald}, G.~H., {Gentile}, G., \&
  {J{\'o}zsa}, G. 2012, \apj, 760, 37

\bibitem[{{Zwicky}(1937)}]{zwicky37}
{Zwicky}, F. 1937, \apj, 86, 217

\end{thebibliography}
\bibliographystyle{aa}

\begin{appendix}

\section{Comparison of the final models to the observed data cubes}

%
%
\begin{figure*}[]
\centering
   \includegraphics[width=\textwidth]{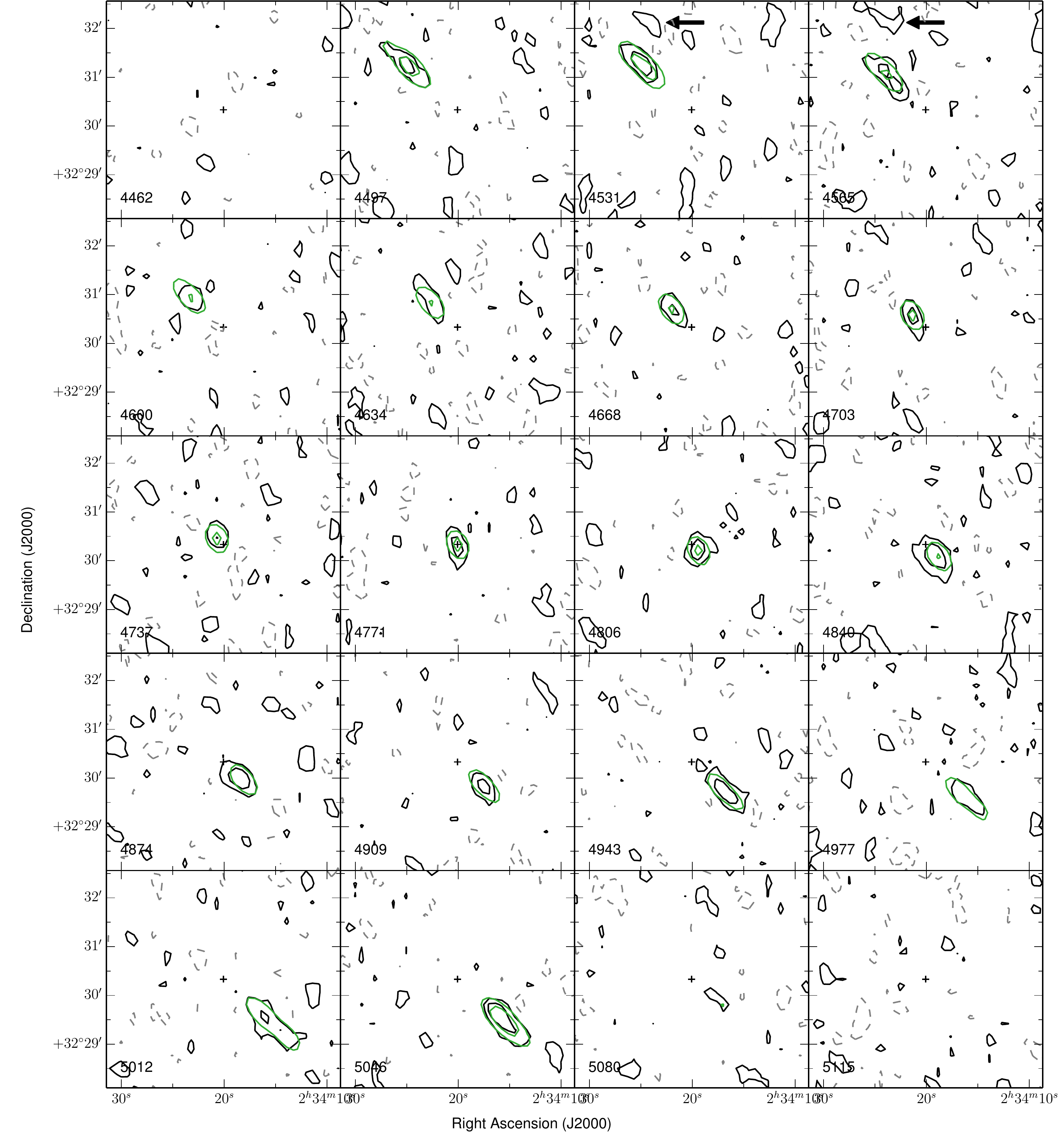}
     \caption{Channel maps from the observed data cube of NGC\,973 (black and gray contours) and our final (\emph{B}) model (green). Contour levels are -2.3, 2.3 (1.5$\sigma$) and 6.2 mJy beam$^{-1}$. The black cross indicates the center of the galaxy. The companion galaxy is indicated by arrows at 4531 and 4565 km s$^{-1}$.}
     \label{fig:N973_final_channels}
\end{figure*}
%
%
\begin{figure*}[]
\centering
   \includegraphics[width=\textwidth]{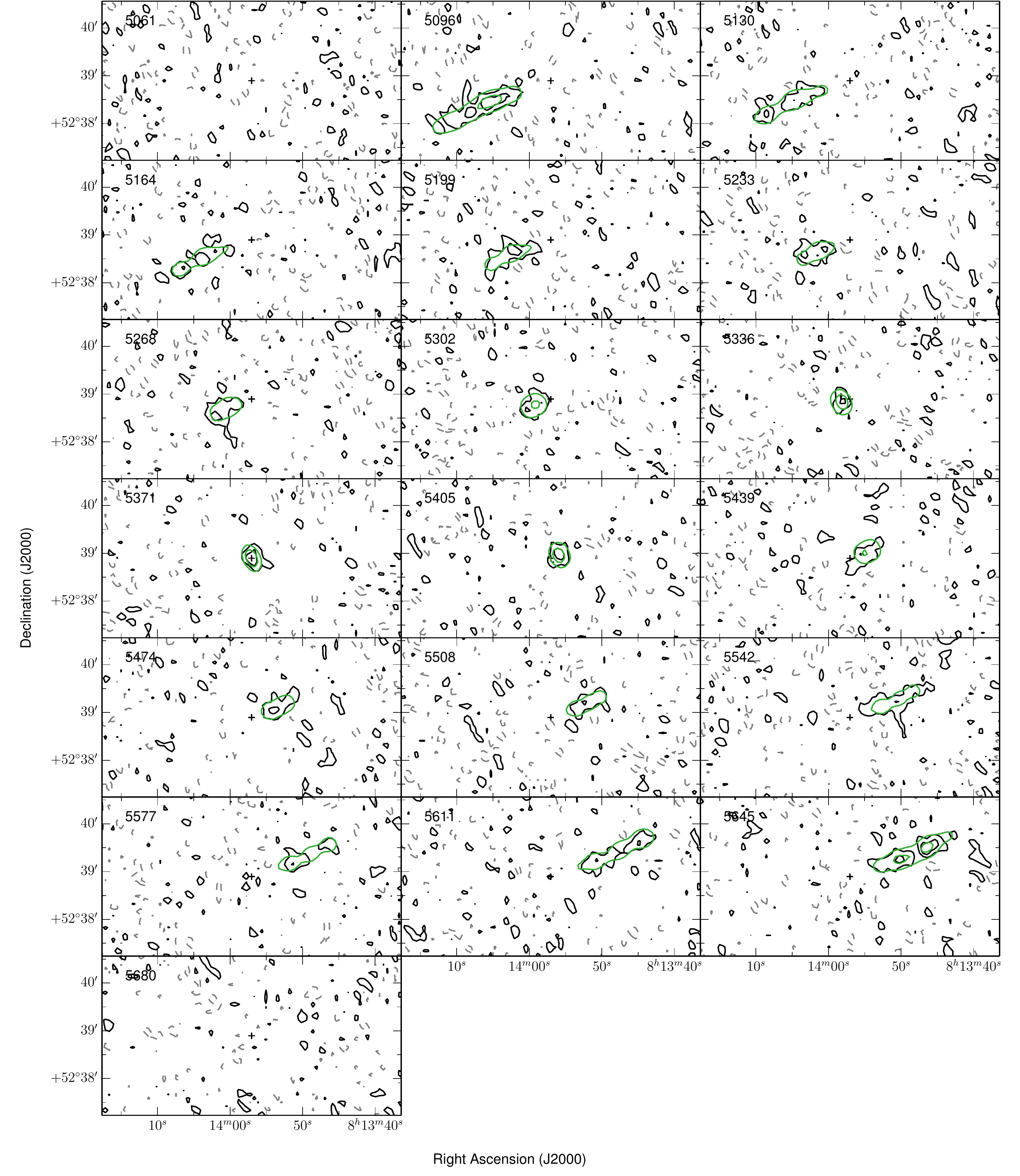}
     \caption{Channel maps from the observed data cube of UGC\,4277 (black and gray contours) and our final (\emph{B}) model (green). Contour levels are -1.5, 1.5 (1.5$\sigma$) and 4.0 mJy beam$^{-1}$. The black cross indicates the center of the galaxy.}
     \label{fig:U4277_final_channels}
\end{figure*}
%
%
\begin{figure*}[]
\centering
   \includegraphics[width=\textwidth]{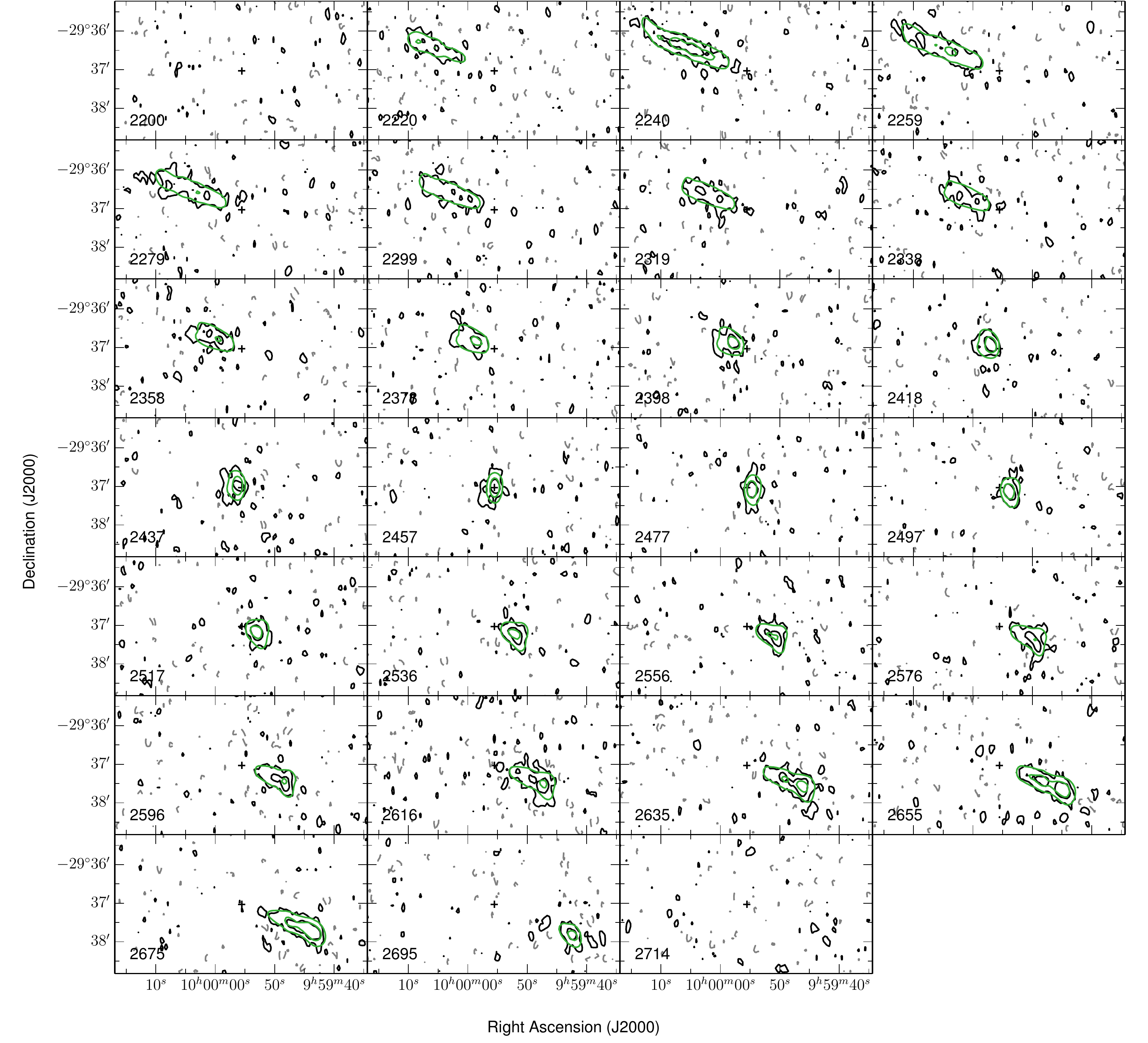}
     \caption{Channel maps from the observed data cube of IC\,2531 (black and gray contours) and our final (\emph{W}) model (green). Contour levels are -1.6, 1.6 (1.5$\sigma$) and 6.5 mJy beam$^{-1}$. The black cross indicates the center of the galaxy.}
     \label{fig:IC2531_final_channels}
\end{figure*}
%
%
\begin{figure*}
\centering
   \includegraphics[width=\textwidth]{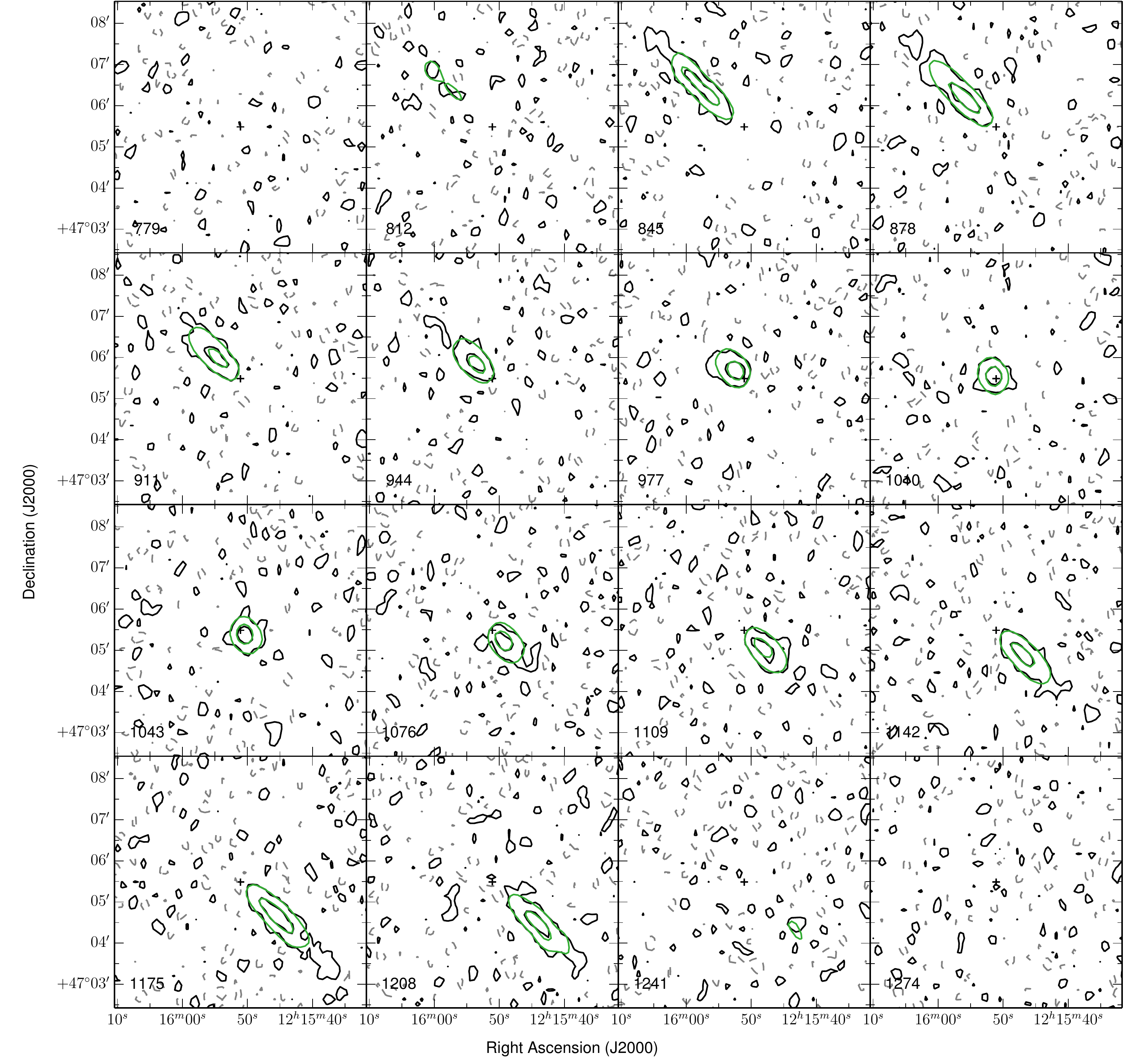}
     \caption{Channel maps from the observed data cube of NGC\,4217 (black and gray contours) and our final (\emph{B}) model of the main disk (green). Contour levels are -1.5, 1.5 (1.5$\sigma$) and 8.0 mJy beam$^{-1}$. The black cross indicates the center of the galaxy.}
     \label{fig:N4217_final_channels}
\end{figure*}
%
%
\begin{figure*}
\centering
   \includegraphics[width=\textwidth]{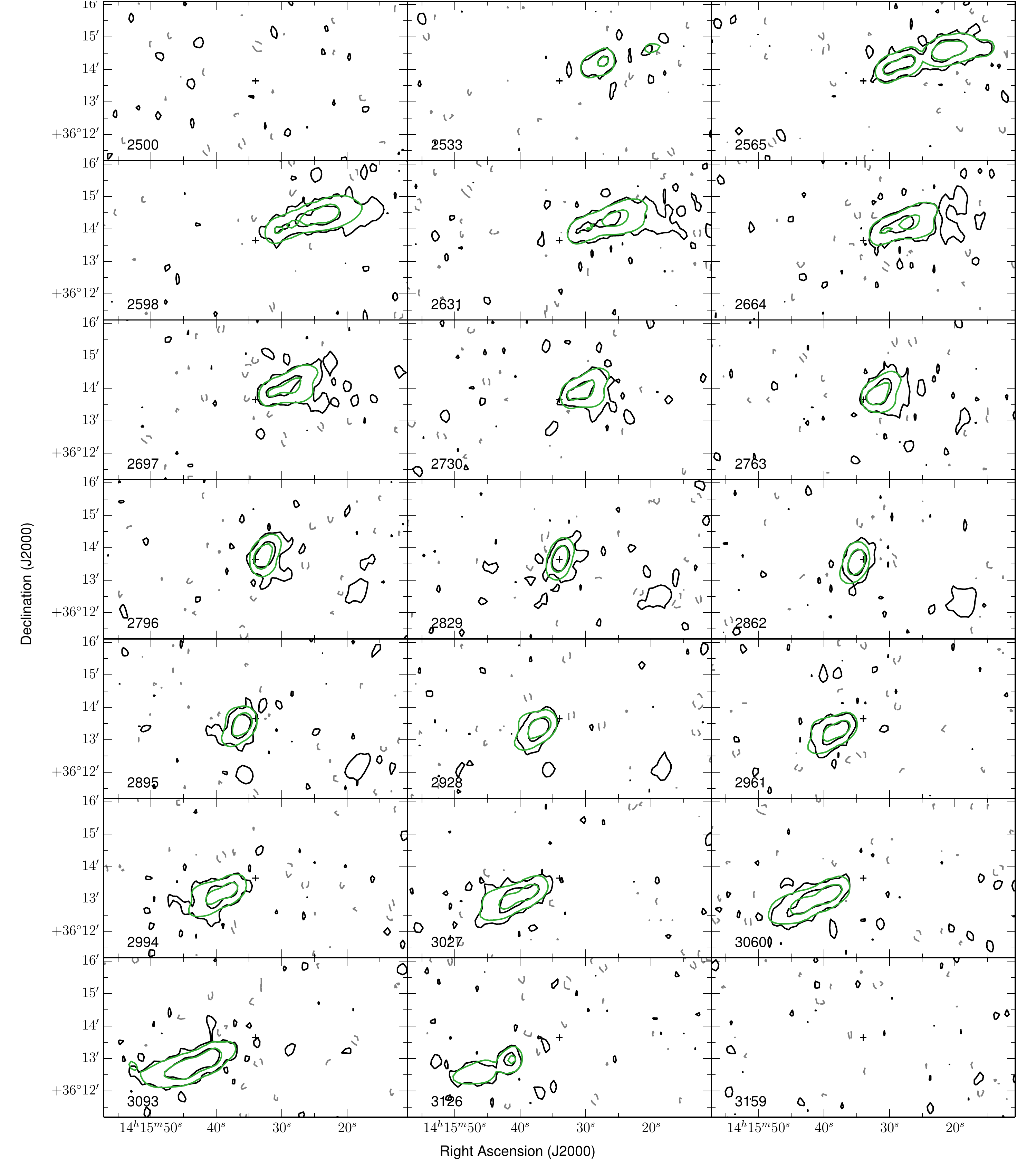}
     \caption{Channel maps from the observed data cube of NGC\,5529 (black and gray contours) and our final (\emph{W2R}) model (green). Contour levels are -0.8, 0.8 (1.5$\sigma$) and 5.6 mJy beam$^{-1}$. The black cross indicates the center of the galaxy.}
     \label{fig:N5529_final_channels}
\end{figure*}
%
%
\begin{figure*}
\centering
   \includegraphics[width=\textwidth]{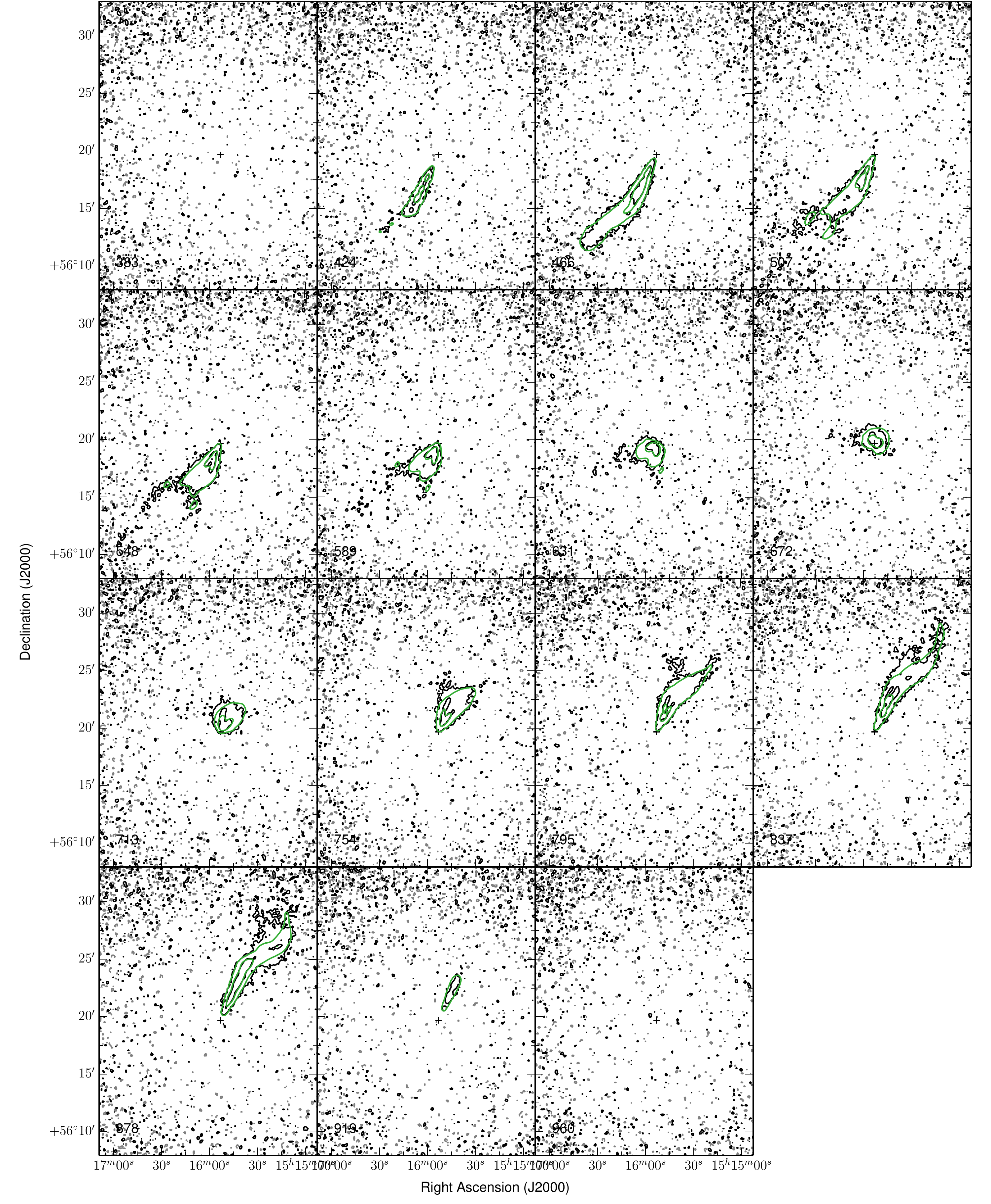}
     \caption{Channel maps from the observed data cube of NGC\,5907 (black and gray contours) and our final (\emph{WF3}) model (green). Contour levels are -0.6, 0.6 (for better readability we take the lowest contour at 2$\sigma$ instead of 1.5$\sigma$) and 6.4 mJy beam$^{-1}$. The black cross indicates the center of the galaxy.}
     \label{fig:N5907_final_channels}
\end{figure*}

\end{appendix}

\end{document}